%% file: main.tex
\documentclass[preprint,12pt]{elsarticle}

\usepackage{amssymb}
\usepackage{amsmath}
\usepackage{paralist}

\usepackage{multirow}

\usepackage[utf8]{inputenc}

\usepackage{lscape}
\usepackage{longtable}
\usepackage[table]{xcolor}

\usepackage{pdflscape}  
\usepackage{xltabular}  
\usepackage{booktabs}  
\usepackage{makecell}  
\usepackage{array}     
\usepackage{geometry}
\journal{Journal of Affective Disorders}

\begin{document}

\begin{frontmatter}



\title{Acoustic and Machine Learning Methods for Speech-Based Suicide Risk Assessment: A Systematic Review} %% Article title
\label{title1}

\author[label1,label2]{Ambre Marie}\corref{cor1}
\cortext[cor1]{Corresponding author:}
\ead{ambre.marie@univ-brest.fr}

\author[label3]{Marine Garnier}
\author[label1,label2]{Thomas Bertin}
\author[label1,label2]{Laura Machart}
\author[label1]{Guillaume Dardenne}
\author[label1]{Gwenolé Quellec}
\author[label1,label3]{Sofian Berrouiguet}

\affiliation[label1]{organization={LaTIM UMR 1101, Inserm},
            %addressline={},
            city={Brest},
            %postcode={},
            %state={},
            country={France}}

\affiliation[label2]{organization={University of Western Brittany},
            %addressline={},
            city={Brest},
            %postcode={},
            %state={},
            country={France}}

\affiliation[label3]{organization={Psychiatry Department, CHU Brest},
            %addressline={},
            city={Brest},
            %postcode={},
            %state={},
            country={France}}

%% Abstract
\begin{abstract}
\label{abstract2}
Suicide remains a public health challenge, necessitating improved detection methods to facilitate timely intervention and treatment. This systematic review evaluates the role of Artificial Intelligence (AI) and Machine Learning (ML) in assessing suicide risk through acoustic analysis of speech. \\
Following PRISMA guidelines, we analyzed 33 articles selected from PubMed, Cochrane, Scopus, and Web of Science databases. The last search was conducted in February 2025. Risk of bias was assessed using the PROBAST tool. Studies analyzing acoustic features between individuals at risk of suicide (RS) and those not at risk (NRS) were included, while studies lacking acoustic data, a suicide-related focus, or sufficient methodological details were excluded. Sample sizes varied widely and were reported in terms of participants or speech segments, depending on the study. Results were synthesized narratively based on acoustic features and classifier performance. \\
Findings consistently showed significant acoustic feature variations between RS and NRS populations, particularly involving jitter, fundamental frequency (F0), Mel-frequency cepstral coefficients (MFCC), and power spectral density (PSD). Classifier performance varied based on algorithms, modalities, and speech elicitation methods, with multimodal approaches integrating acoustic, linguistic, and metadata features demonstrating superior performance. Among the 29 classifier-based studies, reported AUC values ranged from 0.62 to 0.985 and accuracies from 60\% to 99.85\%. Most datasets were imbalanced in favor of NRS, and performance metrics were rarely reported separately by group, limiting clear identification of direction of effect.\\
However, findings were limited by sample size, class imbalance, methodological inconsistency, and population bias. Future research should address these issues to improve clinical validity and generalizability.
This review was not registered. It was funded by the Brittany Region (France) through the doctoral program "Allocations de Recherche Doctorale" (ARED).
\end{abstract}


%\item Acoustic analysis combined with machine learning effectively distinguishes suicide risk groups.
%\item Jitter, fundamental frequency (F0), MFCC, and PSD are key acoustic biomarkers linked to suicide risk.
%\item Multimodal approaches integrating acoustic, linguistic, and metadata features outperform unimodal methods.
%\item Methodological variability, small datasets, and limited demographic diversity restrict current study generalizability.
%\item Future research should standardize protocols and use larger, diverse samples for clinical applicability.

%% Keywords
\begin{keyword}
Suicide \sep Speech \sep Machine Learning \sep Artificial Intelligence \sep Vocal Features \sep Classification


\end{keyword}

\end{frontmatter}



\section*{List of Abbreviations}

\noindent \textbf{AI}: Artificial Intelligence \\
\textbf{ANN}: Artificial Neural Network \\
\textbf{AUC}: Area Under the Receiver-Operating Characteristic curve\\
\textbf{CF}: Center Frequencies \\
\textbf{DNN}: Deep Neural Network \\
\textbf{F0}: Fundamental Frequency \\
\textbf{F1, F2, F3}: Formant Frequencies \\
\textbf{GB}: Gradient Boosting \\
\textbf{GRBASI}: Grade, Roughness, Breathiness, Asthenia, Strain, Instability\\
\textbf{HMM}: Hidden Markov Model \\
\textbf{HNRdB}: Harmonics-to-Noise Ratio \\
\textbf{IR}: Imbalance Ratio \\
\textbf{LLM}: Large Language Model \\
\textbf{MDD}: Major Depressive Disorder \\
\textbf{MFCC}: Mel-Frequency Cepstral Coefficients \\
\textbf{ML}: Machine Learning \\
\textbf{NAQ}: Normalized Amplitude Quotient \\
\textbf{NPV}: Negative Predictive Value \\
\textbf{NRS}: Not at Risk of Suicide \\
\textbf{OQ}: Open Quotient \\
\textbf{PPV}: Positive Predictive Value \\
\textbf{PRISMA}: Preferred Reporting Items for Systematic Reviews and Meta-Analyses \\
\textbf{PSD}: Power Spectral Density \\
\textbf{QOQ}: Quasi-Open Quotient \\
\textbf{RF}: Random Forest \\
\textbf{RS}: At Risk of Suicide \\
\textbf{SVM}: Support Vector Machine \\
\textbf{SW1}: SpeechWellness Challenge 1 \\
\textbf{XGB}: eXtreme Gradient Boosting \\

\section{Introduction}
\label{introduction3}

Suicide remains a major public health concern in many societies, with prevention efforts failing to significantly reduce suicide rates in most countries \citep{Overholser2022Suicide}. It represents one of the main challenges for mental health professionals and public health authorities. According to the 2025 report by the World Health Organization, "suicide occurs throughout the lifespan and was the third leading cause of death among 15-29-year-olds globally in 2021" \citep{WHO2025Suicide}. Current methods for assessing suicide risk primarily rely on rigorous clinical evaluations, which depend on the clinician's experience and the consideration of multiple factors such as the patient's personal and family history, sociodemographic characteristics, and anamnestic data. These assessments are largely based on self-reported information, leaving patients free to disclose or withhold their suicidal thoughts \citep{bolton2015suicide}.

To enhance suicide risk assessment, the development of objective, sensitive, and reproducible tools is essential \citep{cummins2015review}. The emergence of Artificial Intelligence (AI) and, more specifically, Machine Learning (ML) models in the health field has opened new possibilities for improving such assessments. ML models, which rely on statistical, mathematical, and computational techniques, enable the development of predictive algorithms with minimal human intervention by leveraging large-scale datasets \citep{Sidey-Gibbons2019Machine}. These models are particularly effective in classification tasks and are increasingly applied in mental health research, especially in the automated risk assessment of psychiatric disorders such as depression \citep{Koops2021Speech}. AI-driven tools could potentially assist in suicide risk assessment by identifying specific patterns among at-risk individuals, particularly through the analysis of bio-signals or acoustic features of speech.

Depressive and suicidal states are often associated with significant affective disturbances, leading to physiological and cognitive changes that directly influence vocal production and speech characteristics. These disturbances alter vocal output through complex neuromuscular and cognitive coordination, resulting in measurable modifications in acoustic features \citep{cummins2015review}. A study by \citet{Koops2021Speech} compared audio recordings of depressed patients and healthy controls, revealing significant differences in some acoustic features between the two groups. Depressed individuals were found to exhibit a lower speech rate, less pitch variability and lower pitch, increased jitter and shimmer, and a lower harmonics-to-noise ratio (HNRdB). These features are respectively associated with psychomotor slowing, monotony and reduced expressiveness, instability in vocal fold vibration, and reduced vocal clarity.
Various studies have explored the potential of ML models in identifying depression through acoustic, linguistic, and facial expression features \citep{cummins2015review,Cohn2009Detecting,Zhang2019Evaluating}. Similarly, AI-based tools incorporating ML techniques to analyze acoustic features could provide clinicians with additional, reliable, and reproducible support in assessing suicidal risk (RS).

Previous reviews have explored the use of acoustic features in mental health assessment, including suicide risk. \citet{cummins2015review} provided an overview of speech-based markers of affective disorders (e.g., suicidal ideation and depression), and included studies up to 2014. More recently, \citet{Iyer2022Detection} conducted a systematic review addressing suicide risk detection through acoustic analysis, covering literature up to early 2022.

Building on previous work, the present review extends the temporal coverage up to 2025 and applies an expanded set of search terms to capture a broader range of studies. It provides a comprehensive synthesis of recent advances in acoustic-based suicide risk assessment, including the types of vocal features explored, the performance of ML classifiers, and the contribution of multimodal approaches.

Accordingly, this systematic review addresses three research questions:
\label{introduction4}

\begin{enumerate}
\item What acoustic features have been identified as potential indicators of suicide risk in individuals, and how well do they differentiate between those at risk and those not at risk? 
\item How effective are machine learning classifiers based on acoustic features in assessing and predicting suicide risk, and what methodological factors influence their performance? 
\item What is the added value of multimodal approaches - combining acoustic data with other modalities such as linguistic, visual, behavioral, or demographic features - in improving the predictive performance of suicide risk assessment? \end{enumerate}

\section{Methods}

\subsection{Research strategy}
\label{methods6}
This systematic literature review was conducted following PRISMA guidelines and included two main searches covering publications from 2000 to 2025. The first search was performed using bibliographic databases PubMed, Cochrane, Scopus, and Web-of-Science. To reduce publication bias, grey literature was identified through The Grey Literature Report. Initial searches used the keywords: [suicid*] AND ([vocal] OR [acoustic*] OR [prosod*]).

An updated search was conducted using PubMed, Cochrane, Scopus, and Web-of-Science with an expanded keyword set:

\begin{quote}
[suicid*] AND ([speech analysis] OR [voice features] OR [prosody] OR [acoustic markers] OR [speech biometrics] OR [voice quality] OR [phonation] OR [vocal characteristics])
\end{quote}

References from identified articles were manually screened, resulting in the inclusion of three additional studies. 

An additional exploratory search was conducted on Consensus, with the exact query: 
\begin{quote}"What machine learning classifiers or acoustic parameters have been used to assess suicide risk in patients based on voice analysis?"
\end{quote}
The search retrieved six items: four overlapped with existing database searches, leading to the inclusion of two additional unique studies after manual screening of titles and abstracts.

Full search queries and interface details are provided in \ref{appendix:searchstrat}. All searches across databases and exploratory platforms were last conducted in February 2025.

\subsection{Selection of studies}
\label{methods8}
Studies selected addressed the analysis of acoustic features in RS individuals or evaluated classifiers for RS assessment. In this review, RS was defined in a general way to include suicidal ideation, previous suicide attempts, psychometric scale indications, or clinical evaluations of suicide risk. \\
Table~\ref{tab:eligibility} summarizes the inclusion and exclusion criteria used in this review. 
In particular, studies limited to verbal analysis were excluded, as well as those using Deep Neural Networks (DNN) trained directly on raw waveform data without intermediate representation (e.g. spectrogram, acoustic embeddings or transcriptions). Literature reviews and theoretical papers without clinical populations were also excluded. Eligible study designs included experimental, observational, analytical or descriptive studies, and conference papers or abstracts, if they provided sufficient methodological detail: population characteristics, features extracted, classification models used, and performance metrics. For instance, one included abstract (\citet{Yünden2024Examination}) met these criteria by reporting on feature types, classification tasks, model architecture, and quantitative results using standard evaluation metrics.
Studies were grouped for synthesis based on common analytical dimensions such as acoustic features analyzed, statistical comparisons, classifier types, training setups, and performance metrics.
\label{methods5}

\input{table-eligibility}

\subsection{Data extraction and analysis} 
\label{methods9}
Both searches followed a three-step screening process. After removing duplicates, titles were reviewed, followed by abstracts. Finally, potentially eligible full texts were examined to exclude irrelevant studies. Bibliographic references were managed and de-duplicated using Zotero throughout the process.

Marine Garnier (MG) and Sofian Berrouiguet initially designed and conducted the search strategy in November 2021, with the primary objective of assessing the scope and relevance of existing literature on acoustic markers of suicide risk. This first phase served as the conceptual and methodological blueprint for the review. In February 2025, Ambre Marie (AM) retained that framework, enriched it with additional keywords, and applied it systematically across all databases to capture studies published between 2000 and 2025.

Data extraction and screening were conducted manually by one reviewer (AM) without the use of automation tools.
Although parallel extraction by two independent reviewers is the recommended standard, the division of tasks reflects practical constraints and the research team's collective expertise. Any uncertainties or borderline decisions were resolved through team consensus, involving Thomas Bertin, Laura Machart, Guillaume Dardenne, Gwenolé Quellec, and Sofian Berrouiguet, according to PRISMA guidance on dispute resolution in study selection.

Whenever available in the articles, the following information was systematically extracted for each study using a structured spreadsheet template.
\begin{compactitem}
    \item Publication metadata: authors, year, country, institution
    \item Study population characteristics: sample size, age group, gender, clinical group, language
    \item Study methodology: design type (observational, analytical), setting (emergency department, helpline,...), subpopulations (veterans, depressed patients), monocentric/multicentric, prospective/retrospective/transversal
    \item Speech data details: recording conditions, tasks, equipment, speech segment duration
    \item Acoustic features analyzed
    \item Classification models used
    \item Evaluation protocols: validation method, type of analysis, performance metrics. \\All reported Area Under the Receiver-Operating Characteristic curve (AUC) values were systematically extracted. Other performance metrics - including accuracy, specificity, positive predictive value (PPV), negative predictive value (NPV), recall, F1-score - were not collected exhaustively due to substantial heterogeneity in reporting across studies, but were used narratively when relevant to contextualize findings.
    \item Classification modality: unimodal, bimodal, multimodal
    \item Study objectives
\end{compactitem}
Fields not explicitly reported in the original articles were marked as not provided to maintain transparency.

\subsection{Synthesis methods}
\label{methods11}
To facilitate synthesis and comparison across studies, limited data preparation steps were applied. When standard deviation values were not reported but could be derived from available information (e.g., means and sample sizes), they were calculated manually. In such cases, Cohen's \emph{d} effect sizes were also computed to estimate the magnitude of group differences in acoustic features. Additionally, a replication score was calculated for each acoustic feature present in Table~\ref{tab:statcomparisons}, indicating how many studies analyzed that feature. No further data conversions or statistical imputations were performed. All missing information was explicitly marked as not provided ("NP" or "-"). 
Risk of bias and applicability were assessed using the PROBAST tool \citep{Chen2020Introduction}, which is designed to evaluate prediction model studies. The full assessment was conducted manually by one reviewer (AM). No automation tools were used. Results of the PROBAST evaluation are reported in \ref{appendix:probast}.\\
Given the heterogeneity of study populations, acoustic features, and classification approaches, no meta-analysis was performed. However, AUC measures were systematically extracted to allow descriptive comparisons across studies. Data were synthesized narratively and through structured tables based on common analytical dimensions: acoustic features analyzed, statistical comparisons, classifier types, training setups, and performance metrics. Groupings were designed to facilitate cross-study comparisons while preserving methodological diversity. To explore potential causes of heterogeneity, we examined class imbalance between Not at Risk of Suicide (NRS) and RS groups (Figure~\ref{fig:classimbalances}).\\
In addition to tabular synthesis, visualizations were generated to illustrate data distribution across studies. These figures were created using Python (matplotlib, numpy, scipy, pandas) based on the data extraction template.
To further explore potential publication bias, an exploratory funnel plot of accuracy values was generated (\ref{appendix:funnel}). Although no formal recall analyses or assessments of certainty were conducted, robustness of the findings was supported by statistical significance (p-values), effect sizes (Cohen’s \emph{d}), and replication scores. Inclusion of grey literature also aimed to minimize risk of reporting bias.

\section{Results}
\label{results}

Several acoustic features were investigated across the reviewed studies. To facilitate comprehension and comparison, \ref{appendix:acoustic-definition} describes these features, grouped into relevant categories including energy, formants (F1,F2,F3), glottal, or prosodic features, among others.

\subsection{Flow chart}
\label{results16a}
The initial search identified 791 references via database queries, supplemented by 3 manually identified references, and 6 from Consensus. After removing 135 duplicates, 665 unique references remained. Following title screening, 119 articles were selected, while 546 were excluded. After abstract review, 52 articles were retained, with 67 excluded. Finally, 33 articles were included after full-text reading, with 19 additional exclusions. 

\begin{figure*}[ht!]
\centering
\caption{Flowchart summarizing the study selection process}
\includegraphics[width=1\columnwidth]{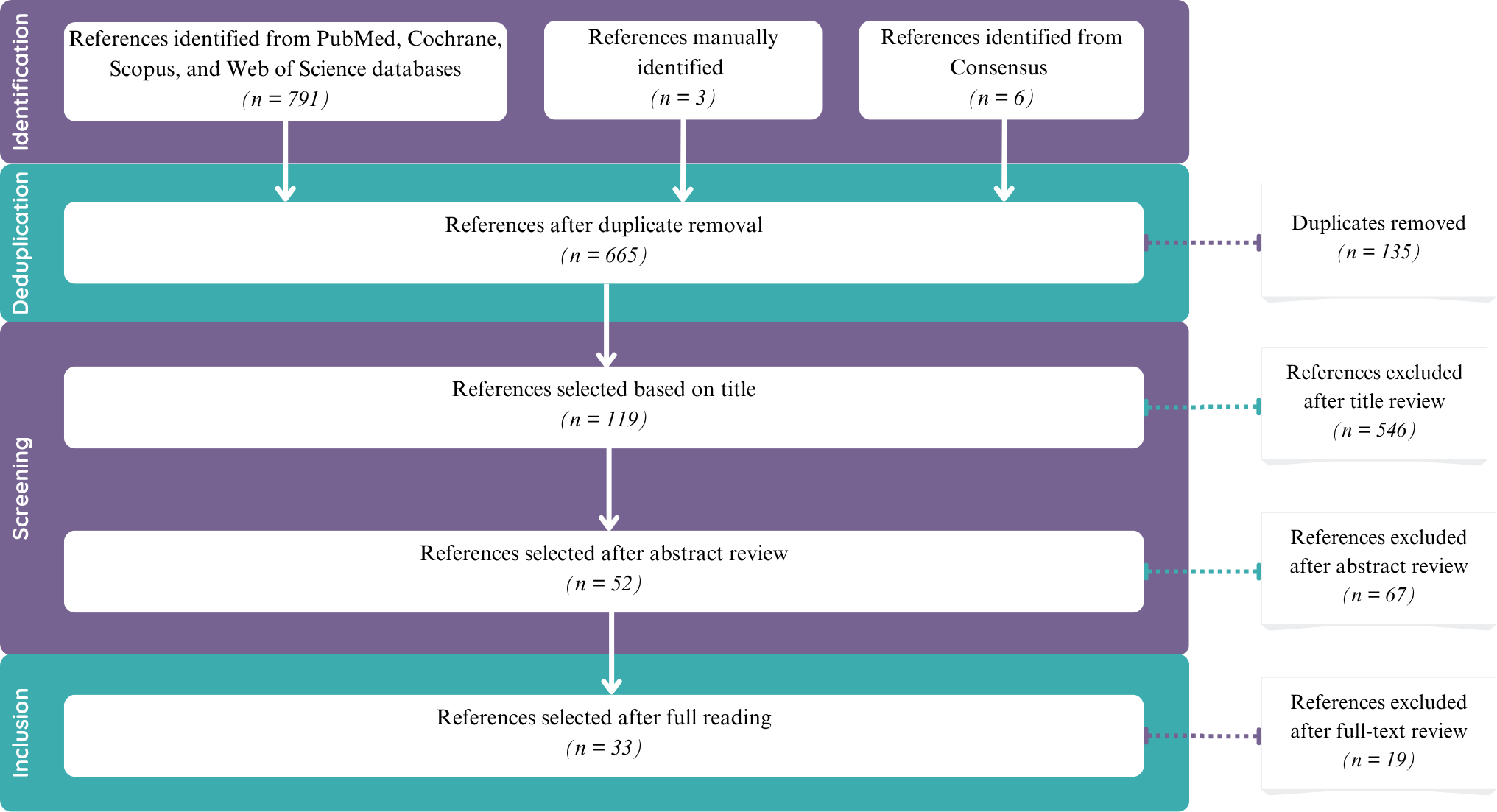}
\label{fig:flowchart}
\end{figure*}

These search results are detailed in Figure~\ref{fig:flowchart}. 

\subsection{Years of publication, authors and countries of origin}

\begin{figure*}[ht!]
\centering
\caption{Publication years of the included studies}
\includegraphics[width=1\columnwidth]{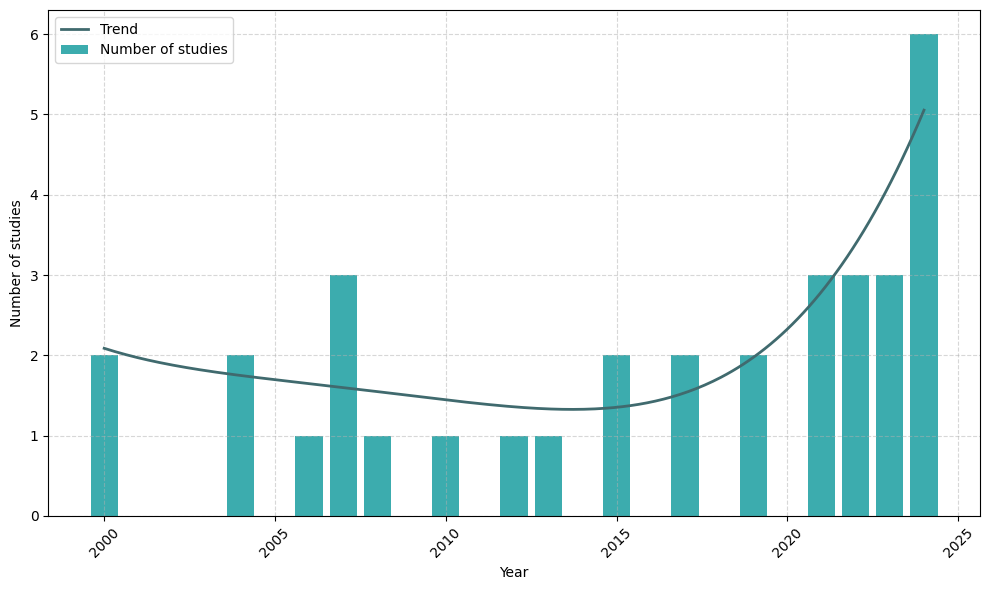}
\label{fig:years}
\end{figure*}

The studies were published between 2000 and 2024. An increasing trend in recent years, shown in Figure~\ref{fig:years}, highlights the growing focus on suicide risk assessment through acoustic analysis.

The included studies were conducted across multiple countries, with the majority originating from the United States (69.70\%; 23/33). Other notable contributors included China (9.09\%; 3/33), Australia, Chile, Germany (each 6.06\%; 2/33), and several other countries, represented by one study. 

\begin{figure*}[ht!]
\centering
\caption{Countries and universities of origin of the included studies}
\includegraphics[width=1\columnwidth]{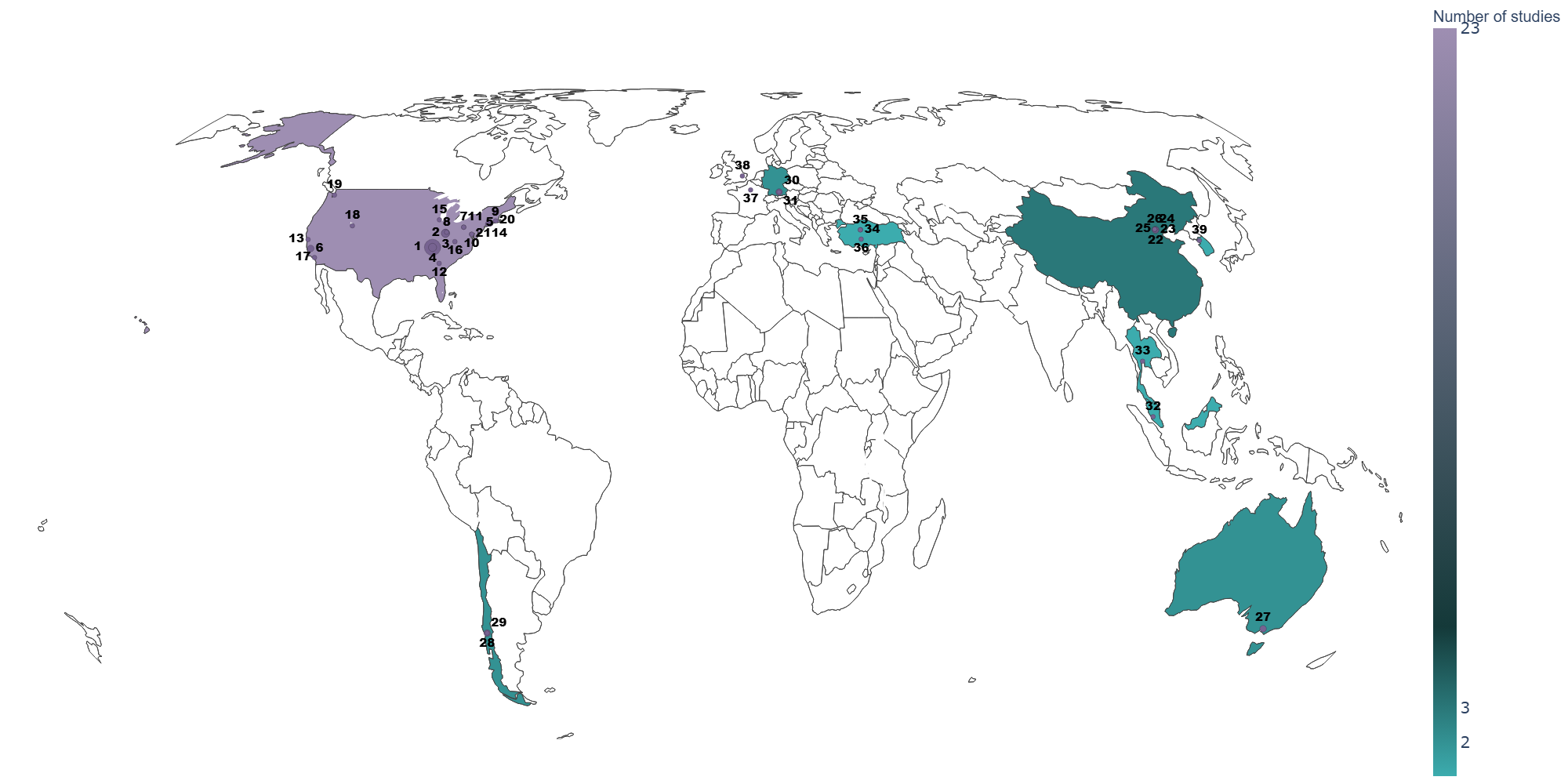}
\scriptsize \\ 1: Vanderbilt University, 2: Cincinnati Children’s Hospital Medical Center, 3: University of Cincinnati, 4: Vanderbilt Psychiatric Hospital, 5: Yale University School of Medicine, 6: University of California, 7: Carnegie Mellon University, 8: Clarigent Health, 9: Dartmouth College, 10: Georgetown University Medical Center, 11: Language Technologies Institute, 12: Mercer University School of Medicine, 13: Modality.AI, 14: Norwalk Hospital, 15: Northwestern University, 16: Princeton Community Hospital, 17: University of Southern California, 18: University of Utah, 19: University of Washington, 20: Warren Alpert Medical School of Brown University, 21: Washington DC VA Medical Center, 22: Beijing University of Technology, 23: WHO Collaborating Center for Research and Training in Suicide Prevention, 24: Beijing Suicide Research and Prevention Center, 25: Peking University, 26: Tsinghua University, 27: Swinburne University of Technology, 28: Universidad Autónoma de Chile, 29: Universidad de La Frontera, 30: University of Augsburg, 31: District Hospital Augsburg, 32: International Islamic University Malaysia, 33: King Mongkut's University of Technology Thonburi, 34: Acıbadem Ankara Hospital, 35: Başkent University, 36: Necmettin Erbakan University, 37: Sorbonne Université, 38: Cambridge University, 39: Seoul National University Hospital
\label{fig:map}
\end{figure*}

At the institutional level, the most prolific contributors were Vanderbilt University (36.36\%; 12/33), followed by the University of Cincinnati, Cincinnati Children’s Hospital Medical Center, and Vanderbilt Psychiatric Hospital (each 9.09\%; 3/33).
A geographical overview of countries and contributing institutions is shown in Figure~\ref{fig:map}.

Most of the articles focused on English-speaking patients (72.73\%; 24/33). Studies involving Mandarin-speaking patients (9.09\%; 3/33), Spanish- and German-speaking patients (6.06\%; 2/33 each), and Korean- and Turkish-speaking patients (3.03\%; 1/33 each) were also included.

\subsection{Sample sizes, study design, and sample characteristics}
\label{results17}
\begin{figure*}[ht!]
\centering
\caption{Distribution of study designs}
\includegraphics[width=1\columnwidth]{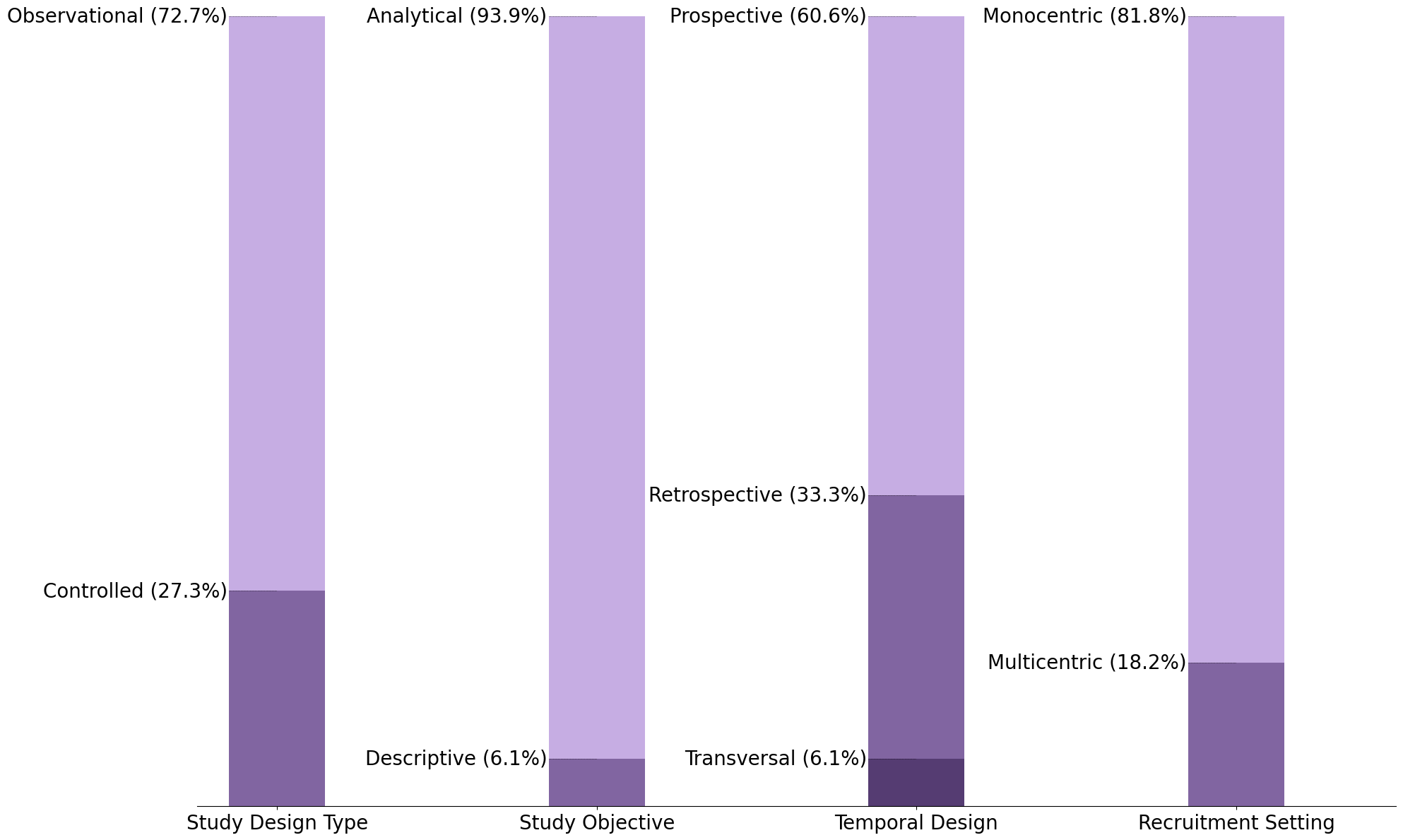}
\label{fig:methodologies}
\end{figure*}

The majority of studies included in this review were observational, analytical, and monocentric (57.57\%, 19/33), with a balanced distribution between retrospective (52.63\%, 10/19) and prospective (47.37\%, 9/19) designs. Controlled prospective studies employing analytical methods were less frequent, with 18.18\% (6/33) conducted in a monocentric setting and 9.09\% (3/33) in a multicentric setting. Detailed methodological characteristics are illustrated in Figure~\ref{fig:methodologies}.

Sample sizes ranged from 16 to 1,179 participants, with a median of 66 (mean 153).
With regard to population characteristics, 66.67\% (22/33) of studies included both men and women, whereas 21.21\% (7/33) focused exclusively on men and 3.03\% (1/33) on women. 

The most frequently studied age group was 25-65 years, included in 27.27\% (9/33) of studies, highlighting a predominant focus on adults. Overall, adults represented 54.55\% (18/33) of the studied populations, while adolescents were included in 15.15\% (5/33) of studies. A substantial proportion of studies (33.33\%, 11/33) did not specify an age category.

\begin{figure*}[ht!]
\centering
\caption{Distribution of study population characteristics}
\includegraphics[width=1\columnwidth]{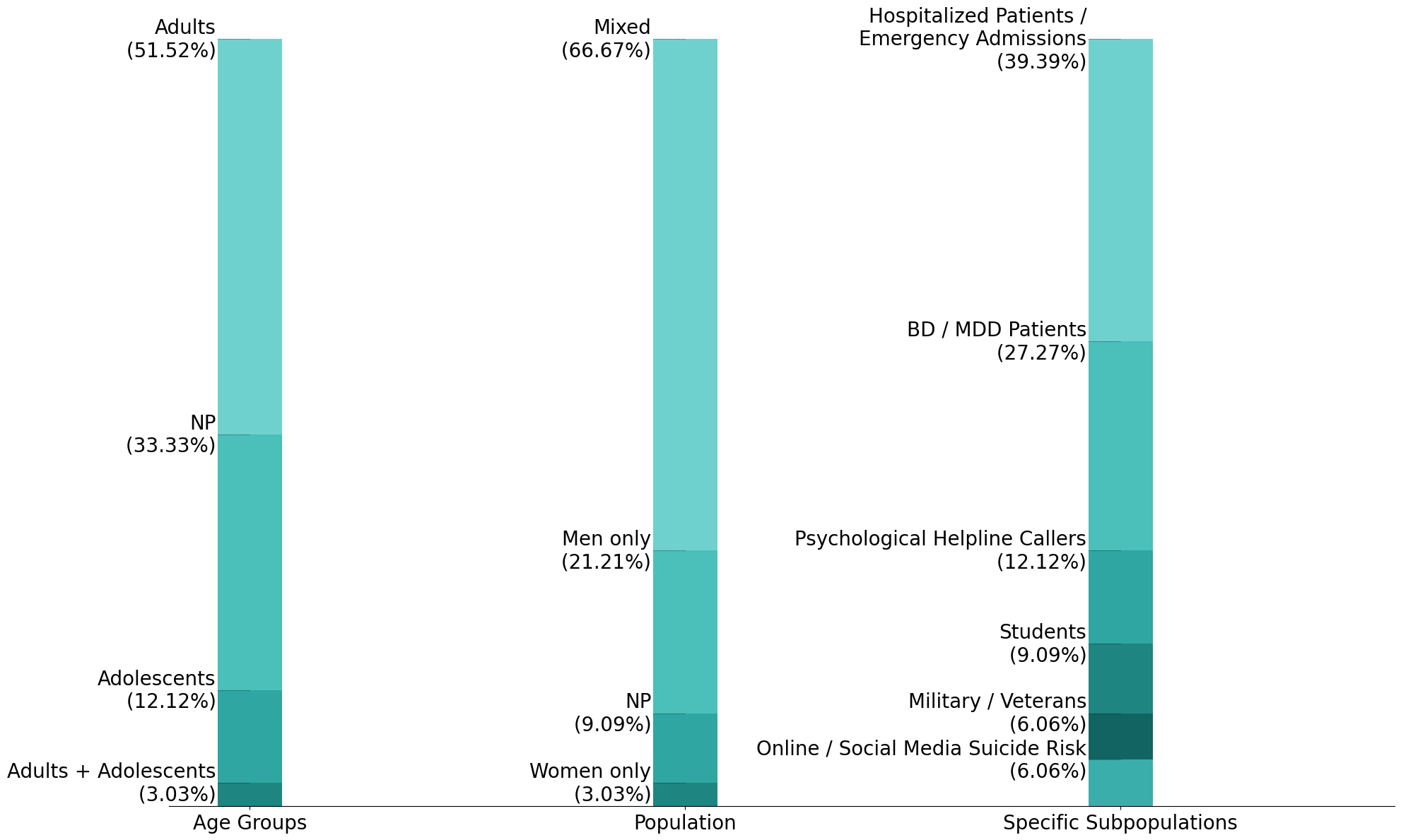}
\scriptsize \\ BD : Bipolar Disorder, MDD : Major Depressive Disorder, NP : Not provided
\label{fig:popchar}
\end{figure*}

Several studies targeted specific subpopulations. Hospitalized patients or those admitted to emergency services were the most represented (39.39\%, 13/33), along with individuals diagnosed with depression, bipolar disorder, or Major Depressive Disorder (MDD) (27.27\%, 9/33). 
Figure~\ref{fig:popchar} displays the distribution of age group representation, population type, and specific subpopulations across the included studies.

A comprehensive summary of study characteristics, sample sizes, and methodologies is provided in \ref{appendix:summary}.

As summarised in \ref{appendix:probast}, PROBAST showed high risk of bias in the Analysis domain for 25 of the 33 included studies (76\%), in Participants for 18 (55\%), and in Outcome for 11 (33\%), whereas the Predictors domain was almost always low risk and free of applicability concerns.

\subsection{Class imbalance between RS and NRS groups}
\label{results20a}
\begin{figure*}[ht!]
\centering
\caption{Class imbalance across the 24 datasets reporting both RS and NRS counts}
\includegraphics[width=1\columnwidth]{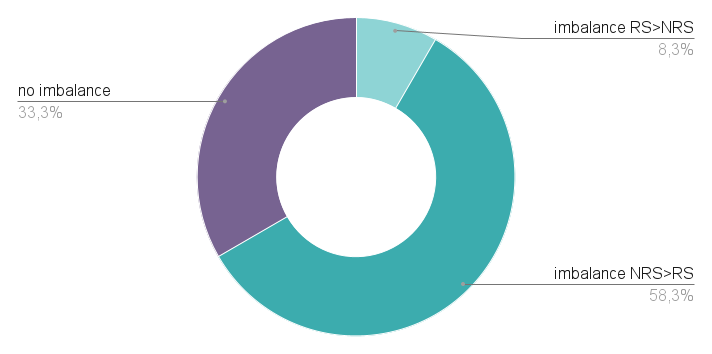}
\label{fig:classimbalances}
\scriptsize{RS: At Risk of Suicide, NRS: Not at Risk of Suicide \\}
\scriptsize{"no imbalance" is defined as IR $\in$ [0.67, 1.5], "imbalance RS $>$ NRS" as IR $>$ 1.5, and "imbalance NRS $>$ RS" as IR $<$ 0.67}
\end{figure*}

In studies reporting both RS and NRS counts, the imbalance ratio (IR) was defined as the ratio of the larger group to the smaller: IR = RS/NRS if RS > NRS, and IR = NRS/RS otherwise.
66.7\% (16/24) displayed a marked class imbalance, most frequently with a surplus of NRS samples (Figure \ref{fig:classimbalances}). 
Only eight datasets were close to balanced (IR $\in$ [0.67, 1.5]). 
Detailed counts for every study are provided in Table~\ref{tab:imbalance}.

\input{table-classimbalances}

\subsection{Objectives of the studies}

The majority of studies (84.85\%, 28/33) focused on evaluating the ability of statistical or ML classifiers to distinguish between RS and NRS individuals and assign them to the appropriate group.

A smaller subset (15.15\%, 5/33) instead aimed to identify statistically significant differences in acoustic features between RS and NRS individuals without employing classification models. These studies sought to highlight relevant acoustic features that could serve as potential indicators of suicide risk.

Beyond these primary objectives, several studies included secondary objectives. Statistical hypothesis testing to assess significant differences in acoustic features between RS and NRS individuals was conducted in 72.73\% (24/33) of studies. Additionally, 15.15\% (5/33) combined acoustic features with additional metadata, including demographics, psychological history, or physiological signals such as Electroencephalography. Real-time or longitudinal monitoring methods for suicide risk detection were explored in 12.12\% (4/33) of studies.

A detailed breakdown of study objectives is provided in \ref{appendix:summary}.

\subsection{Studies using a statistical test}
\label{results19}
Several studies identified significant acoustic differences between RS and NRS individuals. Highly significant differences (p$<$0.001) were observed in spectral and energy-related features such as Delta MFCC11 \citep{Belouali2021Acoustic}, entropy, F1, F2, F3 characteristics, loudness, and spectral slope \citep{iyer2022using, Pillai2023Investigating, Gerczuk2024Exploring}. Moderately significant differences (p$<$0.01 or p$<$0.05) emerged in jitter, fundamental frequency (F0), Normalized Amplitude Quotient (NAQ), Open Quotient (OQ), Quasi-Open Quotient (QOQ), psycho-acoustic sharpness, and MFCCs \citep{Saavedra2021Association, Venek2017Adolescent, Stasak2021Read, figueroa2024comparison}.

Gender-specific patterns were reported, with RS women displaying lower F0 and higher jitter variability, indicative of reduced vocal expressiveness, while RS men showed spectral shifts associated with vocal agitation and lower HNRdB \citep{Saavedra2021Association, figueroa2024comparison, Gerczuk2024Exploring}. Temporal analyses suggest RS individuals have prolonged voiced intervals and speech transitions, reflecting slower speaking rates \citep{Wahidah2015TimingPO}. Despite consistent findings across several studies, generalization of these acoustic markers remains challenging across different datasets \citep{Pillai2023Investigating}.

A detailed examination of statistical comparisons in Table~\ref{tab:statcomparisons} highlights the most relevant acoustic features differentiating RS and NRS individuals. Only acoustic features meeting at least one of the following criteria were retained : (i) an absolute effect size (Cohen's d) greater than 0.5, indicating a moderate or large difference between groups, or (ii) evidence of replication, defined as being reported as significant in at least two independent studies, based on shared acoustic categories (see Table~\ref{tab:acoustic-definition}). Additionally, all retained features had statistically significant group differences (p $\leq$ 0.05). 

\input{table-statcomparisons}

\subsection{Classifiers used}
Twenty-nine studies used classifiers. Sixteen studies (55.17\%) tested a single type of classifier, nine studies (31.03\%) tested two types, one study (3.45\%) tested three types, two studies (6.90\%) tested five types, and one study (3.45\%) tested six types.

The most frequently used classifier was the Support Vector Machine (SVM), appearing in 13 studies (44.83\%). Other commonly tested classifiers included Quadratic Discriminant Analysis in six studies (20.69\%), Gaussian Mixture Model, Linear Discriminant Analysis, Random Forest (RF) and eXtreme Gradient Boosting (XGB), each in four studies (13.79\%). Several other classifiers were each used in three or fewer studies.

Adaboost, Artificial Neural Networks (ANN), LightGBM, transformer-based models, VGGish, and Hidden Markov Models (HMM) were each tested in only one study. While some models may have been limited by the small sample size constraints of most studies, the other models -such as transformer-based approaches- may have been underutilized due to their relatively recent emergence in speech-related tasks and the computational costs associated with their deployment (Table~\ref{tab:classifiers}).

Cross-validation was applied in 82.76\% of studies using a classifier. The most common method was Leave-One-Out Cross-Validation, used in 14 studies (48.28\%). K-Fold Cross-Validation was used in 31.03\% of the validations (9/29), while Holdout Validation was employed in four studies (13.79\%). Only one study (3.45\%) applied Nested Cross-Validation.

\begin{figure*}[ht!]
\centering
\caption{Classifiers used}
\centering
   \includegraphics[width=\linewidth]{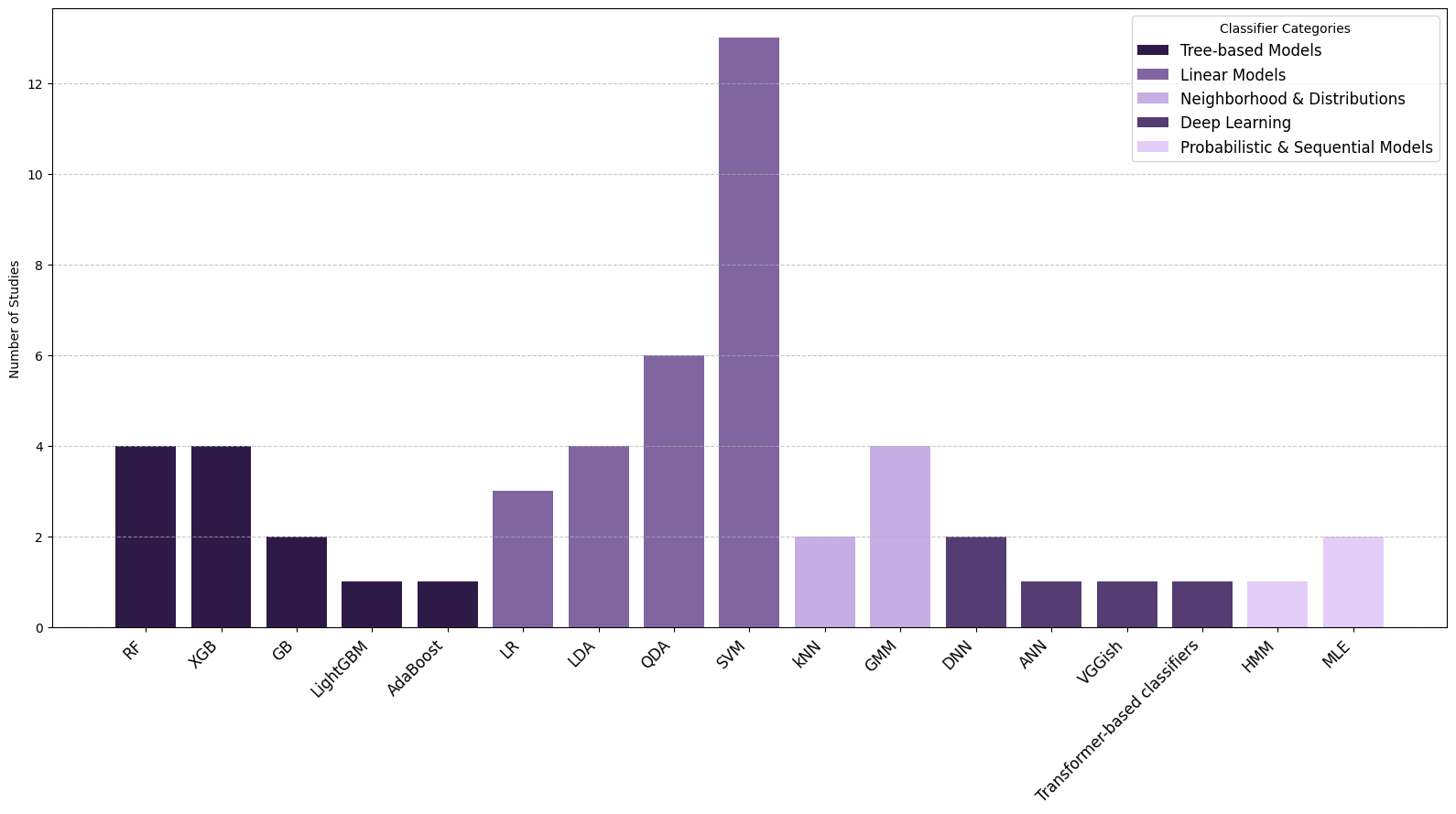}
\hfill
\label{fig:classifiers}
\scriptsize{ANN: Artificial Neural Network, DNN: Deep Neural Network, GB: Gradient Boosting, GMM: Gaussian Mixture Model, HMM: Hidden Markov Model, kNN: k-Nearest Neighbors Algorithm, LDA: Linear Discriminant Analysis, LR: Logistic Regression, MLE: Maximum Likelihood Estimation, QDA: Quadratic Discriminant Analysis, RF: Random Forest, SVM: Support Vector Machine, XGB: eXtreme Gradient Boosting}
\end{figure*}

\begin{figure*}[ht!]
\centering
\caption{Validation techniques used}
\centering
   \includegraphics[width=\linewidth]{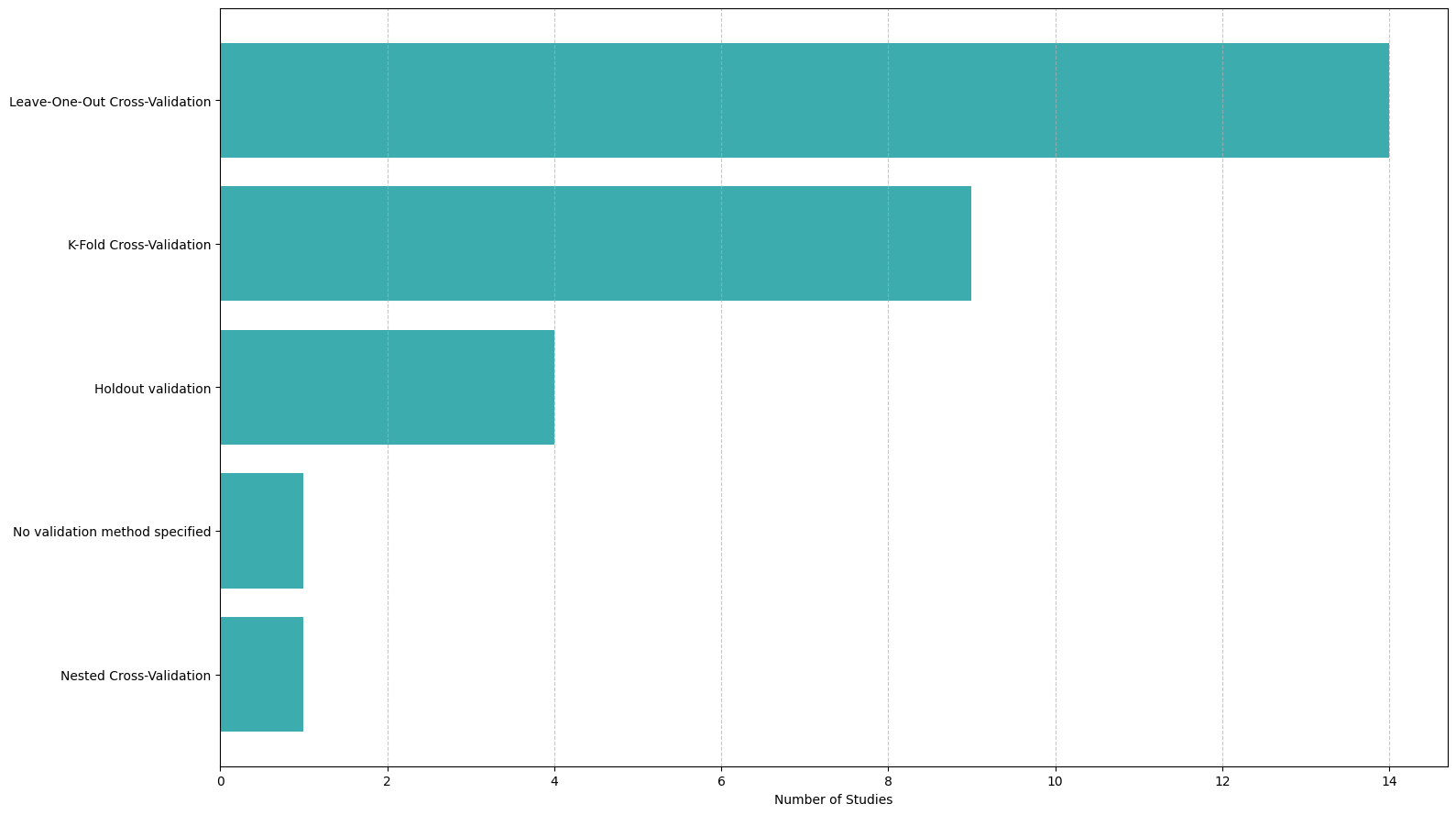}
\label{fig:validation}
\end{figure*}

Table~\ref{tab:classifiers} details the classifiers and cross-validation techniques used by each study, along with the training conditions of the algorithms. Figure~\ref{fig:classifiers} visually summarizes the frequency and categories of classifiers used across studies, while Figure~\ref{fig:validation} illustrates the distribution of cross-validation techniques.

\input{table-classifiers}

\subsection{Performance of classifiers}
\label{results20b}
The discriminative performance of classifiers in identifying suicide risk through vocal analysis was primarily expressed using accuracy (24/29; 82.76\%). AUC was less reported (7/29; 24.14\%), but is presented in Table~\ref{tab:auc}, since it provides a more robust measure of classification performance across different settings.

Statistical heterogeneity in AUC estimates was evaluated using Cochran's Q test and the I² statistic for any subgroup comprising at least two studies with comparable characteristics (e.g. identical classifier algorithm and feature set). Because each subgroup contained only two studies, the statistical power to detect heterogeneity was limited, even when the point estimate of I² was zero. Furthermore, I² is known to be biased and imprecise in meta-analyses with very few studies, and a null point estimate does not preclude the existence of substantive heterogeneity \citep{VonHippel2015The}.
\begin{itemize}
    \item For SVM using acoustic features only (n$=$2), we estimated Q$=$0.27 (p$=$0.61) and I²$=$0\%.
    \item For SVM using acoustic and linguistic features (n$=$2), Q$=$0.72 (p$=$0.40) and I²$=$0\%.
\end{itemize}
Despite these low point estimates, the small number of studies prevents confident inference regarding true heterogeneity. As a consequence, no quantitative meta‑analysis was undertaken, and results are presented through narrative synthesis only.

Previous studies reported strong classifier performances, such as an accuracy of 75\% with SVM and 81.25\% with HMM in longer recordings, with HMM preferred due to its sequential analysis capability \citep{Scherer2013Investigating}. Recent studies achieved AUC values of 0.97 and 0.985 with GB in two distinct studies involving helpline callers \citep{iyer2022using, iyer2022voicebio}, significantly exceeding earlier reported values ranging between 0.62 and 0.78 \citep{Belouali2021Acoustic, Pestian2017A, Shah2019Multimodal, Cohen2023A}.

However, some classifiers demonstrated limitations, particularly in differentiating RS individuals from depressed NRS patients, due to shared acoustic patterns associated with common psychopathological processes, yielding modest performances around 60\% accuracy \citep{Ozdas2004Investigation, Chakravarthula2019Automatic}.

Recent studies have identified additional influencing factors beyond acoustics. \citet{Min2023Acoustic} emphasized longitudinal within-person changes, where acoustic features showed an AUC of 0.67 in baseline assessments, declining to 0.56 after two to four months, indicating variability and temporal dynamics in acoustic features. \citet{amiriparian2024non} demonstrated that integrating metadata such as prior suicide attempts or access to firearms drastically improved classification from a baseline balanced accuracy of 66.2\% (speech alone) to 94.4\% when combined with such demographic data.

Gender-specific models demonstrated improved classifier performance, achieving an enhanced balanced accuracy of 81\% due to differing vocal correlates between RS men and women \citep{Gerczuk2024Exploring}.

The type of speech input notably impacted classifier performance. Speech obtained from open talking was the most common input (54.55\%), closely followed by speech from reading and answering questions (respectively 48.48\% and 45.45\%). Among reading tasks, controlled passages such as the "Rainbow Passage" - used in 21.21\% of studies - were associated with more consistent and reliable results, likely due to reduced variability in linguistic and phonetic content. This finding aligns with earlier studies highlighting the advantages of standardized reading material for acoustic analysis \citep{Kaymaz-Keskinpala2007Screening, T2007Direct, Yingthawornsuk2008Distinguishing, Yingthawornsuk2006Objective}.

Advanced pre-trained speech models (e.g., DNN, VGGish and transformer-based models) achieved impactful results, with accuracies reaching up to 80.7\% and F1-scores of 84.6\% \citep{Cui2024Spontaneous, Chen2024Fine-grained}. However, performance degradation was observed under distribution shifts and cross-dataset evaluations \citep{Pillai2023Investigating}.

Further improvements in classifier performance were observed when analyses separated subjects by gender or by discussed topic \citep{Chakravarthula2019Automatic}. Controlling for inter-individual voice variability using techniques such as vocal tract length normalization also demonstrated a beneficial effect : \citet{Subari2010Comparison} reported a 5\% improvement in classification accuracy when applying such methods. However, controlling for environmental variables was found to negatively impact classifier performance, indicating a measurable negative impact of environmental compensation on classification accuracy, stressing the challenge in designing generalizable classifiers \citep{Kaymaz-Keskinpala2007Screening}.

\input{table-auc}

\subsection{Acoustic data used}

\subsubsection{Relevant acoustic features}

Several studies evaluated classifier performance using a single acoustic feature. \citet{Ozdas2000Analysis} demonstrated that jitter alone differentiated RS from NRS individuals with an accuracy of 80\%. \citet{Yingthawornsuk2006Objective} achieved accuracies of 77\% (maintenance session) and 82\% (reading session) using only power spectral density (PSD) to discriminate between RS and depressive or remitted patients. Similarly, \citet{Wahidah2015Investigating} identified significant differences in spectral density power between treatment sessions using PSD exclusively. \citet{Kaymaz-Keskinpala2007Distinguishing,Ozdas2004Analysis} used only Mel-frequency cepstral coefficients (MFCC), reaching respectively 80\% and 93\% accuracy in distinguishing RS from depressed individuals, with results sometimes lower depending on patient gender and speech task.

Studies employing multiple features such as jitter, shimmer, and related spectral characteristics reported strong performance. By combining jitter with spectral slope, \citet{Ozdas2004Investigation} achieved accuracies of 85\% (RS vs. healthy) and 75\% (RS vs. depressed NRS). \citet{Cohen2023A}, using jitter, shimmer, F0, and intensity, reported an AUC of 0.65 for acoustic features alone, improved to 0.76 through multimodal fusion. \citet{Gerczuk2024Exploring} achieved 81\% balanced accuracy using jitter, shimmer, peakslope, spectral slope, intensity, and pause duration in gender-specific modeling, while \citet{Cui2024Spontaneous} attained 80.7\% accuracy and an F1-score of 84.6\% by combining similar features including MFCC and F1, F2, F3.

Studies using MFCC and F1, F2, F3 produced notable outcomes. \citet{Min2023Acoustic} reported accuracies of 68.7\% (between-person ANN) and up to 79.2\% (within-person XGB) using MFCC, F1, F2, F3, pitch, F0, and energy. \citet{iyer2022using,iyer2022voicebio} demonstrated an AUC of 0.974 and reported an exceptional accuracy of 99.85\% (AUC = 0.985) with similar acoustic features. \citet{Yünden2024Examination} achieved 90\% accuracy primarily using MFCC, F1, F2, F3, pitch, intensity, and speech interval features.

PSD-centered studies also reported high classification accuracies. \citet{Kaymaz-Keskinpala2007Screening} obtained 87\% accuracy using PSD2, while \citet{Yingthawornsuk2007AcousticAO} achieved an accuracy of 88.5\% employing PSD2 and center frequencies (CF3, CF4). In the study by \citet{France2000Acoustical}, features such as amplitude modulation, F1 and F3, and PSD3 led to classification accuracies of 75\% for both RS and control groups, and 71\% for depressed individuals.

Research analyzing voiced-silent intervals and speech timing features also showed promising results. \citet{hashim2012analysis} demonstrated classification accuracies ranging from 75\% to 100\%, and \citet{figueroa2024comparison} identified significant acoustic differences (p$<$0.05) in interval length, pitch, and speech timing. \citet{amiriparian2024non} achieved a baseline accuracy of 66.2\% using acoustic features alone, significantly increasing to 94.4\% with added metadata.

Finally, advanced acoustic features including zero crossing rate and entropy of energy provided valuable insights. \citet{Chen2024Fine-grained} reported an F1-score of 76.96\% for negative emotion recognition, whereas \citet{Pillai2023Investigating} achieved AUC values ranging from 0.62 to 0.68 across multiple datasets.

Table~\ref{tab:acousticparams} lists the acoustic data used in each study.

\input{table-acousticparams}

\subsubsection{Other modalities}

Among the 29 studies evaluating classifiers, 21 (72.41\%) utilized a unimodal acoustic approach, 5 (17.24\%) used bimodal approaches, and 6 (20.68\%) employed multimodal (three or more modalities) methodologies, incorporating linguistic, visual, clinician interaction dynamics, or metadata, alongside acoustic data (Tables ~\ref{tab:classifiers} and ~\ref{tab:auc}).

Several unimodal studies demonstrated strong classification results using acoustic features alone. \citet{Scherer2013Investigating} reported accuracies of 81.25\% (HMM) and 75\% (SVM). \citet{Ozdas2004Analysis} achieved 80\% accuracy distinguishing RS from depressed individuals using MFCC. \citet{hashim2012analysis} obtained 75\%-100\% accuracy using voiced-silent interval features, while \citet{France2000Acoustical} reported accuracies of around 75\% for RS and controls using amplitude modulation, F1, F3, and PSD3. Similarly, \citet{Yingthawornsuk2007AcousticAO} reached 90.33\% accuracy with PSD and CF2, CF3, and \citet{iyer2022voicebio} achieved a very high accuracy of 99.85\% (AUC=0.985).

Bimodal approaches combining acoustic with another modality also showed notable enhancements in classification performance. \citet{Stasak2021Read} improved accuracy from 54-69\% (acoustic features alone) to 77-78\% when incorporating speech disfluency features. \citet{Cui2024Spontaneous}, employing acoustic and linguistic modalities, reported a best accuracy of 80.7\% and an F1-score of 84.6\%.

Multimodal studies frequently showed substantial improvements compared to unimodal approaches. \citet{Venek2017Adolescent} reached 90\% accuracy by combining acoustic and linguistic features with interaction dynamics, such as turn-taking behavior, response latency, and speaking time. \citet{Pestian2017A} improved from an acoustic-only AUC of 0.79 to 0.92 when adding linguistic features, including word and word-pair frequencies derived from interview transcripts. Similarly, \citet{Belouali2021Acoustic} reported AUC improvement from 0.76 (acoustic alone) to 0.80 (acoustic and linguistic features). \citet{Shah2019Multimodal} observed progressive improvements from an AUC of 0.62 (acoustic), 0.72 (acoustic and verbal), to 0.74 with the addition of visual features. These included motion cues (e.g. average displacement of hands and shoulders) and facial behavior (e.g. gaze aversion and action units related to negative emotions).
\citet{amiriparian2024non} significantly increased classification accuracy from 66.2\% (acoustic features alone) to 94.4\% when integrating patient metadata, including history of suicide attempts, access to firearms or lethal medication, hopelessness, trauma, substance abuse, and depressive symptom scores. Likewise, \citet{Cohen2023A} reported an AUC enhancement from 0.65 (single modality) to 0.76 through multimodal decision fusion.

\section{Discussion}

\subsection{Main results}
\label{discussion23a}
This literature review synthesizes current research assessing acoustic features and classifiers for the evaluation of suicidal behavior. Despite the limited number of identified studies, growing interest in AI applications in mental health has emerged, reflected by an increase in publications addressing vocal markers in suicide risk prediction. Findings across reviewed studies indicate substantial promise for employing acoustic features as objective, non-invasive indicators of suicide risk.

Acoustic features such as jitter, MFCC, PSD, F1, F2, F3, F0, pitch, and intensity consistently demonstrated strong discriminative performance. Notably, vocal timing patterns and PSD sub-bands achieved classification accuracies ranging from 90.3\% to 95\%, aligning with findings from \citet{Iyer2022Detection}. In particular, recent studies incorporating multimodal or bimodal approaches reported improved accuracy compared to unimodal methods, aligning with observations by \citet{barua2024artificial}, who observed superior performance from multimodal approaches in mental health disorder detection.

Classifier performance varied, with RF, XGB, and GB models frequently reporting high accuracy and AUC values, supporting recent findings by \citet{ehtemam2024role}. GB classifiers, specifically, achieved AUC scores up to 0.985, highlighting their robustness in acoustic-based suicide risk classification tasks.

Additionally, studies exploring digital and acoustic features, such as those reported by \citet{holmgren2022utilizing}, reinforce the potential of integrating speech analysis with broader digital indicators (e.g., sleep disturbances, heart rate variability) for enhanced suicide risk assessment. The reviewed studies demonstrated comparable recall (up to 86\%) and specificity (up to 70\%) in acoustic-based suicide risk detection, supporting the performance of voice-derived features within digital assessment frameworks.

Gender-specific and subgroup analyses further improved classifier performance, indicating differential acoustic patterns among demographic groups. These results underscore the importance of personalized models to effectively discriminate between risk groups, as observed in multiple reviewed studies.

Recent advancements captured in this review underline the potential utility of acoustic features and sophisticated ML models in suicide risk assessment, supporting their eventual integration into clinical and telehealth settings.

\subsection{Limitations of the studies}
\label{discussion23b}
Several limitations were identified across the studies included in this systematic review. One major limitation is the variability in sample sizes, which ranged from very small samples (fewer than 30 participants) to larger cohorts (over 1,000 participants). This variability introduces heterogeneity that restricts direct comparisons and limits the generalizability of findings. Smaller sample sizes particularly constrain the statistical power and reliability of classifier performance results.

Furthermore, a substantial proportion of studies exhibited marked class imbalance between RS and NRS groups, most frequently characterized by a surplus of NRS samples. This was one of the main drivers of the high risk estimation for the Analysis domain in PROBAST (\ref{appendix:probast}).

This imbalance can significantly affect performance metrics, particularly accuracy, which may remain high even when a model fails to identify most RS cases. In such cases, accuracy becomes misleading, masking poor recall or specificity. Despite this risk, few studies explicitly addressed the impact of class imbalance or adopted appropriate corrective strategies, such as data resampling, class weighting, or cost-sensitive algorithms. Insufficient reporting of complementary metrics (recall, specificity, and predictive values) limits the interpretability and clinical applicability of results. Robust evaluation of suicide risk detection models requires a more comprehensive and transparent use of multiple performance indicators.

Another notable limitation involves methodological heterogeneity among the studies, particularly regarding acoustic feature extraction methods, speech tasks employed (e.g., spontaneous speech, text reading, structured interviews), and recording conditions (controlled laboratory settings versus telehealth or clinical environments). Such methodological diversity complicates comparisons across studies and can introduce biases related to environmental factors or the nature of speech elicitation tasks.

There is also a significant issue with the lack of standardized protocols for assessing and reporting classifier performance metrics. While accuracy and AUC were commonly reported, other important metrics such as recall, specificity, PPV, NPV, and F1-score were inconsistently used or omitted altogether. This inconsistency prevents comprehensive evaluation of model performance and limits comparative assessments across studies.

Moreover, most studies relied on cross-sectional or retrospective designs, restricting their ability to capture dynamic changes in suicidal risk over time. Few studies adopted longitudinal designs, which are essential for understanding temporal variations in vocal features and their practical relevance in real-world clinical scenarios.

An additional limitation concerns the predominant use of English-speaking populations, with limited representation from other linguistic and cultural backgrounds. Language and cultural differences can significantly influence vocal characteristics, such as F0 and prosodic variability, depending on the speaker's language and context \citep{Lee2017The}. Although certain conversational mechanisms like turn-taking appear to be universally structured, cross-cultural studies show quantitative differences in response timing and behavior across languages \citep{Stivers2009Universals}. Such variations may contribute to dataset shift when speech-based classifiers trained in one language are applied to another. For example, performance drops have been observed when models trained on English or German are tested on Amharic \citep{Retta2023Cross-Corpus}, and multilingual speech encoders perform poorly in zero-shot cross-lingual settings \citep{Keller2024SpeechTaxi}. This limitation also affects suicide risk detection: studies report accuracy reductions when applying models across languages or recording conditions \citep{Pillai2023Investigating,cummins2015review}. Future research should address domain adaptation and prioritize the development of multilingual datasets to improve model generalizability. 

Furthermore, most studies did not report whether participants were taking medication. Certain psychotropic drugs, due to their anticholinergic and antidopaminergic effects, can cause speech alterations such as dysarthria, dry mouth, and changes in acoustic features \citep{Sinha2015Predictors}. Similarly, the consumption of drugs, caffeine, or alcohol was rarely documented. However, \citet{France2000Acoustical} noted that patients on medication were no more likely to be misclassified than those not receiving treatment. Additionally, few studies considered potential confounding factors such as speech or language disorders or respiratory conditions.

Finally, exploration into integrating acoustic features with multimodal data remained limited. Despite evidence suggesting superior performance from multimodal approaches, relatively few studies employed comprehensive multimodal methods incorporating visual, linguistic, or contextual metadata. This gap restricts a full understanding of the potential benefits of combined data sources.

In addition, studies with very deep classifiers that directly analyze raw acoustic data without any intermediate representation (e.g. spectrogram, acoustic embeddings or transcriptions) were excluded from this review because of their current lack of interpretability. These advanced deep learning models often do not provide clear explanations of the features analyzed, limiting their clinical applicability. However, studies using deep models based on extracted acoustic representations, which offer potential for downstream interpretability, were retained. Unlike models trained directly on raw audio signals which learn high-dimensional patterns, these models rely on intermediate representations that can be associated with perceptual or physiological correlates (e.g. prosodic or spectral features), or analyzed post hoc to uncover informative acoustic dimensions \citep{Pan2020Acoustic, Dixit2024Explaining}. Future research should prioritize the development of deep yet explainable AI methods to ensure transparency and usability in clinical practice.

\subsubsection{Lack of specificity}

While previous sections have reviewed classifier performance in detecting suicidal risk, one key limitation emerges across multiple studies: the lack of specificity in distinguishing RS from other clinically similar states, such as MDD, anxiety, or general psychological distress.

Several authors have highlighted the limitations of classifiers in distinguishing RS from NRS individuals with MDD\citep{Pestian2017A,Cohen2020A,Ozdas2004Investigation}. These limitations likely reflect the lack of specificity in acoustic feature variations observed among RS individuals. Indeed, acoustic features such as reduced F0, increased jitter, and spectral alterations are commonly observed in both RS and depressive states, complicating the accurate differentiation of these populations \citep{cummins2015review,figueroa2024comparison}.

Recent studies further emphasize the challenge of specificity, demonstrating considerable overlap in acoustic profiles among groups experiencing different types or intensities of psychological distress. For instance, \citet{Cohen2023A} reported that acoustic features alone struggled to discriminate effectively between RS and depressive individuals, with noticeable improvement only through multimodal integration. Similarly, \citet{Min2023Acoustic} found decreased classifier performance when applied across diverse contexts and populations, reflecting limited generalizability and specificity.

Additionally, accurately differentiating varying levels of suicidal severity using acoustic features remains challenging, due to subtle emotional variations and context-dependent speech characteristics \citep{Chen2024Fine-grained,Pillai2023Investigating}. While some studies observed strong discriminative power of acoustic features in high-distress contexts, such as helpline calls \citep{iyer2022using,iyer2022voicebio}, other studies highlighted reduced specificity when dealing with less extreme clinical scenarios or mixed populations \citep{Shen2022Establishment,Gerczuk2024Exploring,Yünden2024Examination}.

However, multimodal approaches consistently enhanced specificity by combining acoustic features with linguistic content, visual signals, or metadata \citep{amiriparian2024non,Cui2024Spontaneous,Gerczuk2024Exploring}. These findings suggest that, although acoustic features alone may lack the granularity required to distinguish RS from other states, they remain valuable components in multimodal frameworks.

\subsubsection{Population bias and confounding variables}

Beyond the acoustic overlap between RS and related psychological conditions, another important limitation concerns population bias and confounding variables. Several studies report high performance scores, which may reflect not only effective classifiers but also restricted or enriched study populations that amplify classification signals.
For instance, \citet{iyer2022using} and \citet{iyer2022voicebio} both report AUC values close to 0.97 and 0.985, respectively. These results were obtained from retrospective analyses of calls to suicide prevention hotlines, in which callers had self-selected into a high-risk group by contacting crisis services. As a result, the classification task often consists of distinguishing among various levels of distress within a uniformly high-risk population, rather than generalizing across the full spectrum of suicide risk in the general population.
Similarly, \citet{amiriparian2024non} report a significant improvement in classification accuracy (from 66.2\% to 94.4\%) when metadata such as previous suicide attempts and access to firearms are added to acoustic features. This highlights the potential for confounding, as these clinical indicators are themselves known risk factors and may dominate the classifier’s decision-making process, overshadowing the contribution of vocal markers.
Such case-mix restrictions limit generalizability and may inflate apparent model performance. Studies using narrowly defined or highly enriched samples, or those combining voice with strong clinical priors, should therefore be interpreted with caution. Consistently, the PROBAST assessment registered major concerns in the Participants domain (\ref{appendix:probast}).

\subsection{Research directions and perspectives}
\label{discussion23d}
The methods used for audio data collection were not standardized and varied across studies. Data were recorded in controlled (laboratory), semi-controlled, or ecological environments, such as home recordings. One of the key challenges for future research remains the development of a standardized approach to audio data collection to ensure consistency across studies. Additionally, ethical constraints associated with recording and storing sensitive speech data necessitate collaborative efforts to build large-scale, well-annotated databases, while ensuring data privacy and patient consent.

Most studies have demonstrated that multimodal approaches significantly enhance classifier performance, offering promising avenues for future research. Integration of acoustic analysis with linguistic, visual, behavioral, or demographic features -particularly during patient-clinician interactions- could improve robustness and classification accuracy \citep{Chakravarthula2019Automatic,Venek2017Adolescent,amiriparian2024non,Cui2024Spontaneous,Cohen2023A}. For instance, \citet{amiriparian2024non} highlighted a significant enhancement in classification accuracy when acoustic features were combined with metadata, such as prior suicide attempts or access to firearms. Similarly, \citet{Cohen2023A} showed improved classification through multimodal decision fusion, suggesting that single-modality models alone might be insufficient for accurate suicide risk detection.

However, these multimodal gains raise important research questions. Future studies should aim to disentangle the respective contributions of acoustic features and additional clinical metadata. Ablation studies and balanced datasets are needed to assess whether performance improvements are induced by vocal characteristics or primarily reflect known risk factors. To improve generalizability, classifiers should be validated on diverse and representative populations, avoiding spectrum bias associated with enriched or narrowly defined samples.

Moreover, the method of speech collection has been shown to impact classification performance. Automatic speech samples obtained from text reading generally yield better results compared to spontaneous speech \citep{Yingthawornsuk2006Objective,Yingthawornsuk2007AcousticAO}. This is because ML models can focus solely on acoustic features rather than semantic variability, leading to more stable predictions. \citet{Yingthawornsuk2008Distinguishing} suggested that standardized text-reading tasks could serve as an alternative to spontaneous speech recordings, particularly for specific patient populations, enhancing reproducibility and comparability across studies. Recent studies further support this notion, emphasizing that controlled speech tasks result in more reliable acoustic features, improving model performance and generalizability \citep{figueroa2024comparison,Yünden2024Examination}.

With the widespread use of smartphones, digital communication tools, and increasing presence on social media, new opportunities for remote and real-time suicide risk monitoring are emerging \citep{quellec_predicting_2024}. ML models embedded in smartphone applications could enable the continuous assessment of speech patterns through phone conversations, voice messages, or video content from social media platforms. Monitoring interactions on suicide prevention hotlines or online support platforms also presents valuable opportunities for early detection \citep{iyer2022using,iyer2022voicebio,Chen2024Fine-grained}. Additionally, longitudinal and within-person analyses of acoustic changes over time, as explored by \citet{Min2023Acoustic}, represent another promising avenue for capturing dynamic risk states and improving temporal sensitivity of classifiers.

A promising direction for future research is the joint analysis of acoustic (prosody, voice quality, rhythm) and semantic content (lexical choice, sentiment analysis, coherence of discourse). By combining speech signal features with automated transcription analysis, ML models could gain a more comprehensive understanding of both how patients speak and what they express, potentially enhancing early detection of suicidal risk \citep{Pillai2023Investigating,Shen2022Establishment}. This hybrid approach may provide clinicians with more interpretable and clinically relevant insights, ultimately improving decision-making and patient outcomes.

The recent 1st SpeechWellness Challenge (SW1) addresses several of these limitations by providing a standardized dataset and common evaluation metrics for suicidal risk detection using acoustic methods. SW1 includes speech data from 600 Chinese adolescents performing diverse tasks : spontaneous speech, reading tasks, emotional expression descriptions. Beyond offering data variety, the challenge implements multimodal approaches combining acoustic and linguistic features, with clearly anonymized and ethically validated data, overcoming previous issues of standardization, limited language representation (primarily English studies), and multimodal integration. SW1 represents an important step towards comprehensive, ethically-sound, and cross-culturally validated methods in suicide risk assessment \citep{wu_1st_2025}. Future datasets should therefore be structured to reduce the specific PROBAST gaps we observed, especially in participant selection and analytic rigour (\ref{appendix:probast}).

\subsection{Ethical questioning and data security}

Despite the significant contributions of AI in healthcare, its integration into clinical practice raises ethical, legal, and security concerns that must be carefully addressed. AI applications offer promising opportunities to enhance risk-assessment and treatment, but also introduce challenges related to data privacy, algorithmic bias, and medico-legal accountability \citep{Morley2020The, Lysaght2019AI-Assisted}. Ensuring robust data anonymization and obtaining explicit patient consent for the collection and use of personal health data are critical challenges that must be tackled before widespread adoption.

Additionally, AI-driven decision-making poses legal and ethical dilemmas, particularly concerning the role of clinicians and medico-legal responsibility in suicide risk assessment. A major concern is the reliance on black-box AI models, whose complexity and lack of transparency interfere with their acceptance in clinical settings. The explainability and interpretability of AI are therefore key factors in ensuring trust and accountability in medical decision-making \citep{Chaddad2023Survey, Amann2020Explainability}.

To safely and effectively implement AI for suicide risk detection, a well-defined medico-legal framework must be established, alongside strict adherence to data protection regulations and ethical guidelines. Moreover, cybersecurity risks remain a critical challenge, as the increasing complexity of AI models exposes healthcare systems to potential data breaches and misuse of sensitive patient information \citep{Jeyaraman2023Unraveling}.

Ultimately, AI should be regarded as a complementary tool, supporting but not replacing clinical expertise. The final decision should remain in the hands of clinicians, who must integrate AI-driven insights alongside recollection, psychological, and contextual factors to ensure comprehensive and patient-centered care.

\section{Conclusion}

This systematic review highlights the significant potential of acoustic features analyzed through AI and ML techniques in the context of suicide risk assessment. Despite promising results demonstrating the discriminative power of various acoustic features and multimodal approaches, current studies are constrained by methodological inconsistencies, small and homogeneous samples, and limited longitudinal data. The specificity of acoustic features in differentiating suicidal risk from depressive disorders and varying risk severity remains challenging, necessitating multimodal integration for improved accuracy. 
Earlier reviews reported varied results. \citet{cummins2015review} linked pitch, rate, jitter and shimmer to affective states with 50–90\% accuracy. When post-2015 datasets and ensemble models are included, performance increases to 90–95\%. The timing pattern and PSD findings of \citet{Iyer2022Detection} (78–85\% accuracy) are confirmed, but the addition of MFCCs, F1, F2, F3, and simple metadata increases median accuracy above 90\% and provides systematic evidence for gender-specific gains. Multimodal improvements noted by \citet{barua2024artificial} in depression studies are amplified in this suicide-focused corpus, where balanced accuracy exceeds 94\% after acoustic, linguistic, and contextual fusion. Compared with the pooled figures of \citet{ehtemam2024role} (accuracy = 0.78, AUC = 0.77), the higher scores observed here reflect the stricter focus on speech features and the prevalent use of GB, RF and XGB. The speech sensitivity of 86\% and specificity of 70\% reported by \citet{holmgren2022utilizing} are complemented by recent acoustic analyses that achieve higher values and document cross-linguistic performance drops. Overall, the synthesis incorporates newer datasets, reports higher accuracies, presents evidence for personalised modelling, and specifies the methodological, ethical and generalisability issues that remain.
Future research must prioritize the development of standardized data collection protocols, larger and culturally diverse participant samples, and longitudinal methodologies. Addressing ethical considerations and data security concerns will also be crucial for the successful translation of AI-driven acoustic analysis tools into practical, reliable clinical aids for suicide prevention.

\section*{Author Contributions}

\textbf{Conceptualization:} 
Ambre Marie, Marine Garnier, Sofian Berrouiguet

\textbf{Methodology:} 
Marine Garnier, Sofian Berrouiguet (initial search strategy), Ambre Marie (updated strategy) 

\textbf{Investigation:} 
Marine Garnier, Sofian Berrouiguet (initial search), Ambre Marie (updated search) 

\textbf{Data curation:} 
Marine Garnier (initial subset extraction), Ambre Marie (full extraction and data management) 

\textbf{Validation:} 
Thomas Bertin, Laura Machart, Guillaume Dardenne, Gwenolé Quellec, Sofian Berrouiguet 

\textbf{Visualization:} 
Ambre Marie (figures and tables)

\textbf{Writing - original draft:} 
Ambre Marie 

\textbf{Writing - review \& editing:} 
Ambre Marie, Thomas Bertin, Gwenolé Quellec, Guillaume Dardenne, Laura Machart

\textbf{Supervision:} 
Sofian Berrouiguet, Gwenolé Quellec, Thomas Bertin, Guillaume Dardenne

\textbf{Final approval of the version to be submitted:} 
All authors

\section*{Statements and Declarations}
\label{other24a}
\begin{itemize}
    
    \item \textbf{Funding:} This work was funded by the Brittany Region (France) through the doctoral program "Allocations de Recherche Doctorale" (ARED).

    \item \textbf{Conflict of Interest:} The authors declare that the research was conducted in the absence of any commercial or financial relationships that could be construed as a potential conflict of interest.

    \item \textbf{Ethics Statement:} This article is a systematic review and does not involve original research with participants. All data used were obtained from previously published studies.
\end{itemize}

\appendix

\section{Supplementary source search strategies}
\label{appendix:searchstrat}

Table~\ref{tab:searchstrat} details the supplementary source search strategies, including full search queries.

\section{Risk of bias and applicability assessment}
\label{appendix:probast}

A quality assessment of the included studies was conducted using the PROBAST tool \citep{Chen2020Introduction}. The full results for each study are presented in Table~\ref{tab:probast}, and visually summarized in Figure~\ref{fig:probast-graph}.

\section{Funnel plot of reported accuracies}
\label{appendix:funnel}
Figure~\ref{fig:funnel} shows a funnel plot of classifier accuracies reported across the included studies. This exploratory analysis aims to visually assess potential reporting bias, following the approach used in some meta-analyses even when heterogeneity prevents formal conclusions.
Due to considerable variability in study designs, dataset sizes, and validation procedures, as well as the known limitations of accuracy as a performance metric in imbalanced classification tasks, this plot is interpreted with caution. In particular, accuracy can mask poor model performance in minority classes, which is critical in suicide risk detection.
The observed asymmetry in the plot is statistically supported by an Egger test (intercept = 1.34, $p$ < 0.001), which suggests potential small-study effects or publication bias. However, this result should not be overinterpreted as it may also reflect methodological differences or variance in reporting standards.
This funnel plot is therefore included for transparency, but it is not used to support any specific conclusions in the main analysis.

\section{Overview of Acoustic Features}

\label{appendix:acoustic-definition}

Table~\ref{tab:acoustic-definition} presents definitions of the included acoustic features, categorized by type.

\section{Summary of Study Characteristics}

\label{appendix:summary}

Tables~\ref{tab:summary_pop}, \ref{tab:summary_data}, and \ref{tab:summary_obj1} present a comprehensive summary of the included studies. 

Table~\ref{tab:summary_pop} describes the characteristics of the studied populations and methodological details. Table~\ref{tab:summary_data} details the conditions of data collection and the acoustic material analyzed. Additionally, Table~\ref{tab:summary_obj1} presents the objectives, results, and conclusions drawn in each study.

\section{Citations for aggregate statistics}

\label{appendix:aggregatestat}

Table~\ref{tab:aggregatestat1} includes the comprehensive dataset underlying the aggregate statistics reported in Section \ref{results}. This table lists each study, the relevant extracted data, and the individual values used to calculate the percentages mentioned in the main text.

\section{PRISMA Checklist}

\label{appendix:prisma}

Table~\ref{tab:prisma}, \ref{tab:prisma2} and \ref{tab:prismaabstract} present the completed PRISMA 2020 checklist for this systematic review. For each of the 27 required items, the location in the main manuscript (page number and/or section heading) is provided to facilitate transparency and reproducibility. The checklist has been completed in accordance with the PRISMA 2020 reporting guidelines.

\bibliographystyle{elsarticle-num-names} 
\bibliography{main}

%% Appendix tables
%%
\setcounter{table}{0}
\renewcommand{\thetable}{A\arabic{table}}
\input{table-searchstrat}

\setcounter{table}{0}
\renewcommand{\thetable}{B\arabic{table}}
\input{table-probast}

\setcounter{figure}{0}
\renewcommand{\thefigure}{B\arabic{figure}}
\begin{figure*}[]
\centering
\caption{Visual summary of PROBAST risk-of-bias and applicability concerns across studies}
\includegraphics[width=\columnwidth]{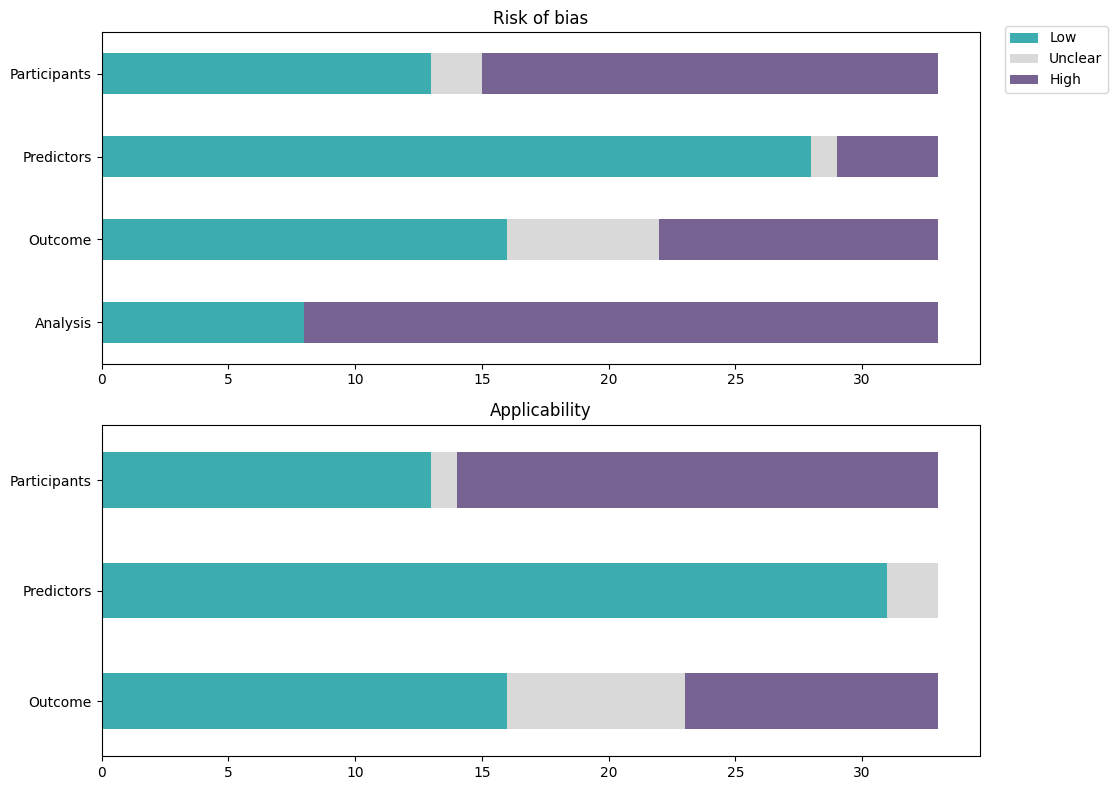}
\label{fig:probast-graph}
\end{figure*}

\setcounter{figure}{0}
\renewcommand{\thefigure}{C\arabic{figure}}
\begin{figure*}[]
\centering
\caption{Funnel plot of reported classification accuracies}
\includegraphics[width=\columnwidth]{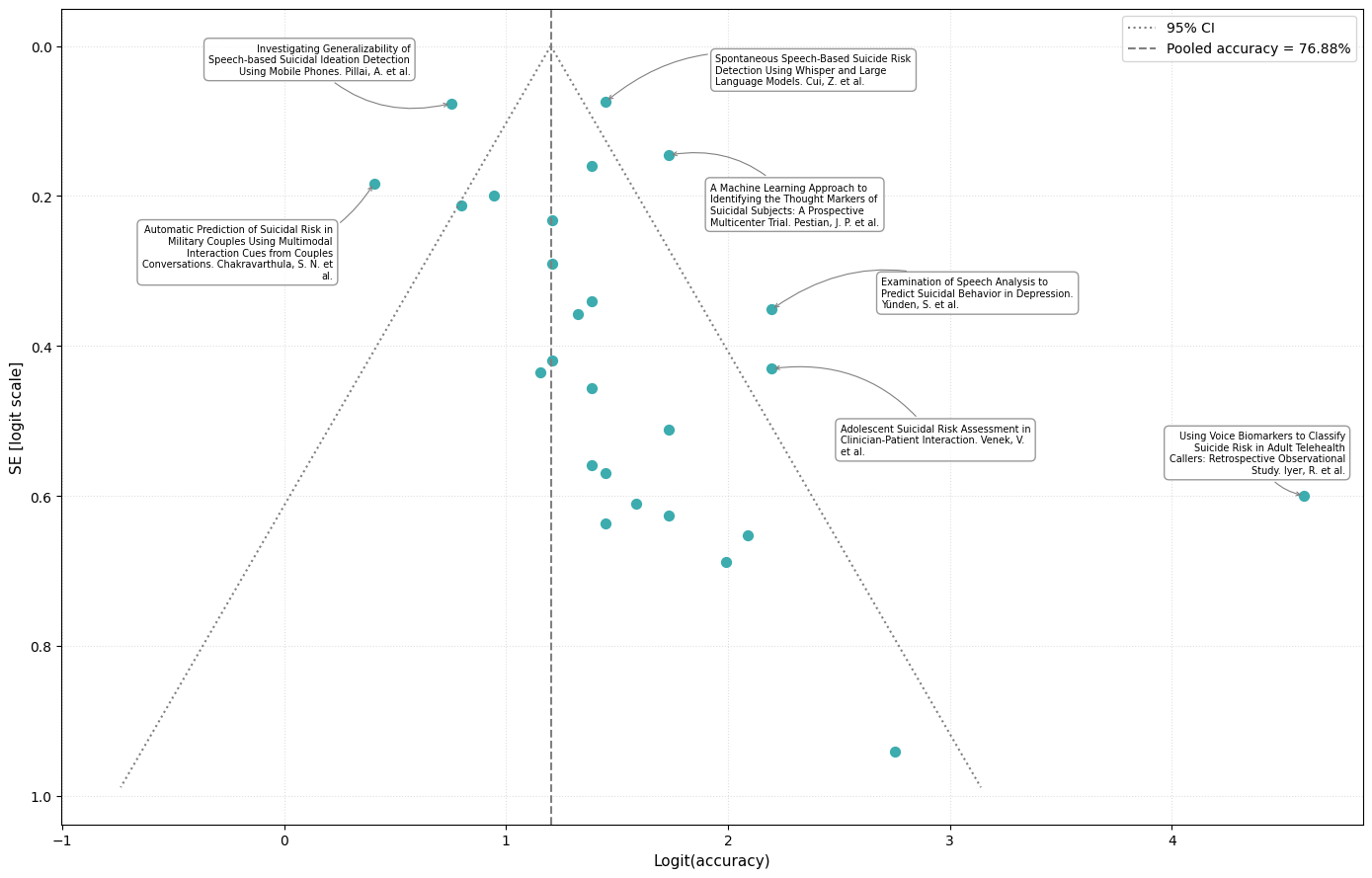}
\scriptsize{Each point represents a study, the vertical dashed line indicates the fixed-effect pooled accuracy. Dashed curves show the 95\% pseudo-confidence limits.}
\label{fig:funnel}
\end{figure*}

\setcounter{table}{0}
\renewcommand{\thetable}{D\arabic{table}}
\input{table-acoustic-definition}

\setcounter{table}{0}
\renewcommand{\thetable}{E\arabic{table}}
\input{table-summary-1}
\input{table-summary-2}
\input{table-summary-31}
\input{table-summary-32}

\setcounter{table}{0}
\renewcommand{\thetable}{F\arabic{table}}
\input{table-aggregatestat1}
\input{table-aggregatestat2}
\input{table-aggregatestat3}

\setcounter{table}{0}
\renewcommand{\thetable}{G\arabic{table}}
\input{table-prismachecklist}
\input{table-prisma2}
\input{table-prismaabstract}

\end{document}

%% file: table-eligibility.tex
% \usepackage{graphicx}
\begin{table*}[h]
\centering
\caption{Eligibility criteria for study inclusion}
\label{tab:eligibility}
\resizebox{\columnwidth}{!}{%
\begin{tabular}{|c|c|}
\hline
\textbf{Inclusion Criteria}                                                                                                                                 & \textbf{Exclusion Criteria}                                                                                                                            \\ \hline
\begin{tabular}[c]{@{}c@{}}Studies analyzing acoustic\\ features in RS individuals\end{tabular}                                                             & \begin{tabular}[c]{@{}c@{}}Studies limited to textual/metadata\\ analysis without acoustic features\end{tabular}                                       \\ \hline
\begin{tabular}[c]{@{}c@{}}RS defined as suicidal ideation, past attempts,\\ psychometric indicators, or clinical evaluation\end{tabular}                   & \begin{tabular}[c]{@{}c@{}}Studies without reference to\\ suicide risk or related conditions\end{tabular}                                              \\ \hline
\begin{tabular}[c]{@{}c@{}}Use of voice recordings with\\ extracted acoustic representations\\ (e.g. spectrograms, embeddings, transcriptions)\end{tabular} & \begin{tabular}[c]{@{}c@{}}Use of deep models trained directly\\ on raw waveform audio\\ (e.g. end-to-end waveform DNNs)\end{tabular}                  \\ \hline
\begin{tabular}[c]{@{}c@{}}Use of classifiers\\ (any type, including deep models\\ applied to intermediate representations)\end{tabular}                    & \begin{tabular}[c]{@{}c@{}}Literature reviews, theoretical\\ papers without clinical populations\end{tabular}                                          \\ \hline
\begin{tabular}[c]{@{}c@{}}Experimental, observational,\\ analytical or descriptive study design\end{tabular}                                               & \begin{tabular}[c]{@{}c@{}}Non-peer-reviewed\\ studies or opinion papers\end{tabular}                                                                  \\ \hline
\begin{tabular}[c]{@{}c@{}}Conference papers or abstracts\\ with sufficient methodological detail \\ (population, features, model, metrics)\end{tabular}    & \begin{tabular}[c]{@{}c@{}}Abstracts lacking methodological information \\ (e.g., no classifier details, \\ features, or population data)\end{tabular} \\ \hline
\end{tabular}%
}
\\
\scriptsize RS: At Risk of Suicide, DNN: Deep Neural Networks
\end{table*}

%% file: table-classimbalances.tex
% \usepackage{graphicx}
\begin{table*}[]
\centering
\caption{Class imbalance across studies}
\label{tab:imbalance}
\resizebox{0.5\columnwidth}{!}{%
\begin{tabular}{c|c|c|c|}
\cline{2-4}
\multicolumn{1}{l|}{\textbf{}}                                             & RS  & NRS & IR    \\ \hline
\multicolumn{1}{|c|}{\textbf{\citet{Ozdas2000Analysis}}}                   & 10  & 10  & 1.00  \\ \hline
\multicolumn{1}{|c|}{\textbf{\citet{Ozdas2004Analysis}}}                   & 10  & 20  & 0.50  \\ \hline
\multicolumn{1}{|c|}{\textbf{\citet{Ozdas2004Investigation}}}              & 10  & 20  & 0.50  \\ \hline
\multicolumn{1}{|c|}{\textbf{\citet{Yingthawornsuk2006Objective}}}         & 10  & 22  & 0.45  \\ \hline
\multicolumn{1}{|c|}{\textbf{\citet{Kaymaz-Keskinpala2007Distinguishing}}} & 20  & 20  & 1.00  \\ \hline
\multicolumn{1}{|c|}{\textbf{\citet{Kaymaz-Keskinpala2007Screening}}}      & 18  & 29  & 0.62  \\ \hline
\multicolumn{1}{|c|}{\textbf{\citet{Yingthawornsuk2007AcousticAO}}}        & 10  & 10  & 1.00  \\ \hline
\multicolumn{1}{|c|}{\textbf{\citet{Yingthawornsuk2008Distinguishing}}}    & 8   & 16  & 0.50  \\ \hline
\multicolumn{1}{|c|}{\textbf{\citet{Subari2010Comparison}}}                & 10  & 20  & 0.50  \\ \hline
\multicolumn{1}{|c|}{\textbf{\citet{hashim2012analysis}}}                  & 7   & 12  & 0.58  \\ \hline
\multicolumn{1}{|c|}{\textbf{\citet{Scherer2013Investigating}}}            & 8   & 8   & 1.00  \\ \hline
\multicolumn{1}{|c|}{\textbf{\citet{Wahidah2015TimingPO}}}                 & 17  & 30  & 0.57  \\ \hline
\multicolumn{1}{|c|}{\textbf{\citet{Pestian2017A}}}                        & 130 & 249 & 0.52  \\ \hline
\multicolumn{1}{|c|}{\textbf{\citet{Venek2017Adolescent}}}                 & 30  & 30  & 1.00  \\ \hline
\multicolumn{1}{|c|}{\textbf{\citet{Chakravarthula2019Automatic}}}         & 59  & 65  & 0.91  \\ \hline
\multicolumn{1}{|c|}{\textbf{\citet{Shah2019Multimodal}}}                  & 45  & 45  & 1.00  \\ \hline
\multicolumn{1}{|c|}{\textbf{\citet{Belouali2021Acoustic}}}                & 84  & 504 & 0.17  \\ \hline
\multicolumn{1}{|c|}{\textbf{\citet{Stasak2021Read}}}                      & 226 & 20  & 11.30 \\ \hline
\multicolumn{1}{|c|}{\textbf{\citet{iyer2022voicebio}}}                    & 77  & 204 & 0.38  \\ \hline
\multicolumn{1}{|c|}{\textbf{\citet{Cohen2023A}}}                          & 26  & 47  & 0.55  \\ \hline
\multicolumn{1}{|c|}{\textbf{\citet{Min2023Acoustic}}}                     & 68  & 30  & 2.27  \\ \hline
\multicolumn{1}{|c|}{\textbf{\citet{Cui2024Spontaneous}}}                  & 631 & 548 & 1.15  \\ \hline
\multicolumn{1}{|c|}{\textbf{\citet{Gerczuk2024Exploring}}}                & 7   & 13  & 0.54  \\ \hline
\multicolumn{1}{|c|}{\textbf{\citet{Yünden2024Examination}}}               & 30  & 60  & 0.50  \\ \hline
\end{tabular}%
}
\\
\scriptsize RS: At Risk of Suicide, NRS: Not at Risk of Suicide.\\The imbalance ratio is defined as: IR $=$ RS/NRS if RS\,>\,NRS, and IR = NRS/RS otherwise.
\end{table*}

%% file: table-statcomparisons.tex
% \usepackage{multirow}
% \usepackage{graphicx}
\begin{table}[]
\centering
\caption{Statistical comparisons of selected acoustic features in NRS and RS subjects}
\label{tab:statcomparisons}
\resizebox{0.92\columnwidth}{!}{%
\begin{tabular}{c|c|c|c|c|c|c|c|}
\cline{2-8}
\multicolumn{1}{l|}{\textbf{}}                                                                     & \textbf{Acoustic feature}                            & \textbf{Category}                                    & \textbf{NRS}       & \textbf{RS}        & \textbf{p-value}         & \textbf{Effect size} & \textbf{\begin{tabular}[c]{@{}c@{}}Replication\\ criteria\end{tabular}} \\ \hline
\multicolumn{1}{|c|}{\textbf{\citet{Ozdas2000Analysis}}}                                           & Jitter                                               & Jitter                                               & 0.017 (0.002)      & 0.022 (0.005)      & 0.007                    & 1.357                & 4                                                                       \\ \hline
\multicolumn{1}{|c|}{\multirow{2}{*}{\textbf{\citet{Ozdas2004Investigation} \textsuperscript{2}}}} & Jitter                                               & Jitter                                               & 0.017 (0.002)      & 0.022 (0.005)      & $<$0.050                   & 1.357                & 4                                                                       \\ \cline{2-8} 
\multicolumn{1}{|c|}{}                                                                             & Spectral slope                                       & Spectral slope                                       & -83.300 (5.460)    & -75.564 (8.531)    & $<$0.050                   & 1.080                & 3                                                                       \\ \hline
\multicolumn{1}{|c|}{\multirow{10}{*}{\textbf{\citet{Scherer2013Investigating}}}}                  & Peak                                                 & Peak Location                                        & -0.230 (0.050)     & -0.250 (0.040)     & \multirow{10}{*}{$<$0.002} & -0.577               & 1                                                                       \\ \cline{2-5} \cline{7-8} 
\multicolumn{1}{|c|}{}                                                                             & NAQ                                                  & Normalized Amplitude Quotient                        & 0.090 (0.040)      & 0.120 (0.050)      &                          & 0.595                & 2                                                                       \\ \cline{2-5} \cline{7-8} 
\multicolumn{1}{|c|}{}                                                                             & Rk                                                   & \multirow{2}{*}{Glottal Closure Instant variability} & 0.300 (0.100)      & 0.360 (0.120)      &                          & 0.529                & 3                                                                       \\ \cline{2-2} \cline{4-5} \cline{7-8} 
\multicolumn{1}{|c|}{}                                                                             & Rg                                                   &                                                      & 1.700 (0.640)      & 1.430 (0.580)      &                          & -0.481               & 3                                                                       \\ \cline{2-5} \cline{7-8} 
\multicolumn{1}{|c|}{}                                                                             & OQ                                                   & Quasi-Open Quotient                                  & 0.310 (0.130)      & 0.420 (0.200)      &                          & 0.709                & 2                                                                       \\ \cline{2-5} \cline{7-8} 
\multicolumn{1}{|c|}{}                                                                             & NAQ Std.                                             & Normalized Amplitude Quotient                        & 0.060 (0.020)      & 0.080 (0.030)      &                          & 0.660                & 2                                                                       \\ \cline{2-5} \cline{7-8} 
\multicolumn{1}{|c|}{}                                                                             & EE Std.                                              & Energy contour variability                           & 0.010 (0.010)      & 0.010 (0.010)      &                          & -0.479               & 2                                                                       \\ \cline{2-5} \cline{7-8} 
\multicolumn{1}{|c|}{}                                                                             & Rk Std.                                              & \multirow{2}{*}{Glottal Closure Instant variability} & 0.100 (0.050)      & 0.120 (0.070)      &                          & 0.423                & 3                                                                       \\ \cline{2-2} \cline{4-5} \cline{7-8} 
\multicolumn{1}{|c|}{}                                                                             & Rg Std.                                              &                                                      & 0.740 (0.340)      & 0.610 (0.240)      &                          & -0.499               & 3                                                                       \\ \cline{2-5} \cline{7-8} 
\multicolumn{1}{|c|}{}                                                                             & OQ Std.                                              & Quasi-Open Quotient                                  & 0.130 (0.050)      & 0.190 (0.100)      &                          & 0.698                & 2                                                                       \\ \hline
\multicolumn{1}{|c|}{\multirow{6}{*}{\textbf{\citet{Venek2017Adolescent}}}}                        & F0                                                   & Fundamental Frequency                                & 149.010            & 224.030            & \multirow{6}{*}{$<$0.010}  & -                    & 4                                                                       \\ \cline{2-5} \cline{7-8} 
\multicolumn{1}{|c|}{}                                                                             & NAQ                                                  & Normalized Amplitude Quotient                        & 0.030              & 0.080              &                          & -                    & 2                                                                       \\ \cline{2-5} \cline{7-8} 
\multicolumn{1}{|c|}{}                                                                             & QOQ                                                  & Quasi-Open Quotient                                  & 0.110              & 0.310              &                          & -                    & 2                                                                       \\ \cline{2-5} \cline{7-8} 
\multicolumn{1}{|c|}{}                                                                             & MDQ                                                  & \multirow{2}{*}{Glottal Closure Instant variability} & 0.110              & 0.130              &                          & -                    & 3                                                                       \\ \cline{2-2} \cline{4-5} \cline{7-8} 
\multicolumn{1}{|c|}{}                                                                             & Rd                                                   &                                                      & 1.090              & 1.650              &                          & -                    & 3                                                                       \\ \cline{2-5} \cline{7-8} 
\multicolumn{1}{|c|}{}                                                                             & F1                                                   & Formant Frequencies                                  & 538.750            & 613.560            &                          & -                    & 6                                                                       \\ \hline
\multicolumn{1}{|c|}{\multirow{10}{*}{\textbf{\citet{Belouali2021Acoustic}}}}                      & Time between GCIs - mean                             & \multirow{2}{*}{Glottal Closure Instant variability} & 0.004 (0.001)      & 0.004 (0.001)      & 0.030                    & 0.498                & 3                                                                       \\ \cline{2-2} \cline{4-8} 
\multicolumn{1}{|c|}{}                                                                             & Time between GCIs - std                              &                                                      & 0.007 (0.003)      & 0.006 (0.002)      & 0.050                    & 0.571                & 3                                                                       \\ \cline{2-8} 
\multicolumn{1}{|c|}{}                                                                             & Energy contour (voiced) - std                        & Energy contour variability                           & 2.881 (0.927)      & 2.361 (0.762)      & $<$0.001                   & 0.612                & 2                                                                       \\ \cline{2-8} 
\multicolumn{1}{|c|}{}                                                                             & Energy contour (voiced) - skewness                   & \multirow{3}{*}{Energy transitions and variations}   & -0.706 (0.329)     & -0.492 (0.261)     & $<$0.001                   & 0.722                & 1                                                                       \\ \cline{2-2} \cline{4-8} 
\multicolumn{1}{|c|}{}                                                                             & Energy contour (voiced) - kurtosis                   &                                                      & 0.158 (0.810)      & -0.361 (0.502)     & $<$0.001                   & 0.771                & 1                                                                       \\ \cline{2-2} \cline{4-8} 
\multicolumn{1}{|c|}{}                                                                             & Energy contour (voiced) - average MSE                &                                                      & 6.029 (3.493)      & 4.335 (2.583)      & 0.010                    & 0.551                & 1                                                                       \\ \cline{2-8} 
\multicolumn{1}{|c|}{}                                                                             & Delta energy entropy - std                           & Entropy of spectral energy                           & 0.363 (0.076)      & 0.324 (0.088)      & 0.040                    & 0.476                & 3                                                                       \\ \cline{2-8} 
\multicolumn{1}{|c|}{}                                                                             & Delta MFCC11 - std                                   & \multirow{3}{*}{Mel-Frequency Cepstral Coefficients} & 0.189 (0.017)      & 0.181 (0.017)      & $<$0.001                   & 0.521                & 2                                                                       \\ \cline{2-2} \cline{4-8} 
\multicolumn{1}{|c|}{}                                                                             & Delta MFCC12 - std                                   &                                                      & 0.183 (0.015)      & 0.176 (0.015)      & 0.010                    & 0.464                & 2                                                                       \\ \cline{2-2} \cline{4-8} 
\multicolumn{1}{|c|}{}                                                                             & Delta MFCC1 - std                                    &                                                      & 1.053 (0.211)      & 0.952 (0.253)      & 0.050                    & 0.434                & 2                                                                       \\ \hline
\multicolumn{1}{|c|}{\multirow{4}{*}{\textbf{\citet{Saavedra2021Association}}}}                    & F0* \textsuperscript{a}                              & Fundamental Frequency                                & 236.030 (33.160)   & 213.880 (39.710)   & 0.029                    & 0.605                & 4                                                                       \\ \cline{2-8} 
\multicolumn{1}{|c|}{}                                                                             & Jitter*** \textsuperscript{a}                        & Jitter                                               & 0.420 (0.340)      & 0.830 (0.440)      & $<$0.010                   & 1.043                & 4                                                                       \\ \cline{2-8} 
\multicolumn{1}{|c|}{}                                                                             & F1*/**/*** \textsuperscript{a}                       & \multirow{2}{*}{Formant Frequencies}                 & -                  & -                  & $<$0.010                   & -                    & 6                                                                       \\ \cline{2-2} \cline{4-8} 
\multicolumn{1}{|c|}{}                                                                             & F2*/**/*** \textsuperscript{a}                       &                                                      & -                  & -                  & $<$0.010                   & -                    & 6                                                                       \\ \hline
\multicolumn{1}{|c|}{\multirow{6}{*}{\textbf{\citet{iyer2022using}}}}                              & Root-mean-squared amplitude\textsuperscript{2}       & Root Mean Squared amplitude                          & -                  & -                  & $<$0.001                   & -                    & 2                                                                       \\ \cline{2-8} 
\multicolumn{1}{|c|}{}                                                                             & First formant frequency\textsuperscript{2}           & Formant Frequencies                                  & -                  & -                  & 0.015                    & -                    & 6                                                                       \\ \cline{2-8} 
\multicolumn{1}{|c|}{}                                                                             & Entropy\textsuperscript{3}                           & Entropy of spectral energy                           & -                  & -                  & $<$0.001                   & -                    & 3                                                                       \\ \cline{2-8} 
\multicolumn{1}{|c|}{}                                                                             & Noise to harmonics ratio\textsuperscript{3}          & Harmonics-to-Noise Ratio                             & -                  & -                  & 0.002                    & -                    & 3                                                                       \\ \cline{2-8} 
\multicolumn{1}{|c|}{}                                                                             & 50th percentile frequency\textsuperscript{3}         & 50th quartile frequency                              & -                  & -                  & 0.009                    & -                    & 3                                                                       \\ \cline{2-8} 
\multicolumn{1}{|c|}{}                                                                             & Spectral slope\textsuperscript{3}                    & Spectral slope                                       & -                  & -                  & 0.009                    & -                    & 3                                                                       \\ \hline
\multicolumn{1}{|c|}{\multirow{13}{*}{\textbf{\citet{iyer2022voicebio}}}}                          & Root mean squared amplitude (dB) \textsuperscript{2} & Root Mean Squared amplitude                          & -                  & -                  & $<$0.001                   & -                    & 2                                                                       \\ \cline{2-8} 
\multicolumn{1}{|c|}{}                                                                             & Entropy \textsuperscript{2}                          & Entropy of spectral energy                           & -                  & -                  & 0.006                    & -                    & 3                                                                       \\ \cline{2-8} 
\multicolumn{1}{|c|}{}                                                                             & Formant2 width (Hz) \textsuperscript{2}              & Formant Bandwidth                                    & -                  & -                  & $<$0.001                   & -                    & 2                                                                       \\ \cline{2-8} 
\multicolumn{1}{|c|}{}                                                                             & Formant3 frequency (Hz) \textsuperscript{2}          & Formant Frequencies                                  & -                  & -                  & $<$0.001                   & -                    & 6                                                                       \\ \cline{2-8} 
\multicolumn{1}{|c|}{}                                                                             & Spectral slope \textsuperscript{2}                   & Spectral slope                                       & -                  & -                  & $<$0.001                   & -                    & 3                                                                       \\ \cline{2-8} 
\multicolumn{1}{|c|}{}                                                                             & Root mean squared amplitude (dB) \textsuperscript{3} & Root Mean Squared amplitude                          & -                  & -                  & $<$0.001                   & -                    & 2                                                                       \\ \cline{2-8} 
\multicolumn{1}{|c|}{}                                                                             & Dominant frequency (Hz) \textsuperscript{3}          & Fundamental Frequency                                & -                  & -                  & 0.010                    & -                    & 4                                                                       \\ \cline{2-8} 
\multicolumn{1}{|c|}{}                                                                             & Formant1 frequency (Hz) \textsuperscript{3}          & Formant Frequencies                                  & -                  & -                  & 0.020                    & -                    & 6                                                                       \\ \cline{2-8} 
\multicolumn{1}{|c|}{}                                                                             & Formant1 width (Hz) \textsuperscript{3}              & Formant Bandwidth                                    & -                  & -                  & 0.070                    & -                    & 2                                                                       \\ \cline{2-8} 
\multicolumn{1}{|c|}{}                                                                             & Formant2 frequency (Hz) \textsuperscript{3}          & Formant Frequencies                                  & -                  & -                  & 0.020                    & -                    & 6                                                                       \\ \cline{2-8} 
\multicolumn{1}{|c|}{}                                                                             & Formant2 width (Hz) \textsuperscript{3}              & Formant Bandwidth                                    & -                  & -                  & $<$0.001                   & -                    & 2                                                                       \\ \cline{2-8} 
\multicolumn{1}{|c|}{}                                                                             & 50th quartile (Hz) \textsuperscript{3}               & 50th quartile frequency                              & -                  & -                  & $<$0.001                   & -                    & 3                                                                       \\ \cline{2-8} 
\multicolumn{1}{|c|}{}                                                                             & Spectral slope \textsuperscript{3}                   & Spectral slope                                       & -                  & -                  & $<$0.001                   & -                    & 3                                                                       \\ \hline
\multicolumn{1}{|c|}{\multirow{8}{*}{\textbf{\citet{Pillai2023Investigating}}}}                    & Kurtosis F0                                          & F0 percentiles                                       & -                  & -                  & \multirow{8}{*}{$<$0.050}  & -                    & 2                                                                       \\ \cline{2-5} \cline{7-8} 
\multicolumn{1}{|c|}{}                                                                             & IQR 1-2 band (250-650Hz)                             & \multirow{2}{*}{Formant Frequencies}                 & -                  & -                  &                          & -                    & 6                                                                       \\ \cline{2-2} \cline{4-5} \cline{7-8} 
\multicolumn{1}{|c|}{}                                                                             & IQR 1-3 band (250-650Hz)                             &                                                      & -                  & -                  &                          & -                    & 6                                                                       \\ \cline{2-5} \cline{7-8} 
\multicolumn{1}{|c|}{}                                                                             & 1st quartile frequency band (250-650Hz)              & 50th quartile frequency                              & -                  & -                  &                          & -                    & 3                                                                       \\ \cline{2-5} \cline{7-8} 
\multicolumn{1}{|c|}{}                                                                             & Spectral harmonicity                                 & Harmonics-to-Noise Ratio                             & -                  & -                  &                          & -                    & 3                                                                       \\ \cline{2-5} \cline{7-8} 
\multicolumn{1}{|c|}{}                                                                             & MFCC\_2                                               & \multirow{3}{*}{Mel-Frequency Cepstral Coefficients} & -                  & -                  &                          & -                    & 2                                                                       \\ \cline{2-2} \cline{4-5} \cline{7-8} 
\multicolumn{1}{|c|}{}                                                                             & MFCC\_4                                               &                                                      & -                  & -                  &                          & -                    & 2                                                                       \\ \cline{2-2} \cline{4-5} \cline{7-8} 
\multicolumn{1}{|c|}{}                                                                             & Skewness of MFCC\_12                                  &                                                      & -                  & -                  &                          & -                    & 2                                                                       \\ \hline
\multicolumn{1}{|c|}{\multirow{27}{*}{\textbf{\citet{figueroa2024comparison}}}}                    & F1 \textsuperscript{i}                               & \multirow{3}{*}{Formant Frequencies}                 & 370.260 (49.530)   & 403.130 (42.420)   & \multirow{3}{*}{0.001}   & -0.704               & 6                                                                       \\ \cline{2-2} \cline{4-5} \cline{7-8} 
\multicolumn{1}{|c|}{}                                                                             & F2 \textsuperscript{a}                               &                                                      & 1545.540 (100.990) & 1623.130 (141.100) &                          & -0.649               & 6                                                                       \\ \cline{2-2} \cline{4-5} \cline{7-8} 
\multicolumn{1}{|c|}{}                                                                             & F3 \textsuperscript{i}                               &                                                      & 2792.070 (207.790) & 2944.410 (199.660) &                          & -0.745               & 6                                                                       \\ \cline{2-8} 
\multicolumn{1}{|c|}{}                                                                             & F0 \textsuperscript{a}                               & \multirow{3}{*}{Fundamental Frequency}               & 182.110 (49.770)   & 214.000 (46.800)   & \multirow{7}{*}{0.002}   & -0.657               & 4                                                                       \\ \cline{2-2} \cline{4-5} \cline{7-8} 
\multicolumn{1}{|c|}{}                                                                             & F0 \textsuperscript{i}                               &                                                      & 177.190 (50.910)   & 208.720 (44.580)   &                          & -0.652               & 4                                                                       \\ \cline{2-2} \cline{4-5} \cline{7-8} 
\multicolumn{1}{|c|}{}                                                                             & F0 \textsuperscript{u}                               &                                                      & 179.330 (48.680)   & 212.060 (46.900)   &                          & -0.683               & 4                                                                       \\ \cline{2-5} \cline{7-8} 
\multicolumn{1}{|c|}{}                                                                             & F1 \textsuperscript{a}                               & \multirow{4}{*}{Formant Frequencies}                 & 637.620 (87.190)   & 694.910 (87.320)   &                          & -0.657               & 6                                                                       \\ \cline{2-2} \cline{4-5} \cline{7-8} 
\multicolumn{1}{|c|}{}                                                                             & F2 \textsuperscript{e}                               &                                                      & 1846.920 (138.940) & 1941.680 (148.190) &                          & -0.663               & 6                                                                       \\ \cline{2-2} \cline{4-5} \cline{7-8} 
\multicolumn{1}{|c|}{}                                                                             & F2 \textsuperscript{o}                               &                                                      & 1276.630 (120.260) & 1351.380 (104.070) &                          & -0.657               & 6                                                                       \\ \cline{2-2} \cline{4-5} \cline{7-8} 
\multicolumn{1}{|c|}{}                                                                             & F3 \textsuperscript{e}                               &                                                      & 2609.370 (198.670) & 2751.100 (204.730) &                          & -0.704               & 6                                                                       \\ \cline{2-8} 
\multicolumn{1}{|c|}{}                                                                             & HNRdB \textsuperscript{a}                            & \multirow{2}{*}{Harmonics-to-Noise Ratio}            & 15.790 (2.250)     & 17.170 (2.470)     & \multirow{2}{*}{0.003}   & -0.590               & 3                                                                       \\ \cline{2-2} \cline{4-5} \cline{7-8} 
\multicolumn{1}{|c|}{}                                                                             & HNRdB \textsuperscript{u}                            &                                                      & 18.200 (3.150)     & 19.690 (3.000)     &                          & -0.481               & 3                                                                       \\ \cline{2-8} 
\multicolumn{1}{|c|}{}                                                                             & F0 \textsuperscript{e}                               & Fundamental Frequency                                & 180.180 (49.640)   & 212.330 (44.980)   & 0.004                    & -0.674               & 4                                                                       \\ \cline{2-8} 
\multicolumn{1}{|c|}{}                                                                             & F1 \textsuperscript{o}                               & \multirow{3}{*}{Formant Frequencies}                 & 516.760 (55.410)   & 550.880 (58.400)   & \multirow{3}{*}{0.005}   & -0.602               & 6                                                                       \\ \cline{2-2} \cline{4-5} \cline{7-8} 
\multicolumn{1}{|c|}{}                                                                             & F2 \textsuperscript{i}                               &                                                      & 2224.550 (201.050) & 2353.120 (222.760) &                          & -0.611               & 6                                                                       \\ \cline{2-2} \cline{4-5} \cline{7-8} 
\multicolumn{1}{|c|}{}                                                                             & F3 \textsuperscript{o}                               &                                                      & 2575.020 (208.850) & 2720.190 (232.010) &                          & -0.663               & 6                                                                       \\ \cline{2-8} 
\multicolumn{1}{|c|}{}                                                                             & Jitterlocal \textsuperscript{u}                      & \multirow{2}{*}{Jitter}                              & 0.990 (0.430)      & 0.800 (0.330)      & 0.006                    & -0.511               & 4                                                                       \\ \cline{2-2} \cline{4-8} 
\multicolumn{1}{|c|}{}                                                                             & Jitterlocal \textsuperscript{e}                      &                                                      & 0.800 (0.230)      & 0.700 (0.210)      & \multirow{3}{*}{0.008}   & -0.472               & 4                                                                       \\ \cline{2-5} \cline{7-8} 
\multicolumn{1}{|c|}{}                                                                             & F0 \textsuperscript{o}                               & Fundamental Frequency                                & 183.510 (50.580)   & 214.370 (47.190)   &                          & -0.628               & 4                                                                       \\ \cline{2-5} \cline{7-8} 
\multicolumn{1}{|c|}{}                                                                             & F3 \textsuperscript{a}                               & Formant Frequencies                                  & 2460.540 (189.470) & 2601.870 (257.110) &                          & -0.641               & 6                                                                       \\ \cline{2-8} 
\multicolumn{1}{|c|}{}                                                                             & Jitterlocal \textsuperscript{a}                      & Jitter                                               & 0.730 (0.200)      & 0.630 (0.190)      & 0.009                    & -0.508               & 4                                                                       \\ \cline{2-8} 
\multicolumn{1}{|c|}{}                                                                             & F3 \textsuperscript{u}                               & \multirow{2}{*}{Formant Frequencies}                 & 2598.360 (170.640) & 2691.690 (171.650) & 0.011                    & -0.546               & 6                                                                       \\ \cline{2-2} \cline{4-8} 
\multicolumn{1}{|c|}{}                                                                             & F1 \textsuperscript{e}                               &                                                      & 481.480 (57.590)   & 508.210 (52.330)   & 0.017                    & -0.482               & 6                                                                       \\ \cline{2-8} 
\multicolumn{1}{|c|}{}                                                                             & Jitterlocal \textsuperscript{i}                      & Jitter                                               & 0.900 (0.490)      & 0.740 (0.270)      & 0.024                    & -0.391               & 4                                                                       \\ \cline{2-8} 
\multicolumn{1}{|c|}{}                                                                             & HNRdB \textsuperscript{o}                            & \multirow{2}{*}{Harmonics-to-Noise Ratio}            & 18.100 (2.400)     & 19.090 (2.590)     & 0.025                    & -0.400               & 3                                                                       \\ \cline{2-2} \cline{4-8} 
\multicolumn{1}{|c|}{}                                                                             & HNRdB \textsuperscript{e}                            &                                                      & 17.110 (2.320)     & 18.060 (2.890)     & 0.026                    & -0.370               & 3                                                                       \\ \cline{2-8} 
\multicolumn{1}{|c|}{}                                                                             & Jitterppq5 \textsuperscript{a}                       & Jitter                                               & 0.300 (0.100)      & 0.270 (0.100)      & 0.048                    & -0.366               & 4                                                                       \\ \hline
\multicolumn{1}{|c|}{\multirow{3}{*}{\textbf{\citet{Gerczuk2024Exploring}}}}                       & SlopeV0-500Hzmean                                    & SlopeV0-500Hz mean                                   & -                  & -                  & \multirow{3}{*}{$<$0.010}  & -0,783               & 1                                                                       \\ \cline{2-5} \cline{7-8} 
\multicolumn{1}{|c|}{}                                                                             & F0 80th percentile                                   & F0 percentiles                                       & -                  & -                  &                          & -0,661               & 2                                                                       \\ \cline{2-5} \cline{7-8} 
\multicolumn{1}{|c|}{}                                                                             & F1 Bandwidthmean                                     & Formant Bandwidth                                    & -                  & -                  &                          & -0,249               & 2                                                                       \\ \hline
\end{tabular}%
}
\scriptsize{\\Mean (standard deviation), \textsuperscript{1}: Median (IQR), \textsuperscript{2}: Men only, NRS : Not at Risk of Suicide, RS: At Risk of Suicide, \textsuperscript{3}: Women only, *: reading text task, **: response to an open-ended question / interview, ***: vocalization of the vowel /a/, \textsuperscript{a}: Parameter analyzed on vowel /a/, \textsuperscript{e}: Parameter analyzed on vowel /e/, \textsuperscript{i}: Parameter analyzed on vowel /i/, \textsuperscript{o}: Parameter analyzed on vowel /o/, \textsuperscript{u}: Parameter analyzed on vowel /u/}
\end{table}

%% file: table-classifiers.tex
% \usepackage{graphicx}
\begin{table*}[]
\centering
\caption{Classifiers and validation applied in each study, training conditions, performance expression and features used}
\label{tab:classifiers}
\resizebox{\columnwidth}{!}{%
                         & Unimodal                                                            & Acoustic                                                                                & 2024                                                                   \\ \hline
\end{tabular}%
}
\scriptsize{ANN: Artificial Neural Network, AUC: Area Under the Receiver-Operating Characteristic curve, AVH: Auditory Verbal Hallucinations, CS: Control Subjects, CV: Cross-Validation, DNN: Deep Neural Network, GB: Gradient Boosting, GMM: Gaussian Mixture Model, HMM: Hidden Markov Model, kNN: k-Nearest Neighbors Algorithm, LDA: Linear Discriminant Analysis, LOOCV: Leave-one-out Cross-Validation, LR: Logistic Regression, M: Men, MDD: Major Depressive Disorder, MLE: Maximum Likelihood Estimation, NPV: Negative Predictive Value, NRS: Not at Risk of Suicide, PPV: Positive Predictive Value, PT: Persecutory Thoughts, QDA: Quadratic Discriminant Analysis, RF: Random Forest, RS: At Risk of Suicide, SVM: Support Vector Machine, W: Women, XGB: eXtreme Gradient Boosting}
\end{table*}

%% file: table-auc.tex
% \usepackage{multirow}
% \usepackage{graphicx}
\begin{table}[]
\centering
\caption{Areas under the curve (AUC) obtained according to the algorithms and as a function of the features used as input to the classifier}
\label{tab:auc}
\resizebox{\columnwidth}{!}{%
\begin{tabular}{l|l|l|l|l|l|l|l|l|l|}
\cline{2-10}
\cellcolor[HTML]{FFFFFF}\textbf{}                                             & \textbf{Features}  & \textbf{SVM} & \textbf{GB} & \textbf{XGB} & \textbf{DNN} & \textbf{kNN} & \textbf{LR} & \textbf{RF} & \textbf{ANN} \\ \hline
\multicolumn{1}{|l|}{}                                                        & \textbf{A}         & 0.76 (0.03)  & -           & -            & -            & -            & -           & -           & -            \\ \cline{2-10} 
\multicolumn{1}{|l|}{\multirow{-2}{*}{\textbf{\citet{Pestian2017A}}}}         & \textbf{A + L}     & 0.87 (0.02)  & -           & -            & -            & -            & -           & -           & -            \\ \hline
\multicolumn{1}{|l|}{}                                                        & \textbf{A}         & 0.62         & -           & -            & -            & -            & -           & -           & -            \\ \cline{2-10} 
\multicolumn{1}{|l|}{}                                                        & \textbf{A + L}     & 0.72         & -           & -            & -            & -            & -           & -           & -            \\ \cline{2-10} 
\multicolumn{1}{|l|}{\multirow{-3}{*}{\textbf{\citet{Shah2019Multimodal}}}}   & \textbf{A + L + V} & -            & -           & -            & -            & -            & -           & 0.67        & -            \\ \hline
\multicolumn{1}{|l|}{}                                                        & \textbf{A}         & 0.63 (0.25)  & -           & 0.77 (0.08)  & 0.75 (0.06)  & 0.69 (0.11)  & 0.78 (0.12) & 0.76 (0.06) & -            \\ \cline{2-10} 
\multicolumn{1}{|l|}{\multirow{-2}{*}{\textbf{\citet{Belouali2021Acoustic}}}} & \textbf{A + L}     & 0.64 (0.27)  & -           & 0.77 (0.05)  & 0.77 (0.08)  & 0.69 (0.15)  & 0.77 (0.12) & 0.80 (0.06) & -            \\ \hline
\multicolumn{1}{|l|}{\textbf{\citet{iyer2022using}}}                          & \textbf{A}         & -            & 0.97        & -            & -            & -            & -           & -           & -            \\ \hline
\multicolumn{1}{|l|}{\textbf{\citet{iyer2022voicebio}}}                       & \textbf{A}         & -            & 0.985       & -            & -            & -            & -           & -           & -            \\ \hline
\multicolumn{1}{|l|}{}                                                        & \textbf{L}         & 0.65         & -           & -            & -            & -            & -           & -           & -            \\ \cline{2-10} 
\multicolumn{1}{|l|}{}                                                        & \textbf{V}         & -            & -           & -            & -            & -            & 0.62        & -           & -            \\ \cline{2-10} 
\multicolumn{1}{|l|}{\multirow{-3}{*}{\textbf{\citet{Cohen2023A}}}}           & \textbf{A + L + V} & 0.76         & -           & -            & -            & -            & 0.76        & -           & -            \\ \hline
\multicolumn{1}{|l|}{}                                                        & \textbf{A + D}     & -            & -           & -            & -            & -            & -           & -           & 0.62         \\ \cline{2-10} 
\multicolumn{1}{|l|}{}                                                        & \textbf{A*}        & -            & -           & 0.67         & -            & -            & -           & -           & -            \\ \cline{2-10} 
\multicolumn{1}{|l|}{\multirow{-3}{*}{\textbf{\citet{Min2023Acoustic}}}}      & \textbf{A**}       & -            & -           & 0.56         & -            & -            & -           & -           & -            \\ \hline
\end{tabular}%
}
\scriptsize{AUC (std), *: baseline to 2 months, **: baseline 2 to 4 months, A: acoustic, ANN: Artificial Neural Network, D: demographic, DNN: Deep Neural Network, GB: Gradient Boosting, kNN: k-Nearest Neighbors Algorithm, L: linguistic, LR: Logistic Regression, RF: Random Forest, SVM: Support Vector Machine, V: visual, XGB: eXtreme Gradient Boosting}

\end{table}

%% file: table-acousticparams.tex
% \usepackage{graphicx}
\begin{table*}[]
\centering
\caption{Main acoustic features used in each study}
\label{tab:acousticparams}
\resizebox{1\columnwidth}{!}{%
\begin{tabular}{c|lllll|l|llll|lllll|ll|c|}
\cline{2-19}
\multicolumn{1}{l|}{\textbf{}}                                             & \multicolumn{5}{c|}{\textbf{Source features}}                                                                                                     & \multicolumn{1}{c|}{\textbf{\begin{tabular}[c]{@{}c@{}}Filter\\  features\end{tabular}}} & \multicolumn{4}{c|}{\textbf{Spectral features}}                                                                                                                                                                                    & \multicolumn{5}{c|}{\textbf{Prosodic features}}                                                                                                                                                                                                                                                          & \multicolumn{2}{c|}{\textbf{Timing patterns}}                                                                                                                         & \multicolumn{1}{l|}{} \\ \cline{2-19} 
\multicolumn{1}{l|}{}                                                      & \multicolumn{1}{c|}{Jitter} & \multicolumn{1}{c|}{Shimmer} & \multicolumn{1}{c|}{QOQ} & \multicolumn{1}{c|}{NAQ} & \multicolumn{1}{c|}{Peakslope} & \multicolumn{1}{c|}{\begin{tabular}[c]{@{}c@{}}Formants\\  (Hz)\end{tabular}}            & \multicolumn{1}{c|}{MFCC} & \multicolumn{1}{c|}{PSD} & \multicolumn{1}{c|}{\begin{tabular}[c]{@{}c@{}}Glottal\\  flow\\  spectral\\  slope\end{tabular}} & \multicolumn{1}{c|}{\begin{tabular}[c]{@{}c@{}}F0\\  (Hz)\end{tabular}} & \multicolumn{1}{c|}{\begin{tabular}[c]{@{}c@{}}Intensity\\  (dB)\end{tabular}} & \multicolumn{1}{c|}{\begin{tabular}[c]{@{}c@{}}Speech\\  time\end{tabular}} & \multicolumn{1}{c|}{\begin{tabular}[c]{@{}c@{}}Pause\\  duration\end{tabular}} & \multicolumn{1}{c|}{Pitch} & \multicolumn{1}{c|}{Energy} & \multicolumn{1}{c|}{\begin{tabular}[c]{@{}c@{}}Interval\\  Length\end{tabular}} & \multicolumn{1}{c|}{\begin{tabular}[c]{@{}c@{}}Transition\\  features\end{tabular}} & Other                 \\ \hline
\multicolumn{1}{|c|}{\textbf{\citet{France2000Acoustical}}}                & \multicolumn{1}{l|}{}       & \multicolumn{1}{l|}{}        & \multicolumn{1}{l|}{}    & \multicolumn{1}{l|}{}    &                                & \multicolumn{1}{c|}{X}                                                                   & \multicolumn{1}{l|}{}     & \multicolumn{1}{c|}{X}   & \multicolumn{1}{l|}{}                                                                             & \multicolumn{1}{c|}{X}                                                  & \multicolumn{1}{l|}{}                                                          & \multicolumn{1}{l|}{}                                                       & \multicolumn{1}{l|}{}                                                          & \multicolumn{1}{l|}{}      &                             & \multicolumn{1}{l|}{}                                                           &                                                                                     & \multicolumn{1}{l|}{} \\ \hline
\multicolumn{1}{|c|}{\textbf{\citet{Ozdas2000Analysis}}}                   & \multicolumn{1}{c|}{X}      & \multicolumn{1}{l|}{}        & \multicolumn{1}{l|}{}    & \multicolumn{1}{l|}{}    &                                &                                                                                          & \multicolumn{1}{l|}{}     & \multicolumn{1}{l|}{}    & \multicolumn{1}{l|}{}                                                                             &                                                                         & \multicolumn{1}{l|}{}                                                          & \multicolumn{1}{l|}{}                                                       & \multicolumn{1}{l|}{}                                                          & \multicolumn{1}{l|}{}      &                             & \multicolumn{1}{l|}{}                                                           &                                                                                     & \multicolumn{1}{l|}{} \\ \hline
\multicolumn{1}{|c|}{\textbf{\citet{Ozdas2004Analysis}}}                   & \multicolumn{1}{l|}{}       & \multicolumn{1}{l|}{}        & \multicolumn{1}{l|}{}    & \multicolumn{1}{l|}{}    &                                &                                                                                          & \multicolumn{1}{c|}{X}    & \multicolumn{1}{l|}{}    & \multicolumn{1}{l|}{}                                                                             &                                                                         & \multicolumn{1}{l|}{}                                                          & \multicolumn{1}{l|}{}                                                       & \multicolumn{1}{l|}{}                                                          & \multicolumn{1}{l|}{}      &                             & \multicolumn{1}{l|}{}                                                           &                                                                                     & \multicolumn{1}{l|}{} \\ \hline
\multicolumn{1}{|c|}{\textbf{\citet{Ozdas2004Investigation}}}              & \multicolumn{1}{c|}{X}      & \multicolumn{1}{l|}{}        & \multicolumn{1}{l|}{}    & \multicolumn{1}{l|}{}    &                                &                                                                                          & \multicolumn{1}{l|}{}     & \multicolumn{1}{l|}{}    & \multicolumn{1}{c|}{X}                                                                            &                                                                         & \multicolumn{1}{l|}{}                                                          & \multicolumn{1}{l|}{}                                                       & \multicolumn{1}{l|}{}                                                          & \multicolumn{1}{l|}{}      &                             & \multicolumn{1}{l|}{}                                                           &                                                                                     & \multicolumn{1}{l|}{} \\ \hline
\multicolumn{1}{|c|}{\textbf{\citet{Yingthawornsuk2006Objective}}}         & \multicolumn{1}{l|}{}       & \multicolumn{1}{l|}{}        & \multicolumn{1}{l|}{}    & \multicolumn{1}{l|}{}    &                                &                                                                                          & \multicolumn{1}{l|}{}     & \multicolumn{1}{c|}{X}   & \multicolumn{1}{l|}{}                                                                             &                                                                         & \multicolumn{1}{l|}{}                                                          & \multicolumn{1}{l|}{}                                                       & \multicolumn{1}{l|}{}                                                          & \multicolumn{1}{l|}{}      &                             & \multicolumn{1}{l|}{}                                                           &                                                                                     & \multicolumn{1}{l|}{} \\ \hline
\multicolumn{1}{|c|}{\textbf{\citet{Kaymaz-Keskinpala2007Distinguishing}}} & \multicolumn{1}{l|}{}       & \multicolumn{1}{l|}{}        & \multicolumn{1}{l|}{}    & \multicolumn{1}{l|}{}    &                                &                                                                                          & \multicolumn{1}{c|}{X}    & \multicolumn{1}{l|}{}    & \multicolumn{1}{l|}{}                                                                             &                                                                         & \multicolumn{1}{l|}{}                                                          & \multicolumn{1}{l|}{}                                                       & \multicolumn{1}{l|}{}                                                          & \multicolumn{1}{l|}{}      &                             & \multicolumn{1}{l|}{}                                                           &                                                                                     & \multicolumn{1}{l|}{} \\ \hline
\multicolumn{1}{|c|}{\textbf{\citet{Kaymaz-Keskinpala2007Screening}}}      & \multicolumn{1}{l|}{}       & \multicolumn{1}{l|}{}        & \multicolumn{1}{l|}{}    & \multicolumn{1}{l|}{}    &                                &                                                                                          & \multicolumn{1}{c|}{X}    & \multicolumn{1}{c|}{X}   & \multicolumn{1}{l|}{}                                                                             &                                                                         & \multicolumn{1}{l|}{}                                                          & \multicolumn{1}{l|}{}                                                       & \multicolumn{1}{l|}{}                                                          & \multicolumn{1}{l|}{}      & \multicolumn{1}{c|}{X}      & \multicolumn{1}{l|}{}                                                           &                                                                                     & \multicolumn{1}{l|}{} \\ \hline
\multicolumn{1}{|c|}{\textbf{\citet{Yingthawornsuk2007AcousticAO}}}        & \multicolumn{1}{l|}{}       & \multicolumn{1}{l|}{}        & \multicolumn{1}{l|}{}    & \multicolumn{1}{l|}{}    &                                &                                                                                          & \multicolumn{1}{l|}{}     & \multicolumn{1}{c|}{X}   & \multicolumn{1}{l|}{}                                                                             &                                                                         & \multicolumn{1}{l|}{}                                                          & \multicolumn{1}{l|}{}                                                       & \multicolumn{1}{l|}{}                                                          & \multicolumn{1}{l|}{}      &                             & \multicolumn{1}{l|}{}                                                           &                                                                                     & b                     \\ \hline
\multicolumn{1}{|c|}{\textbf{\citet{Yingthawornsuk2008Distinguishing}}}    & \multicolumn{1}{l|}{}       & \multicolumn{1}{l|}{}        & \multicolumn{1}{l|}{}    & \multicolumn{1}{l|}{}    &                                &                                                                                          & \multicolumn{1}{l|}{}     & \multicolumn{1}{c|}{X}   & \multicolumn{1}{l|}{}                                                                             &                                                                         & \multicolumn{1}{l|}{}                                                          & \multicolumn{1}{l|}{}                                                       & \multicolumn{1}{l|}{}                                                          & \multicolumn{1}{l|}{}      &                             & \multicolumn{1}{l|}{}                                                           &                                                                                     & b                     \\ \hline
\multicolumn{1}{|c|}{\textbf{\citet{Subari2010Comparison}}}                & \multicolumn{1}{l|}{}       & \multicolumn{1}{l|}{}        & \multicolumn{1}{l|}{}    & \multicolumn{1}{l|}{}    &                                & \multicolumn{1}{c|}{X}                                                                   & \multicolumn{1}{c|}{X}    & \multicolumn{1}{l|}{}    & \multicolumn{1}{l|}{}                                                                             &                                                                         & \multicolumn{1}{l|}{}                                                          & \multicolumn{1}{l|}{}                                                       & \multicolumn{1}{l|}{}                                                          & \multicolumn{1}{l|}{}      &                             & \multicolumn{1}{l|}{}                                                           &                                                                                     & b                     \\ \hline
\multicolumn{1}{|c|}{\textbf{\citet{hashim2012analysis}}}                  & \multicolumn{1}{l|}{}       & \multicolumn{1}{l|}{}        & \multicolumn{1}{l|}{}    & \multicolumn{1}{l|}{}    &                                &                                                                                          & \multicolumn{1}{l|}{}     & \multicolumn{1}{l|}{}    & \multicolumn{1}{l|}{}                                                                             &                                                                         & \multicolumn{1}{l|}{}                                                          & \multicolumn{1}{l|}{}                                                       & \multicolumn{1}{l|}{}                                                          & \multicolumn{1}{l|}{}      &                             & \multicolumn{1}{c|}{X}                                                          & \multicolumn{1}{c|}{X}                                                              & \multicolumn{1}{l|}{} \\ \hline
\multicolumn{1}{|c|}{\textbf{\citet{Scherer2013Investigating}}}            & \multicolumn{1}{l|}{}       & \multicolumn{1}{l|}{}        & \multicolumn{1}{c|}{X}   & \multicolumn{1}{c|}{X}   & \multicolumn{1}{c|}{X}         &                                                                                          & \multicolumn{1}{c|}{X}    & \multicolumn{1}{c|}{X}   & \multicolumn{1}{l|}{}                                                                             & \multicolumn{1}{c|}{X}                                                  & \multicolumn{1}{l|}{}                                                          & \multicolumn{1}{l|}{}                                                       & \multicolumn{1}{l|}{}                                                          & \multicolumn{1}{l|}{}      & \multicolumn{1}{c|}{X}      & \multicolumn{1}{l|}{}                                                           &                                                                                     & \multicolumn{1}{l|}{} \\ \hline
\multicolumn{1}{|c|}{\textbf{\citet{Wahidah2015Investigating}}}            & \multicolumn{1}{l|}{}       & \multicolumn{1}{l|}{}        & \multicolumn{1}{l|}{}    & \multicolumn{1}{l|}{}    &                                &                                                                                          & \multicolumn{1}{l|}{}     & \multicolumn{1}{c|}{X}   & \multicolumn{1}{l|}{}                                                                             &                                                                         & \multicolumn{1}{l|}{}                                                          & \multicolumn{1}{l|}{}                                                       & \multicolumn{1}{l|}{}                                                          & \multicolumn{1}{l|}{}      &                             & \multicolumn{1}{l|}{}                                                           &                                                                                     & \multicolumn{1}{l|}{} \\ \hline
\multicolumn{1}{|c|}{\textbf{\citet{Wahidah2015TimingPO}}}                 & \multicolumn{1}{l|}{}       & \multicolumn{1}{l|}{}        & \multicolumn{1}{l|}{}    & \multicolumn{1}{l|}{}    &                                &                                                                                          & \multicolumn{1}{c|}{X}    & \multicolumn{1}{c|}{X}   & \multicolumn{1}{l|}{}                                                                             &                                                                         & \multicolumn{1}{l|}{}                                                          & \multicolumn{1}{l|}{}                                                       & \multicolumn{1}{l|}{}                                                          & \multicolumn{1}{l|}{}      &                             & \multicolumn{1}{c|}{X}                                                          & \multicolumn{1}{c|}{X}                                                              & \multicolumn{1}{l|}{} \\ \hline
\multicolumn{1}{|c|}{\textbf{\citet{Pestian2017A}}}                        & \multicolumn{1}{l|}{}       & \multicolumn{1}{l|}{}        & \multicolumn{1}{c|}{X}   & \multicolumn{1}{c|}{X}   & \multicolumn{1}{c|}{X}         & \multicolumn{1}{c|}{X}                                                                   & \multicolumn{1}{l|}{}     & \multicolumn{1}{l|}{}    & \multicolumn{1}{l|}{}                                                                             & \multicolumn{1}{c|}{X}                                                  & \multicolumn{1}{l|}{}                                                          & \multicolumn{1}{l|}{}                                                       & \multicolumn{1}{l|}{}                                                          & \multicolumn{1}{l|}{}      &                             & \multicolumn{1}{l|}{}                                                           &                                                                                     & \multicolumn{1}{l|}{} \\ \hline
\multicolumn{1}{|c|}{\textbf{\citet{Venek2017Adolescent}}}                 & \multicolumn{1}{l|}{}       & \multicolumn{1}{l|}{}        & \multicolumn{1}{c|}{X}   & \multicolumn{1}{c|}{X}   & \multicolumn{1}{c|}{X}         & \multicolumn{1}{c|}{X}                                                                   & \multicolumn{1}{l|}{}     & \multicolumn{1}{l|}{}    & \multicolumn{1}{l|}{}                                                                             & \multicolumn{1}{c|}{X}                                                  & \multicolumn{1}{l|}{}                                                          & \multicolumn{1}{c|}{X}                                                      & \multicolumn{1}{c|}{X}                                                         & \multicolumn{1}{l|}{}      &                             & \multicolumn{1}{l|}{}                                                           &                                                                                     & \multicolumn{1}{l|}{} \\ \hline
\multicolumn{1}{|c|}{\textbf{\citet{Chakravarthula2019Automatic}}}         & \multicolumn{1}{c|}{X}      & \multicolumn{1}{c|}{X}       & \multicolumn{1}{l|}{}    & \multicolumn{1}{l|}{}    &                                &                                                                                          & \multicolumn{1}{c|}{X}    & \multicolumn{1}{l|}{}    & \multicolumn{1}{l|}{}                                                                             &                                                                         & \multicolumn{1}{c|}{X}                                                         & \multicolumn{1}{l|}{}                                                       & \multicolumn{1}{l|}{}                                                          & \multicolumn{1}{c|}{X}     &                             & \multicolumn{1}{l|}{}                                                           &                                                                                     & \multicolumn{1}{l|}{} \\ \hline
\multicolumn{1}{|c|}{\textbf{\citet{Shah2019Multimodal}}}                  & \multicolumn{1}{l|}{}       & \multicolumn{1}{l|}{}        & \multicolumn{1}{l|}{}    & \multicolumn{1}{l|}{}    &                                & \multicolumn{1}{c|}{X}                                                                   & \multicolumn{1}{l|}{}     & \multicolumn{1}{l|}{}    & \multicolumn{1}{l|}{}                                                                             & \multicolumn{1}{c|}{X}                                                  & \multicolumn{1}{l|}{}                                                          & \multicolumn{1}{l|}{}                                                       & \multicolumn{1}{c|}{X}                                                         & \multicolumn{1}{c|}{X}     &                             & \multicolumn{1}{l|}{}                                                           &                                                                                     & \multicolumn{1}{l|}{} \\ \hline
\multicolumn{1}{|c|}{\textbf{\citet{Belouali2021Acoustic}}}                & \multicolumn{1}{c|}{X}      & \multicolumn{1}{c|}{X}       & \multicolumn{1}{c|}{X}   & \multicolumn{1}{c|}{X}   &                                &                                                                                          & \multicolumn{1}{c|}{X}    & \multicolumn{1}{l|}{}    & \multicolumn{1}{l|}{}                                                                             & \multicolumn{1}{c|}{X}                                                  & \multicolumn{1}{l|}{}                                                          & \multicolumn{1}{l|}{}                                                       & \multicolumn{1}{l|}{}                                                          & \multicolumn{1}{l|}{}      & \multicolumn{1}{c|}{X}      & \multicolumn{1}{l|}{}                                                           &                                                                                     & a                     \\ \hline
\multicolumn{1}{|c|}{\textbf{\citet{Saavedra2021Association}}}             & \multicolumn{1}{c|}{X}      & \multicolumn{1}{l|}{}        & \multicolumn{1}{l|}{}    & \multicolumn{1}{l|}{}    &                                & \multicolumn{1}{c|}{X}                                                                   & \multicolumn{1}{l|}{}     & \multicolumn{1}{l|}{}    & \multicolumn{1}{l|}{}                                                                             & \multicolumn{1}{c|}{X}                                                  & \multicolumn{1}{c|}{X}                                                         & \multicolumn{1}{l|}{}                                                       & \multicolumn{1}{l|}{}                                                          & \multicolumn{1}{l|}{}      &                             & \multicolumn{1}{l|}{}                                                           &                                                                                     & \multicolumn{1}{l|}{} \\ \hline
\multicolumn{1}{|c|}{\textbf{\citet{Stasak2021Read}}}                      & \multicolumn{1}{l|}{}       & \multicolumn{1}{l|}{}        & \multicolumn{1}{l|}{}    & \multicolumn{1}{l|}{}    &                                &                                                                                          & \multicolumn{1}{l|}{}     & \multicolumn{1}{l|}{}    & \multicolumn{1}{l|}{}                                                                             &                                                                         & \multicolumn{1}{l|}{}                                                          & \multicolumn{1}{l|}{}                                                       & \multicolumn{1}{l|}{}                                                          & \multicolumn{1}{l|}{}      &                             & \multicolumn{1}{l|}{}                                                           &                                                                                     & c                     \\ \hline
\multicolumn{1}{|c|}{\textbf{\citet{iyer2022using}}}                       & \multicolumn{1}{l|}{}       & \multicolumn{1}{l|}{}        & \multicolumn{1}{l|}{}    & \multicolumn{1}{l|}{}    &                                & \multicolumn{1}{c|}{X}                                                                   & \multicolumn{1}{l|}{}     & \multicolumn{1}{l|}{}    & \multicolumn{1}{c|}{X}                                                                            & \multicolumn{1}{c|}{X}                                                  & \multicolumn{1}{c|}{X}                                                         & \multicolumn{1}{l|}{}                                                       & \multicolumn{1}{l|}{}                                                          & \multicolumn{1}{c|}{X}     & \multicolumn{1}{c|}{X}      & \multicolumn{1}{l|}{}                                                           &                                                                                     & a                     \\ \hline
\multicolumn{1}{|c|}{\textbf{\citet{iyer2022voicebio}}}                    & \multicolumn{1}{l|}{}       & \multicolumn{1}{l|}{}        & \multicolumn{1}{l|}{}    & \multicolumn{1}{l|}{}    &                                & \multicolumn{1}{c|}{X}                                                                   & \multicolumn{1}{l|}{}     & \multicolumn{1}{l|}{}    & \multicolumn{1}{c|}{X}                                                                            & \multicolumn{1}{c|}{X}                                                  & \multicolumn{1}{c|}{X}                                                         & \multicolumn{1}{l|}{}                                                       & \multicolumn{1}{l|}{}                                                          & \multicolumn{1}{c|}{X}     & \multicolumn{1}{c|}{X}      & \multicolumn{1}{l|}{}                                                           &                                                                                     & a                     \\ \hline
\multicolumn{1}{|c|}{\textbf{\citet{Cohen2023A}}}                          & \multicolumn{1}{c|}{X}      & \multicolumn{1}{c|}{X}       & \multicolumn{1}{l|}{}    & \multicolumn{1}{l|}{}    &                                &                                                                                          & \multicolumn{1}{l|}{}     & \multicolumn{1}{l|}{}    & \multicolumn{1}{l|}{}                                                                             & \multicolumn{1}{c|}{X}                                                  & \multicolumn{1}{c|}{X}                                                         & \multicolumn{1}{l|}{}                                                       & \multicolumn{1}{l|}{}                                                          & \multicolumn{1}{l|}{}      &                             & \multicolumn{1}{l|}{}                                                           &                                                                                     & c                     \\ \hline
\multicolumn{1}{|c|}{\textbf{\citet{Min2023Acoustic}}}                     & \multicolumn{1}{l|}{}       & \multicolumn{1}{l|}{}        & \multicolumn{1}{l|}{}    & \multicolumn{1}{l|}{}    &                                & \multicolumn{1}{c|}{X}                                                                   & \multicolumn{1}{c|}{X}    & \multicolumn{1}{l|}{}    & \multicolumn{1}{l|}{}                                                                             & \multicolumn{1}{c|}{X}                                                  & \multicolumn{1}{l|}{}                                                          & \multicolumn{1}{l|}{}                                                       & \multicolumn{1}{l|}{}                                                          & \multicolumn{1}{c|}{X}     & \multicolumn{1}{c|}{X}      & \multicolumn{1}{l|}{}                                                           &                                                                                     & a, b                  \\ \hline
\multicolumn{1}{|c|}{\textbf{\citet{Pillai2023Investigating}}}             & \multicolumn{1}{l|}{}       & \multicolumn{1}{l|}{}        & \multicolumn{1}{l|}{}    & \multicolumn{1}{l|}{}    &                                &                                                                                          & \multicolumn{1}{c|}{X}    & \multicolumn{1}{l|}{}    & \multicolumn{1}{l|}{}                                                                             &                                                                         & \multicolumn{1}{l|}{}                                                          & \multicolumn{1}{l|}{}                                                       & \multicolumn{1}{l|}{}                                                          & \multicolumn{1}{l|}{}      &                             & \multicolumn{1}{l|}{}                                                           &                                                                                     & a                     \\ \hline
\multicolumn{1}{|c|}{\textbf{\citet{amiriparian2024non}}}                  & \multicolumn{1}{c|}{X}      & \multicolumn{1}{c|}{X}       & \multicolumn{1}{l|}{}    & \multicolumn{1}{l|}{}    &                                & \multicolumn{1}{c|}{X}                                                                   & \multicolumn{1}{c|}{X}    & \multicolumn{1}{l|}{}    & \multicolumn{1}{l|}{}                                                                             & \multicolumn{1}{c|}{X}                                                  & \multicolumn{1}{c|}{X}                                                         & \multicolumn{1}{c|}{X}                                                      & \multicolumn{1}{c|}{X}                                                         & \multicolumn{1}{c|}{X}     & \multicolumn{1}{c|}{X}      & \multicolumn{1}{l|}{}                                                           &                                                                                     & a                     \\ \hline
\multicolumn{1}{|c|}{\textbf{\citet{Chen2024Fine-grained}}}                & \multicolumn{1}{l|}{}       & \multicolumn{1}{l|}{}        & \multicolumn{1}{l|}{}    & \multicolumn{1}{l|}{}    &                                &                                                                                          & \multicolumn{1}{l|}{}     & \multicolumn{1}{l|}{}    & \multicolumn{1}{l|}{}                                                                             &                                                                         & \multicolumn{1}{l|}{}                                                          & \multicolumn{1}{l|}{}                                                       & \multicolumn{1}{l|}{}                                                          & \multicolumn{1}{l|}{}      &                             & \multicolumn{1}{l|}{}                                                           &                                                                                     & a                     \\ \hline
\multicolumn{1}{|c|}{\textbf{\citet{Cui2024Spontaneous}}}                  & \multicolumn{1}{c|}{X}      & \multicolumn{1}{c|}{X}       & \multicolumn{1}{l|}{}    & \multicolumn{1}{l|}{}    &                                & \multicolumn{1}{c|}{X}                                                                   & \multicolumn{1}{c|}{X}    & \multicolumn{1}{l|}{}    & \multicolumn{1}{l|}{}                                                                             & \multicolumn{1}{c|}{X}                                                  & \multicolumn{1}{l|}{}                                                          & \multicolumn{1}{l|}{}                                                       & \multicolumn{1}{l|}{}                                                          & \multicolumn{1}{l|}{}      & \multicolumn{1}{c|}{X}      & \multicolumn{1}{l|}{}                                                           &                                                                                     & c                     \\ \hline
\multicolumn{1}{|c|}{\textbf{\citet{figueroa2024comparison}}}              & \multicolumn{1}{c|}{X}      & \multicolumn{1}{c|}{X}       & \multicolumn{1}{l|}{}    & \multicolumn{1}{l|}{}    &                                & \multicolumn{1}{c|}{X}                                                                   & \multicolumn{1}{l|}{}     & \multicolumn{1}{l|}{}    & \multicolumn{1}{l|}{}                                                                             & \multicolumn{1}{c|}{X}                                                  & \multicolumn{1}{l|}{}                                                          & \multicolumn{1}{c|}{X}                                                      & \multicolumn{1}{l|}{}                                                          & \multicolumn{1}{c|}{X}     & \multicolumn{1}{c|}{X}      & \multicolumn{1}{c|}{X}                                                          &                                                                                     & b                     \\ \hline
\multicolumn{1}{|c|}{\textbf{\citet{Gerczuk2024Exploring}}}                & \multicolumn{1}{c|}{X}      & \multicolumn{1}{c|}{X}       & \multicolumn{1}{l|}{}    & \multicolumn{1}{l|}{}    & \multicolumn{1}{c|}{X}         & \multicolumn{1}{c|}{X}                                                                   & \multicolumn{1}{c|}{X}    & \multicolumn{1}{l|}{}    & \multicolumn{1}{c|}{X}                                                                            &                                                                         & \multicolumn{1}{c|}{X}                                                         & \multicolumn{1}{l|}{}                                                       & \multicolumn{1}{c|}{X}                                                         & \multicolumn{1}{c|}{X}     & \multicolumn{1}{c|}{X}      & \multicolumn{1}{l|}{}                                                           &                                                                                     & a                     \\ \hline
\multicolumn{1}{|c|}{\textbf{\citet{Yünden2024Examination}}}               & \multicolumn{1}{l|}{}       & \multicolumn{1}{l|}{}        & \multicolumn{1}{l|}{}    & \multicolumn{1}{l|}{}    &                                & \multicolumn{1}{c|}{X}                                                                   & \multicolumn{1}{c|}{X}    & \multicolumn{1}{l|}{}    & \multicolumn{1}{l|}{}                                                                             & \multicolumn{1}{c|}{X}                                                  & \multicolumn{1}{c|}{X}                                                         & \multicolumn{1}{c|}{X}                                                      & \multicolumn{1}{c|}{X}                                                         & \multicolumn{1}{c|}{X}     &                             & \multicolumn{1}{c|}{X}                                                          &                                                                                     & a                     \\ \hline
\end{tabular}%
}

\scriptsize{a: (zero crossing rate, entropy of energy, chroma vector and deviation, other spectral and energy-based features), b: (bandwidth, weight coefficient, center frequency), c: (voice quality (GRBASI), disfluency features), F0: Fundamental Frequency, MFCC: Mel-Frequency Cepstral Coefficients, NAQ: Normalized Amplitude Quotient, PSD: Power Spectral Density, QOQ: Quasi-Open Quotient}
\end{table*}

%% file: table-searchstrat.tex
% \usepackage{multirow}
% \usepackage{graphicx}
\begin{table*}[]
\centering
\caption{Source search strategies}
\label{tab:searchstrat}
\resizebox{\columnwidth}{!}{%
\begin{tabular}{|c|c|c|c|c|}
\hline
\textbf{Database}                 & \textbf{Platform \& Interface}                                                                                               & \textbf{Search string}                                                                                                                                                                                                                                                                                                                                                                                                 & \textbf{Limits}                                                           & \textbf{Fields searched}                                            \\ \hline
\multirow{2}{*}{Cochrane Library} & \multirow{2}{*}{\begin{tabular}[c]{@{}c@{}}Cochrane Library\\ advanced search tab\end{tabular}}                              & \begin{tabular}[c]{@{}c@{}}suicid* AND \\ (vocal \\ OR acoustic* \\ OR prosod*)\end{tabular}                                                                                                                                                                                                                                                                                                                           & \begin{tabular}[c]{@{}c@{}}Publication\\ year \textless 2021\end{tabular} & \begin{tabular}[c]{@{}c@{}}Title, Abstract,\\ Keywords\end{tabular} \\ \cline{3-5} 
                                  &                                                                                                                              & \begin{tabular}[c]{@{}c@{}}suicid* AND \\ ("speech analysis" \\ OR "voice features"\\ OR prosody \\ OR "acoustic markers"\\ OR "speech biometrics"\\ OR "voice quality"\\ OR phonation\\ OR "vocal characteristics")\end{tabular}                                                                                                                                                                                      & \begin{tabular}[c]{@{}c@{}}No publication\\ year limits\end{tabular}      & \begin{tabular}[c]{@{}c@{}}Title, Abstract,\\ Keywords\end{tabular} \\ \hline
\multirow{2}{*}{PubMed}           & \multirow{2}{*}{\begin{tabular}[c]{@{}c@{}}PubMed (NCBI)\\ web interface\end{tabular}}                                       & \begin{tabular}[c]{@{}c@{}}suicid*{[}Title/Abstract{]} AND \\ (vocal{[}Title/Abstract{]}\\ OR acoustic*{[}Title/Abstract{]} \\ OR prosod*{[}Title/Abstract{]})\end{tabular}                                                                                                                                                                                                                                            & \begin{tabular}[c]{@{}c@{}}Publication\\ year \textless 2021\end{tabular} & Title, Abstract                                                     \\ \cline{3-5} 
                                  &                                                                                                                              & \begin{tabular}[c]{@{}c@{}}suicid*{[}Title/Abstract{]} AND\\ ("speech analysis"{[}Title/Abstract{]} \\ OR "voice features"{[}Title/Abstract{]}\\ OR prosody{[}Title/Abstract{]} \\ OR "acoustic markers"{[}Title/Abstract{]}\\ OR "speech biometrics"{[}Title/Abstract{]} \\ OR "voice quality"{[}Title/Abstract{]}\\ OR phonation{[}Title/Abstract{]} \\ OR "vocal characteristics"{[}Title/Abstract{]})\end{tabular} & \begin{tabular}[c]{@{}c@{}}No publication\\ year limits\end{tabular}      & Title, Abstract                                                     \\ \hline
\multirow{2}{*}{Scopus}           & \multirow{2}{*}{\begin{tabular}[c]{@{}c@{}}Scopus (Elsevier)\\ advanced search\end{tabular}}                                 & \begin{tabular}[c]{@{}c@{}}TITLE-ABS(suicid* AND \\ (vocal \\ OR acoustic* \\ OR prosod*))\end{tabular}                                                                                                                                                                                                                                                                                                                & \begin{tabular}[c]{@{}c@{}}Publication\\ year \textless 2021\end{tabular} & Title, Abstract                                                     \\ \cline{3-5} 
                                  &                                                                                                                              & \begin{tabular}[c]{@{}c@{}}TITLE-ABS-KEY(suicid* AND\\ ("speech analysis" \\ OR "voice features" \\ OR prosody \\ OR "acoustic markers"\\ OR "speech biometrics" \\ OR "voice quality"\\ OR phonation \\ OR "vocal characteristics"))\end{tabular}                                                                                                                                                                     & \begin{tabular}[c]{@{}c@{}}No publication\\ year limits\end{tabular}      & \begin{tabular}[c]{@{}c@{}}Title, Abstract,\\ Keywords\end{tabular} \\ \hline
The Grey Literature Report        & \begin{tabular}[c]{@{}c@{}}GreyLit Report\\ (NY Academy archive\\ via Wayback Machine)\end{tabular}                          & \begin{tabular}[c]{@{}c@{}}suicid* AND \\ (vocal \\ OR acoustic* \\ OR prosod*)\end{tabular}                                                                                                                                                                                                                                                                                                                           & \begin{tabular}[c]{@{}c@{}}Publication\\ year \textless 2021\end{tabular} & Title, Abstract                                                     \\ \hline
\multirow{2}{*}{Web of Science}   & \multirow{2}{*}{\begin{tabular}[c]{@{}c@{}}Web of Science\\ Core Collection\end{tabular}}                                    & \begin{tabular}[c]{@{}c@{}}TS$=$(suicid* AND \\ (vocal \\ OR acoustic* \\ OR prosod*))\end{tabular}                                                                                                                                                                                                                                                                                                                      & \begin{tabular}[c]{@{}c@{}}Publication\\ year \textless 2021\end{tabular} & \begin{tabular}[c]{@{}c@{}}Title, Abstract,\\ Keywords\end{tabular} \\ \cline{3-5} 
                                  &                                                                                                                              & \begin{tabular}[c]{@{}c@{}}TS$=$(suicid* AND\\ ("speech analysis" \\ OR "voice features" \\ OR prosody \\ OR "acoustic markers" \\ OR  "speech biometrics" \\ OR "voice quality" \\ OR phonation \\ OR "vocal characteristics"))\end{tabular}                                                                                                                                                                            & \begin{tabular}[c]{@{}c@{}}No publication\\ year limits\end{tabular}      & \begin{tabular}[c]{@{}c@{}}Title, Abstract,\\ Keywords\end{tabular} \\ \hline
Consensus                         & \begin{tabular}[c]{@{}c@{}}Consensus.ai \textbackslash{}textsuperscript\{1\}\\ free web version (February 2025)\end{tabular} & \begin{tabular}[c]{@{}c@{}}"What machine learning classifiers \\ or acoustic parameters\\ have been used to assess suicide risk \\ in patients based on voice analysis?"\end{tabular}                                                                                                                                                                                                                                  & \begin{tabular}[c]{@{}c@{}}No publication\\ year limits\end{tabular}      & Not applicable                                                      \\ \hline
\end{tabular}%
}

\scriptsize \textsuperscript{1}: Consensus.ai is an AI-based tool, not a structured bibliographic database. Details of its use are as follows: accessed on February 2025; query as given; returned 6 records; 4 duplicates with database searches; 2 unique studies included; titles and abstracts manually screened. Although this method may not be standard in PRISMA workflows, it is fully documented here for transparency and reproducibility.

\end{table*}

%% file: table-probast.tex
% \usepackage{multirow}
% \usepackage{graphicx}
\begin{table}[]
\centering
\caption{ Risk of Bias and Applicability Concerns using PROBAST}
\label{tab:probast}
\resizebox{\columnwidth}{!}{%
\begin{tabular}{|c|cc|cc|cc|c|}
\hline
\multirow{2}{*}{\textbf{Study}}             & \multicolumn{2}{c|}{\textbf{Participants}}                  & \multicolumn{2}{c|}{\textbf{Predictors}}                    & \multicolumn{2}{c|}{\textbf{Outcome}}                       & \textbf{Analysis} \\ \cline{2-8} 
                                            & \multicolumn{1}{c|}{\textbf{Risk}} & \textbf{Applicability} & \multicolumn{1}{c|}{\textbf{Risk}} & \textbf{Applicability} & \multicolumn{1}{c|}{\textbf{Risk}} & \textbf{Applicability} & \textbf{Risk}     \\ \hline
\citet{France2000Acoustical}                & \multicolumn{1}{c|}{High}          & High                   & \multicolumn{1}{c|}{High}          & Unclear                & \multicolumn{1}{c|}{High}          & High                   & High              \\ \hline
\citet{Ozdas2000Analysis}                   & \multicolumn{1}{c|}{High}          & High                   & \multicolumn{1}{c|}{High}          & Unclear                & \multicolumn{1}{c|}{Unclear}       & Low                    & High              \\ \hline
\citet{Ozdas2004Analysis}                   & \multicolumn{1}{c|}{High}          & High                   & \multicolumn{1}{c|}{High}          & Low                    & \multicolumn{1}{c|}{Unclear}       & Low                    & High              \\ \hline
\citet{Ozdas2004Investigation}              & \multicolumn{1}{c|}{High}          & High                   & \multicolumn{1}{c|}{High}          & Low                    & \multicolumn{1}{c|}{Unclear}       & Low                    & High              \\ \hline
\citet{Yingthawornsuk2006Objective}         & \multicolumn{1}{c|}{High}          & High                   & \multicolumn{1}{c|}{Low}           & Low                    & \multicolumn{1}{c|}{Unclear}       & Unclear                & High              \\ \hline
\citet{Kaymaz-Keskinpala2007Distinguishing} & \multicolumn{1}{c|}{High}          & High                   & \multicolumn{1}{c|}{Low}           & Low                    & \multicolumn{1}{c|}{High}          & High                   & High              \\ \hline
\citet{Kaymaz-Keskinpala2007Screening}      & \multicolumn{1}{c|}{Low}           & Low                    & \multicolumn{1}{c|}{Low}           & Low                    & \multicolumn{1}{c|}{Low}           & Low                    & High              \\ \hline
\citet{Yingthawornsuk2007AcousticAO}        & \multicolumn{1}{c|}{Low}           & Low                    & \multicolumn{1}{c|}{Low}           & Low                    & \multicolumn{1}{c|}{High}          & Low                    & High              \\ \hline
\citet{Yingthawornsuk2008Distinguishing}    & \multicolumn{1}{c|}{Low}           & Low                    & \multicolumn{1}{c|}{Low}           & Low                    & \multicolumn{1}{c|}{High}          & Low                    & High              \\ \hline
\citet{Subari2010Comparison}                & \multicolumn{1}{c|}{Low}           & Low                    & \multicolumn{1}{c|}{Low}           & Low                    & \multicolumn{1}{c|}{High}          & Low                    & High              \\ \hline
\citet{hashim2012analysis}                  & \multicolumn{1}{c|}{High}          & High                   & \multicolumn{1}{c|}{Low}           & Low                    & \multicolumn{1}{c|}{High}          & Unclear                & High              \\ \hline
\citet{Scherer2013Investigating}            & \multicolumn{1}{c|}{High}          & High                   & \multicolumn{1}{c|}{Low}           & Low                    & \multicolumn{1}{c|}{Low}           & Low                    & High              \\ \hline
\citet{Wahidah2015Investigating}            & \multicolumn{1}{c|}{High}          & High                   & \multicolumn{1}{c|}{Low}           & Low                    & \multicolumn{1}{c|}{Low}           & Unclear                & High              \\ \hline
\citet{Wahidah2015TimingPO}                 & \multicolumn{1}{c|}{High}          & High                   & \multicolumn{1}{c|}{Low}           & Low                    & \multicolumn{1}{c|}{Low}           & Low                    & Low               \\ \hline
\citet{Pestian2017A}                        & \multicolumn{1}{c|}{Low}           & Low                    & \multicolumn{1}{c|}{Low}           & Low                    & \multicolumn{1}{c|}{Low}           & Low                    & Low               \\ \hline
\citet{Venek2017Adolescent}                 & \multicolumn{1}{c|}{Low}           & Low                    & \multicolumn{1}{c|}{Low}           & Low                    & \multicolumn{1}{c|}{Low}           & Low                    & Low               \\ \hline
\citet{Chakravarthula2019Automatic}         & \multicolumn{1}{c|}{Low}           & Low                    & \multicolumn{1}{c|}{Low}           & Low                    & \multicolumn{1}{c|}{High}          & High                   & Low               \\ \hline
\citet{Shah2019Multimodal}                  & \multicolumn{1}{c|}{High}          & High                   & \multicolumn{1}{c|}{Low}           & Low                    & \multicolumn{1}{c|}{High}          & High                   & Low               \\ \hline
\citet{Belouali2021Acoustic}                & \multicolumn{1}{c|}{Low}           & Low                    & \multicolumn{1}{c|}{Low}           & Low                    & \multicolumn{1}{c|}{Low}           & Low                    & Low               \\ \hline
\citet{Saavedra2021Association}             & \multicolumn{1}{c|}{Low}           & High                   & \multicolumn{1}{c|}{Low}           & Low                    & \multicolumn{1}{c|}{Low}           & Low                    & High              \\ \hline
\citet{Stasak2021Read}                      & \multicolumn{1}{c|}{High}          & High                   & \multicolumn{1}{c|}{Low}           & Low                    & \multicolumn{1}{c|}{Low}           & Low                    & High              \\ \hline
\citet{iyer2022using}                       & \multicolumn{1}{c|}{Unclear}       & Low                    & \multicolumn{1}{c|}{Low}           & Low                    & \multicolumn{1}{c|}{Low}           & Unclear                & High              \\ \hline
\citet{iyer2022voicebio}                    & \multicolumn{1}{c|}{Unclear}       & Low                    & \multicolumn{1}{c|}{Low}           & Low                    & \multicolumn{1}{c|}{Low}           & Unclear                & High              \\ \hline
\citet{Shen2022Establishment}               & \multicolumn{1}{c|}{High}          & High                   & \multicolumn{1}{c|}{Low}           & Low                    & \multicolumn{1}{c|}{Unclear}       & High                   & High              \\ \hline
\citet{Cohen2023A}                          & \multicolumn{1}{c|}{Low}           & Low                    & \multicolumn{1}{c|}{Low}           & Low                    & \multicolumn{1}{c|}{Low}           & Low                    & High              \\ \hline
\citet{Min2023Acoustic}                     & \multicolumn{1}{c|}{Low}           & Low                    & \multicolumn{1}{c|}{Low}           & Low                    & \multicolumn{1}{c|}{Low}           & Low                    & High              \\ \hline
\citet{Pillai2023Investigating}             & \multicolumn{1}{c|}{High}          & High                   & \multicolumn{1}{c|}{Low}           & Low                    & \multicolumn{1}{c|}{Low}           & Unclear                & High              \\ \hline
\citet{amiriparian2024non}                  & \multicolumn{1}{c|}{High}          & High                   & \multicolumn{1}{c|}{Low}           & Low                    & \multicolumn{1}{c|}{High}          & High                   & High              \\ \hline
\citet{Chen2024Fine-grained}                & \multicolumn{1}{c|}{High}          & High                   & \multicolumn{1}{c|}{Low}           & Low                    & \multicolumn{1}{c|}{High}          & High                   & High              \\ \hline
\citet{Cui2024Spontaneous}                  & \multicolumn{1}{c|}{Low}           & High                   & \multicolumn{1}{c|}{Low}           & Low                    & \multicolumn{1}{c|}{Low}           & High                   & Low               \\ \hline
\citet{figueroa2024comparison}              & \multicolumn{1}{c|}{Low}           & High                   & \multicolumn{1}{c|}{Low}           & Low                    & \multicolumn{1}{c|}{Low}           & High                   & Low               \\ \hline
\citet{Gerczuk2024Exploring}                & \multicolumn{1}{c|}{High}          & Low                    & \multicolumn{1}{c|}{Low}           & Low                    & \multicolumn{1}{c|}{Unclear}       & High                   & High              \\ \hline
\citet{Yünden2024Examination}               & \multicolumn{1}{c|}{High}          & Unclear                & \multicolumn{1}{c|}{Unclear}       & Low                    & \multicolumn{1}{c|}{High}          & Unclear                & High              \\ \hline
\end{tabular}%
}
\end{table}

%% file: table-acoustic-definition.tex
% \usepackage{multirow}
% \usepackage{graphicx}
% \usepackage[normalem]{ulem}
\begin{table*}[]
\centering
\caption{Acoustic features Grouped by Feature Category}
\label{tab:acoustic-definition}
\resizebox{0.95\columnwidth}{!}{%
%
}
\end{table*}

%% file: table-summary-1.tex
% \usepackage{graphicx}
\begin{table}[]
\centering
\caption{Summary of study populations and methodologies}
\label{tab:summary_pop}
\resizebox{\columnwidth}{!}{%
%
       & 90 (30 RS, 30 MDD, 30 CS)                                                                                & NP            & NP                                                                                                                                 & NP                                                                                                                                                   & 2024          \\ \hline
\end{tabular}%
}
\scriptsize{AVH: Auditory Verbal Hallucinations, BBRS: Butler-Brown Read Speech, CCHMC: Cincinnati Children's Hospital Medical Center (Ohio, USA), CS: Control Subjects, M: Men, MDD: Major Depressive Disorder, NRS: Not at Risk of Suicide, NP: Non Provided, PCH: Princeton Community Hospital, PT: Persecutory Thoughts, RS: At Risk of Suicide, sd: standard deviation, SNUH: Seoul National University Hospital, UC: University of Cincinnati Medical Center, W: Women, X: Mixed}
\end{table}

%% file: table-summary-2.tex
% \usepackage{graphicx}
\begin{table}[]
\centering
\caption{Summary of data collection settings and acoustic material}
\label{tab:summary_data}
\resizebox{\columnwidth}{!}{%
                                                                                                                              & Turkish                                                    & NP                                                                                                                                            & Controlled environment                                                                                                                       & Pre-determined text read aloud                                                                                                                             & NP                                                                                          & NP                                                                                                                                                                & NP                                                                                                                              & 2024          \\ \hline
\end{tabular}%
}
\scriptsize{AVH: Auditory Verbal Hallucinations, BDI: Beck Depression Inventory, BSSI: Beck Scale for Suicide Ideation, C-SSRS: Columbia suicide severity rating scale, CS: Control Subjects, HDRS: Hamilton Depression Rating Scale, HIPAA: Health Insurance Portability and Accountability Act, MDD: Major Depressive Disorder, MDS: Multimodal Diagnostic System, MINI: Mini International Neuropsychiatric Interview, NP: Non Provided, NRS: Not at Risk of Suicide, PHQ-9: Patient Health Questionnaire 9, PT: Persecutory Thoughts, RS: At Risk of Suicide, SIQ-Jr: suicidal ideation Questionnaire Junior, SIS: Pierce Suicide Intent Scale, SNUH: Seoul National University Hospital, std: standard deviation, UQ: ubiquitous questionnaire (semi-structured interview with 5 open-ended questions)}
\end{table}

%% file: table-summary-31.tex
% \usepackage{graphicx}
% \usepackage[normalem]{ulem}
\begin{table}[]
\centering
\caption{Summary of study objectives, results, and conclusions}
\label{tab:summary_obj1}
\resizebox{\columnwidth}{!}{%
\begin{tabular}{|c|c|c|c|c|}
\hline
\cellcolor[HTML]{FFFFFF}\textbf{Study}      & \textbf{Objective}                                                                                                                                                                                                                                                                                                                                            & \textbf{Results}                                                                                                                                                                                                                                                                                                                                                                                                                     & \textbf{Conclusion}                                                                                                                                                                                                 & \textbf{Year} \\ \hline
\citet{France2000Acoustical}                & \begin{tabular}[c]{@{}c@{}}Evaluate the ability of classifiers to discriminate\\ between RS, depressed, and control subjects\\ based on acoustic features.\end{tabular}                                                                                                                                                                                       & \begin{tabular}[c]{@{}c@{}}- AM, F1, F3, and PSD were the most effective discriminators.\\ - Quadratic discriminant function \\ achieved classification accuracy\\ of 75\% (RS), 71\% (depression), and 75\% (control).\end{tabular}                                                                                                                                                                                                 & \begin{tabular}[c]{@{}c@{}}Multimodal classifiers\\ significantly improve classification\\ performance over single-feature models.\end{tabular}                                                                     & 2000          \\ \hline
\citet{Ozdas2000Analysis}                   & \begin{tabular}[c]{@{}c@{}}Assess the effectiveness of jitter in distinguishing\\ RS from NRS individuals, \\ with statistical comparisons\\ between groups.\end{tabular}                                                                                                                                                                                     & \begin{tabular}[c]{@{}c@{}}- ANOVA test: Jitter showed a significant difference\\ between RS and NRS (p = 0.0069).\\ Classifier performance: 80\% accuracy in differentiating\\ suicidal from non-suicidal individuals.\end{tabular}                                                                                                                                                                                                 & \begin{tabular}[c]{@{}c@{}}Jitter-based classification \\ effectively discriminates\\ suicidal speech from healthy speech.\end{tabular}                                                                             & 2000          \\ \hline
\citet{Ozdas2004Analysis}                   & \begin{tabular}[c]{@{}c@{}}Assess the discrimination performance between\\ RS, depressed, and control subjects using\\ MFCC acoustic features.\end{tabular}                                                                                                                                                                                                   & \begin{tabular}[c]{@{}c@{}}80\% accuracy in classifying RS vs. \\ depressed using the first\\ four MFCC with a GMM classifier.\end{tabular}                                                                                                                                                                                                                                                                                          & \begin{tabular}[c]{@{}c@{}}MFCC-based classification \\ performed well, \\ with GMM significantly \\ outperforming the unimodal\\ Gaussian approach.\end{tabular}                                                   & 2004          \\ \hline
\citet{Ozdas2004Investigation}              & \begin{tabular}[c]{@{}c@{}}Evaluate the discrimination performance between\\ RS, depressed, and control subjects using\\ multiple acoustic features.\end{tabular}                                                                                                                                                                                             & \begin{tabular}[c]{@{}c@{}}T-test: Jitter was significantly different only between RS and CS;\\ spectral slope differed across all groups.\\ Classifier performance: \\ - Jitter alone: 80\% accuracy for RS vs. CS,\\ but poor for RS vs. depressed (60\%).\\ - Spectral slope: 74\% accuracy for RS vs. depressed.\\ - Combined jitter + slope: 85\% accuracy \\ (RS vs. CS), 75\% (RS vs. depressed).\end{tabular}                & \begin{tabular}[c]{@{}c@{}}Multiparametric classification \\ significantly improves\\ discrimination accuracy, \\ highlighting the superiority of\\ feature combination over \\ single-feature models.\end{tabular} & 2004          \\ \hline
\citet{Yingthawornsuk2006Objective}         & \begin{tabular}[c]{@{}c@{}}Use a ML model to discriminate RS from NRS\\ individuals based on acoustic features.\end{tabular}                                                                                                                                                                                                                                  & \begin{tabular}[c]{@{}c@{}}Maintenance session: 77\% accuracy for differentiating\\ depressed from RS patients, 85\% for RS vs. remitted patients.\\ Reading session: 82\% accuracy for RS vs. depressed patients.\\ -> PSD was the most effective feature for distinguishing\\ remitted speech from depressive and RS speech.\end{tabular}                                                                                          & \begin{tabular}[c]{@{}c@{}}Reading-based classification \\ showed strong performance\\ and may be a valuable approach \\ for suicide risk assessment.\end{tabular}                                                  & 2006          \\ \hline
\citet{Kaymaz-Keskinpala2007Distinguishing} & \begin{tabular}[c]{@{}c@{}}Distinguish RS from MDD patients\\ using MFCC and cross-validation\end{tabular}                                                                                                                                                                                                                                                    & \begin{tabular}[c]{@{}c@{}}Best : 93\% accuracy for RS classification in men reading task\\ Higher results for only RS or only MDD patients, lower for both\end{tabular}                                                                                                                                                                                                                                                             & \begin{tabular}[c]{@{}c@{}}MFCC can help discriminate RS\\ behavior in MDD patients.\end{tabular}                                                                                                                   & 2007          \\ \hline
\citet{Kaymaz-Keskinpala2007Screening}      & \begin{tabular}[c]{@{}c@{}}Evaluate the classification performance between RS, \\ depressed, and remitted individuals using spectral\\ acoustic features.\end{tabular}                                                                                                                                                                                        & \begin{tabular}[c]{@{}c@{}}- 87\% accuracy for women (reading session) using PSD2.\\ - Classification improved from 69.23\% \\ to 72.53\% (maintenance session)\\ and 67.82\% to 73.37\% (reading session) \\ when using 8 MFCC instead of 4.\end{tabular}                                                                                                                                                                           & \begin{tabular}[c]{@{}c@{}}MFCC and PSDs effectively \\ distinguish RS from NRS,\\ with better performance \\ in controlled reading sessions\\ than in interviews.\end{tabular}                                     & 2007          \\ \hline
\citet{Yingthawornsuk2007AcousticAO}        & \begin{tabular}[c]{@{}c@{}}Evaluate the classification performance between RS\\ and depressed NRS women \\ using spectral acoustic features.\end{tabular}                                                                                                                                                                                                     & \begin{tabular}[c]{@{}c@{}}Interview session: PSD1, PSD2, CF2, and BW2\\ were the most discriminating features.\\ Reading session: PSD1, PSD2, and CF2\\ achieved 90.33\% accuracy.\end{tabular}                                                                                                                                                                                                                                     & \begin{tabular}[c]{@{}c@{}}PSD1, PSD2, CF2, and BW2 \\ effectively distinguish RS\\ from NRS women, \\ with better classification\\ performance in the reading session.\end{tabular}                                & 2007          \\ \hline
\citet{Yingthawornsuk2008Distinguishing}    & \begin{tabular}[c]{@{}c@{}}Evaluate the classification performance between RS,\\ depressed, and remitted men using spectral\\ acoustic features with a GMM classifier.\end{tabular}                                                                                                                                                                           & \begin{tabular}[c]{@{}c@{}}Automatic (reading) session:\\ - Depressed vs. RS: 88.5\% accuracy\\ - Remitted vs. RS: 90.25\% accuracy\\ Spontaneous (interview) session:\\ - Depressed vs. RS: 85.58\% accuracy\\ - Remitted vs. RS: 81.08\% accuracy\\ Most discriminating features: CF4, CF3, PSD2\end{tabular}                                                                                                                      & \begin{tabular}[c]{@{}c@{}}Classification accuracy \\ varies by speech type, \\ with better performance \\ in automatic (reading)\\ speech samples than in \\ spontaneous (interview) speech.\end{tabular}          & 2008          \\ \hline
\citet{Subari2010Comparison}                & \begin{tabular}[c]{@{}c@{}}Compare speaker normalization techniques\\ for classifying RS, MDD, CS patients\end{tabular}                                                                                                                                                                                                                                       & Accuracy from 75\% to 90\% depending on technique                                                                                                                                                                                                                                                                                                                                                                                    & \begin{tabular}[c]{@{}c@{}}ML and formant slope methods improve\\ classification rates of RS individuals\end{tabular}                                                                                               & 2010          \\ \hline
\citet{hashim2012analysis}                  & \begin{tabular}[c]{@{}c@{}}Identify speech signals most effective in\\ distinguishing RS from NRS depressed\\ patients using temporal models\\ (transition features and interval \\ Power Density Functions - PDFs).\end{tabular}                                                                                                                             & \begin{tabular}[c]{@{}c@{}}- Transition feature alone achieved 74\% accuracy.\\ - Some PDF combinations (voiced/silent) \\ reached 75\%-100\% accuracy.\\ Best performance:\\ 94\% accuracy using a single transition feature +\\ one interval feature PDF.\end{tabular}                                                                                                                                                             & \begin{tabular}[c]{@{}c@{}}Transition and interval feature \\ PDFs effectively\\ classify high-risk suicidal \\ and depressive patients.\end{tabular}                                                               & 2012          \\ \hline
\citet{Scherer2013Investigating}            & \begin{tabular}[c]{@{}c@{}}Evaluate ML's ability to classify RS and\\ NRS adolescents using acoustic features from\\ text reading and clinical interviews.\end{tabular}                                                                                                                                                                                       & \begin{tabular}[c]{@{}c@{}}- HMM: 81.25\% accuracy,\\ SVM: 75\% accuracy (interview analysis). \\ - T-test revealed significant\\ vocal differences between RS and NRS adolescents.\end{tabular}                                                                                                                                                                                                                                     & \begin{tabular}[c]{@{}c@{}}HMM and SVM classifiers \\ performed well in\\ distinguishing suicidal \\ from non-suicidal adolescents.\end{tabular}                                                                    & 2013          \\ \hline
\citet{Wahidah2015Investigating}            & \begin{tabular}[c]{@{}c@{}}Assess whether patients' vocal \\ characteristics change significantly\\ from the initial high-risk \\ (RS) recording to follow-up \\ sessions after treatment.\end{tabular}                                                                                                                                                       & \begin{tabular}[c]{@{}c@{}}T-test showed a statistically significant difference in\\ spectral density power (SDP) between the first (RS) and\\ second post-treatment session.\end{tabular}                                                                                                                                                                                                                                           & \begin{tabular}[c]{@{}c@{}}SDP features can track psychological\\ severity progression in RS patients after\\ hospitalization and treatment.\end{tabular}                                                           & 2015          \\ \hline
\citet{Wahidah2015TimingPO}                 & \begin{tabular}[c]{@{}c@{}}Study 1: Assess the ability of timing patterns and\\ spectral features to differentiate \\ RS from depressed individuals.\\ - Phase 1: Train classifiers on Database A.\\ - Phase 2: Use best-performing \\ features to classify RS in Database B.\\ Study 2: Determine if these \\ features can predict HAMD scores.\end{tabular} & \begin{tabular}[c]{@{}c@{}}- 91\% accuracy in RS classification \\ using the interval feature "voiced16:20".\\ - Acoustic features showed potential for \\ predicting HAMD scores (linear regression).\end{tabular}                                                                                                                                                                                                                  & \begin{tabular}[c]{@{}c@{}}Combining two acoustic \\ features yielded the\\ best classification performance, \\ confirming their effectiveness\\ in suicide risk and depression assessment.\end{tabular}            & 2015          \\ \hline
\citet{Pestian2017A}                        & \begin{tabular}[c]{@{}c@{}}Use a ML model to discriminate \\ between RS and NRS individuals\\ based on acoustic and/or linguistic features.\end{tabular}                                                                                                                                                                                                      & \begin{tabular}[c]{@{}c@{}}- Acoustic features alone: AUC = 0.79 (±0.03)\\ - Acoustic + linguistic features: AUC = 0.92 (±0.02)\end{tabular}                                                                                                                                                                                                                                                                                         & \begin{tabular}[c]{@{}c@{}}Combining acoustic and linguistic features\\ significantly improves \\ classification performance.\end{tabular}                                                                          & 2017          \\ \hline
\citet{Venek2017Adolescent}                 & \begin{tabular}[c]{@{}c@{}}Assess a classification algorithm's ability\\ to distinguish RS from NRS individuals\\ and differentiate RS with or without a recent act\\ using verbal, non-verbal, \\ and interactional features.\end{tabular}                                                                                                                   & \begin{tabular}[c]{@{}c@{}}- Multiparametric classifier (18 multimodal features)\\ achieved 90\% accuracy for RS vs. NRS classification.\\ - 60\% accuracy for distinguishing \\ suicidal from non-suicidal recidivism.\\ - Nonverbal acoustic features were the most useful for\\ recidivism prediction (73.3\% accuracy, \\ hierarchical classification).\\ - ANOVA confirmed significant differences between groups.\end{tabular} & \begin{tabular}[c]{@{}c@{}}Verbal, non-verbal, and interactional \\ features effectively classify\\ RS/NRS individuals and \\ predict recidivism.\end{tabular}                                                      & 2017          \\ \hline
\end{tabular}%
}
\scriptsize{\\AM: Amplitude Modulation, AUC: Area Under the Receiver-Operating Characteristic curve, BW: Bandwidth, CF: Center Frequencies, CS: Control Subjects, F0: Fundamental Frequency, F1,F2,F3: Formant Frequencies, GMM: Gaussian Mixture Model, HAMD: Hamilton Depression Rating Scale, HMM: Hidden Markov Models, MDD: Major Depressive Disorder, MFCC: Mel-Frequency Cepstral Coefficients, ML: Machine Learning, NRS: Not at Risk of Suicide, PDFs: Power Density Functions, PSD: Power Spectral Density, RS: At Risk of Suicide, SDP: Spectral Density Power, SVM: Support Vector Machine}
\end{table}

%% file: table-summary-32.tex
% \usepackage{graphicx}
% \usepackage[normalem]{ulem}
\begin{table}[]
\centering

\label{tab:summary_obj2}
\resizebox{\columnwidth}{!}{%
\begin{tabular}{|c|c|c|c|c|}
\hline
\cellcolor[HTML]{FFFFFF}\textbf{Study} & \textbf{Objective}                                                                                                                                                                                                                                                                                                                                                                        & \textbf{Results}                                                                                                                                                                                                                                                                                                                    & \textbf{Conclusion}                                                                                                                                                                                                                                                                                     & \textbf{Year} \\ \hline
\citet{Chakravarthula2019Automatic}    & \begin{tabular}[c]{@{}c@{}}Assess whether acoustic, lexical, \\ and interaction dynamics features\\ can classify individuals as RS vs. NRS\\ or predict their degree of suicidal risk.\end{tabular}                                                                                                                                                                                       & \begin{tabular}[c]{@{}c@{}}- Interaction dynamics were the most informative features.\\ - Partitioning data by gender and other factors\\ improved classification performance, \\ especially for RS vs. NRS. \\ - Classification was more effective for RS vs. NRS than for\\ differentiating levels of suicidal risk.\end{tabular} & \begin{tabular}[c]{@{}c@{}}Partition-based approaches, particularly gender-based,\\ enhance classification performance, \\ but predicting the severity\\ of suicidal risk remains challenging\end{tabular}                                                                                              & 2019          \\ \hline
\citet{Shah2019Multimodal}             & \begin{tabular}[c]{@{}c@{}}Assess the discriminative \\ power of verbal, acoustic,\\ and visual behavioral markers \\ for classifying RS and NRS\\ individuals using social media audio samples.\end{tabular}                                                                                                                                                                             & \begin{tabular}[c]{@{}c@{}}Statistical test: No significant difference in\\ formants between RS and NRS.\\ Classifier performance: - XGB: AUC = 0.62 (acoustic only), \\ 0.72 (acoustic + verbal), 0.74 (trimodal).\\ - RF: AUC = 0.70 (trimodal).\end{tabular}                                                                     & \begin{tabular}[c]{@{}c@{}}Combining acoustic and linguistic features improves\\ classification performance across different models.\end{tabular}                                                                                                                                                       & 2019          \\ \hline
\citet{Belouali2021Acoustic}           & \begin{tabular}[c]{@{}c@{}}Evaluate the performance of different ML models\\ using acoustic and linguistic features\\ to classify RS and NRS individuals,\\ with statistical comparisons between groups.\end{tabular}                                                                                                                                                                     & \begin{tabular}[c]{@{}c@{}}- RF with acoustic features: AUC = 0.76 ($\pm$0.06)\\ - RF with acoustic + linguistic features: AUC = 0.80 ($\pm$0.06)\\ - Combining acoustics and linguistics\\ improved classification performance.\end{tabular}                                                                                               & \begin{tabular}[c]{@{}c@{}}Audio data collected via a mobile app can\\ classify RS and NRS individuals in US Army veterans\\ with AUC = 0.80, 86\% recall, and 70\% specificity.\end{tabular}                                                                                                           & 2021          \\ \hline
\citet{Saavedra2021Association}        & \begin{tabular}[c]{@{}c@{}}Examine the link between acoustic features\\ (F0, F1, F2, F3, Energy, Jitter)\\ and suicidal ideation in \\ university students in Temuco, Chile.\end{tabular}                                                                                                                                                                                                 & \begin{tabular}[c]{@{}c@{}}- Significant differences in F0, F1, F2, and Jitter\\ between RS and NRS groups (p < .05).\\ - Higher suicide intention intensity correlates with increased \\ Jitter and decreased F0, F1, and F2.\end{tabular}                                                                                         & \begin{tabular}[c]{@{}c@{}}Acoustic features vary significantly in Chilean\\ university students with suicidal behavior.\end{tabular}                                                                                                                                                                   & 2021          \\ \hline
\citet{Stasak2021Read}                 & \begin{tabular}[c]{@{}c@{}}Evaluation of GRBASI voice quality attributes\\ ("roughness," "breathiness") \\ and speech disfluencies\\ for automatic discrimination \\ between psychiatric inpatients (RS)\\ and individuals with no suicide history.\end{tabular}                                                                                                                          & \begin{tabular}[c]{@{}c@{}}RS had lower manually annotated voice quality and higher\\ speech disfluency than CS. Classification performance:\\ - GRBASI alone: 54-69\% accuracy (SVM/kNN). \\  - GRBASI + disfluency: 77-78\% accuracy.\end{tabular}                                                                                & \begin{tabular}[c]{@{}c@{}}Automatic classification achieved up to 80\% accuracy.\\ Higher speech disfluency correlated with more\\ severe depression (BDI-II) and increased suicide risk.\end{tabular}                                                                                                 & 2021          \\ \hline
\citet{iyer2022using}                  & \begin{tabular}[c]{@{}c@{}}Classify psychological distress in helpline callers\\ using vocal characteristics and machine learning.\end{tabular}                                                                                                                                                                                                                                           & \begin{tabular}[c]{@{}c@{}}AUC = 0.974 (95\% CI: 0.962-0.985)\\ Accuracy: Low distress: 93.39\% High distress: 94.64\%\end{tabular}                                                                                                                                                                                                 & \begin{tabular}[c]{@{}c@{}}AI models accurately classify \\ distress in helpline callers,\\ enabling real-time triage and intervention.\end{tabular}                                                                                                                                                    & 2022          \\ \hline
\citet{iyer2022voicebio}               & \begin{tabular}[c]{@{}c@{}}Automatically classify suicide risk\\ in telehealth callers using voice biomarkers\\ and machine learning.\end{tabular}                                                                                                                                                                                                                                        & Classification accuracy of 99.85\%, AUC of 0.985                                                                                                                                                                                                                                                                                    & \begin{tabular}[c]{@{}c@{}}Feasibility of voice-based suicide risk classification\\ with high accuracy, potential for real-time application\end{tabular}                                                                                                                                                & 2022          \\ \hline
\citet{Shen2022Establishment}          & \begin{tabular}[c]{@{}c@{}}Develop and validate an EWLSRA,\\ evaluate the psychometric characteristics\\ (reliability and validity) of the list,\\ determine associations between emotional words\\ and suicide risk\end{tabular}                                                                                                                                                         & \begin{tabular}[c]{@{}c@{}}Despair ($\rho$=0.54, P<0.001), sadness ($\rho$=0.37, P=0.006),\\ helplessness ($\rho$=0.45, P=0.001), \\ numbness ($\rho$=0.35, P=0.009)\\ correlated with suicide risk. Higher numbness index in suicide\\ attempters vs. non-attempters (P=0.049).\end{tabular}                                       & \begin{tabular}[c]{@{}c@{}}EWLSRA: adequate psychometric properties.\\ Supports AI-based suicide risk assessment.\\ Future refinement and broader application needed.\end{tabular}                                                                                                                      & 2022          \\ \hline
\citet{Cohen2023A}                     & \begin{tabular}[c]{@{}c@{}}Evaluate Multimodal Diagnostic \\ System (MDS) feasibility, \\ effectiveness, and interpretability\\ for remote monitoring of depression,\\ anxiety, and RS. Analyze mental disorders\\ using multimodal features \\ (speech, language, facial behavior) \\ and assess ML models' \\ discriminative ability for mental \\ state characterization.\end{tabular} & \begin{tabular}[c]{@{}c@{}}RS identification achieved AUC = 0.65 with text features,\\ improving to AUC = 0.76 using decision fusion.\\ Single-modality models performed near chance, \\ but fusion enhanced discrimination.\end{tabular}                                                                                           & \begin{tabular}[c]{@{}c@{}}MDS was found feasible, effective, and interpretable \\ for remote monitoring of depression,\\ anxiety, and RS. Modality combinations improved \\ ML-based mental state characterization.\\ MDS showed promise as a scalable, \\ cost-effective solution.\end{tabular}       & 2023          \\ \hline
\citet{Min2023Acoustic}                & \begin{tabular}[c]{@{}c@{}}Voice changes over time were analyzed to predict\\ suicidality worsening using ML classifiers\\ on acoustic features of psychiatric patients.\end{tabular}                                                                                                                                                                                                     & \begin{tabular}[c]{@{}c@{}}Between-Person Model: \\ ANN (68.7\%) outperformed XGB (63.7\%) \\ and LightGBM (61.1\%),\\ while RF and SVM performed worse. \\ Within-Person Model: \\ XGB achieved 79.2\% (internal validation, 2M) \\ and 61.8\% (external validation, 2M $\rightarrow$ 4M).\end{tabular}                                        & \begin{tabular}[c]{@{}c@{}}A within-person model outperformed a\\ between-person model in detecting high RS\\ from acoustic features, \\ emphasizing individual variability\\ and supporting personalized \\ ML-based RS assessment.\end{tabular}                                                       & 2023          \\ \hline
\citet{Pillai2023Investigating}        & \begin{tabular}[c]{@{}c@{}}Investigate the generalizability of\\ speech-based suicidal ideation\\ detection models using mobile phone recordings\end{tabular}                                                                                                                                                                                                                             & \begin{tabular}[c]{@{}c@{}}Speech-based suicidal ideation\\ detection ranges from 0.62 to 0.68 for different datasets\end{tabular}                                                                                                                                                                                                  & \begin{tabular}[c]{@{}c@{}}Data similarity $\neq$ better transferability. \\ Domain adaptation (Minimax Entropy,\\ Contrastive Learning) improves generalization.\end{tabular}                                                                                                                               & 2023          \\ \hline
\citet{amiriparian2024non}             & \begin{tabular}[c]{@{}c@{}}Develop a non-invasive, speech-based method\\ for RS assessment using ML\end{tabular}                                                                                                                                                                                                                                                                          & \begin{tabular}[c]{@{}c@{}}Speech-based classification accuracy:\\ 66.2\% (best using eGeMAPS features on vowels)\\ Speech + metadata classification accuracy: 94.4\%\end{tabular}                                                                                                                                                  & \begin{tabular}[c]{@{}c@{}}Speech alone is a moderate predictor of suicide risk,\\ but when combined with metadata, \\ classification accuracy\\ significantly improves.\end{tabular}                                                                                                                   & 2024          \\ \hline
\citet{Chen2024Fine-grained}           & \begin{tabular}[c]{@{}c@{}}Develop and assess speech emotion recognition\\ models for fine-grained negative emotions in\\ hotline calls to support suicide prevention.\end{tabular}                                                                                                                                                                                                       & \begin{tabular}[c]{@{}c@{}}Negative emotion recognition: F1 = 76.96\% (Wav2vec2.0). \\ Fine-grained classification: \\ Weighted F1 = 41.74\% (Whisper-large-v3). \\ Challenges: subtle emotions, context-dependent emotions.\end{tabular}                                                                                           & \begin{tabular}[c]{@{}c@{}}Deep Learning models effective \\ for binary classification,\\ struggled with fine-grained multi-label.\\ Performance affected by dataset size, architecture.\end{tabular}                                                                                                   & 2024          \\ \hline
\citet{Cui2024Spontaneous}             & \begin{tabular}[c]{@{}c@{}}Investigate RS detection\\ using speech and LLMs\end{tabular}                                                                                                                                                                                                                                                                                                  & \begin{tabular}[c]{@{}c@{}}Best accuracy: 0.807\\ Best F1-score: 0.846\end{tabular}                                                                                                                                                                                                                                                 & \begin{tabular}[c]{@{}c@{}}Speech and LLM-based fusion methods show\\ potential for real-world \\ suicide risk detection applications\end{tabular}                                                                                                                                                      & 2024          \\ \hline
\citet{figueroa2024comparison}         & \begin{tabular}[c]{@{}c@{}}Analyze acoustic variations by vowel type across\\ suicidal risk levels in adolescents.\end{tabular}                                                                                                                                                                                                                                                           & \begin{tabular}[c]{@{}c@{}}Acoustic differences in RS vs. NRS adolescents.\\ Women: F0, HNRdB, jitter, SDP. Men: F0, HNRdB (P < 0.05).\end{tabular}                                                                                                                                                                                 & \begin{tabular}[c]{@{}c@{}}Acoustic features of voice and speech\\ may serve as indicators of suicide risk\end{tabular}                                                                                                                                                                                 & 2024          \\ \hline
\citet{Gerczuk2024Exploring}           & \begin{tabular}[c]{@{}c@{}}Investigate gender differences\\ in speech for suicide risk,\\ traditional vs. deep-learning features, \\ impact of gender-based\\ modeling on classification.\end{tabular}                                                                                                                                                                                    & \begin{tabular}[c]{@{}c@{}}Gender-based modeling improved accuracy,\\ best: Emotion-finetuned Wav2vec2.0, \\ gender-exclusive modeling,\\ 81\% balanced accuracy (subject-level)\end{tabular}                                                                                                                                       & \begin{tabular}[c]{@{}c@{}}Gender-based modeling improves classification. \\ Suicidal speech differs by gender. \\ Men: risk $\leftrightarrow$ agitation. \\ Women: risk $\leftrightarrow$ depression\\ ($\downarrow$ pitch, $\downarrow$ energy). \\ DL models (Wav2vec2.0) effective. \\ Larger datasets needed.\end{tabular} & 2024          \\ \hline
\citet{Yünden2024Examination}          & \begin{tabular}[c]{@{}c@{}}Analyze speech characteristics to\\ predict RS using AI-based models\end{tabular}                                                                                                                                                                                                                                                                              & MFCC more successful with rates of 90\% accuracy                                                                                                                                                                                                                                                                                    & \begin{tabular}[c]{@{}c@{}}Speech features (MFCC) can help predict suicidal\\ behavior in depression.\end{tabular}                                                                                                                                                                                      & 2024          \\ \hline
\end{tabular}%
}
\scriptsize{\\AI: Artificial Intelligence, ANN: Artificial Neural Network, AUC: Area Under the Receiver-Operating Characteristic curve, BDI-II: Beck Depression Inventory II, CF: Center Frequencies, CI: Confidence Interval, CS: Control Subjects, DL: Deep Learning, EWLSRA: Enhanced Weighted Linear Spectral Regression Analysis, eGeMAPS: extended Geneva Minimalistic Acoustic Parameter Set, F0: Fundamental Frequency, F1,F2,F3: Formant Frequencies, GRBASI: Grade Roughness Breathiness Asthenia Strain Instability, HNRdB: Harmonics-to-Noise Ratio, kNN: k-Nearest Neighbors Algorithm, LightGBM: Light Gradient Boosting Machine, LLM: Large Language Model, MDD: Major Depressive Disorder, MDS: Multimodal Diagnostic System, MFCC: Mel-Frequency Cepstral Coefficients, ML: Machine Learning, NRS: Not at Risk of Suicide, RF: Random Forest, RS: At Risk of Suicide, SDP: Spectral Density Power, SVM: Support Vector Machine, XGB: eXtreme Gradient Boosting}

\end{table}

%% file: table-aggregatestat1.tex
% \usepackage{multirow}
% \usepackage{graphicx}
\begin{table}[]
\centering
\caption{Citations for aggregate statistics}
\label{tab:aggregatestat1}
\resizebox{\columnwidth}{!}{%
                  & Unimodal                                                                               \\ \hline
\end{tabular}%
}
\scriptsize{BD: Bipolar Disorder, CV: Cross-Validation, GMM: Gaussian Mixture Model, LDA: Linear Discriminant Analysis, LOOCV: Leave-one-out Cross-Validation, M: Men, MDD: Major Depressive Disorder, ML: Machine Learning, MLE: Maximum Likelihood Estimation, NP: Not Provided, NPV: Negative Predictive Value, NRS: Not at Risk of Suicide, PPV: Positive Predictive Value, QDA: Quadratic Discriminant Analysis, RS: At Risk of Suicide, W: Women, X: Mixed}
\end{table}

%% file: table-aggregatestat2.tex
% \usepackage{multirow}
% \usepackage{graphicx}
\begin{table}[]
\centering
\label{tab:aggregatestat2}
\resizebox{\columnwidth}{!}{%
                         & Unimodal                                                                               \\ \hline
\end{tabular}%
}
\scriptsize{AUC: Area Under the Receiver-Operating Characteristic curve, BD: Bipolar Disorder, CV: Cross-Validation, DNN: Deep Neural Network, EEG: Electroencephalography, GB: Gradient Boosting, HMM: Hidden Markov Model, kNN: k-Nearest Neighbors Algorithm, LDA: Linear Discriminant Analysis, LOOCV: Leave-one-out Cross-Validation, LR: Logistic Regression, MDD: Major Depressive Disorder, ML: Machine Learning, NP: Not Provided, NPV: Negative Predictive Value, NRS: Not at Risk of Suicide, PPV: Positive Predictive Value, QDA: Quadratic Discriminant Analysis, RF: Random Forest, RS: At Risk of Suicide, sd: standard deviation, SVM: Support Vector Machine, X: Mixed, XGB: eXtreme Gradient Boosting}
\end{table}

%% file: table-aggregatestat3.tex
% \usepackage{multirow}
% \usepackage{graphicx}
\begin{table}[]
\centering
\label{tab:aggregatestat3}
\resizebox{\columnwidth}{!}{%
                                                                          & NP                                                                                        & Unimodal                                                                               \\ \hline
\end{tabular}%
}
\scriptsize{ANN: Artificial Neural Network, AUC: Area Under the Receiver-Operating Characteristic curve, BD: Bipolar Disorder, CV: Cross-Validation, DNN: Deep Neural Network, EEG: Electroencephalography, GB: Gradient Boosting, LightGBM: Light Gradient Boosting Machine, LOOCV: Leave-one-out Cross-Validation, LR: Logistic Regression, MDD: Major Depressive Disorder, ML: Machine Learning, NP: Not Provided, NPV: Negative Predictive Value, NRS: Not at Risk of Suicide, PPV: Positive Predictive Value, RF: Random Forest, RS: At Risk of Suicide, sd: standard deviation, SVM: Support Vector Machine, VGGish: Visual Geometry Group-ish (audio embedding model derived from VGG), X: Mixed, XGB: eXtreme Gradient Boosting}

\end{table}

%% file: table-prismachecklist.tex
% \usepackage{multirow}
% \usepackage{graphicx}
\begin{table}[]
\centering
\caption{PRISMA 2020 Checklist}
\label{tab:prisma}
\resizebox{\columnwidth}{!}{%
\begin{tabular}{|cccc|}
\hline
\multicolumn{1}{|c|}{Section and Topic}                  & \multicolumn{1}{c|}{Item} & \multicolumn{1}{c|}{Checklist Item}                                                                                                                                                                                                                                                                                                                                           & Location where item is reported                           \\ \hline
\multicolumn{4}{|l|}{TITLE}                                                                                                                                                                                                                                                                                                                                                                                                                                                                                                      \\ \hline
\multicolumn{1}{|c|}{Title}                              & \multicolumn{1}{c|}{1}    & \multicolumn{1}{c|}{Identify the report as a systematic review.}                                                                                                                                                                                                                                                                                                              & Page \pageref{title1}                                     \\ \hline
\multicolumn{4}{|l|}{ABSTRACT}                                                                                                                                                                                                                                                                                                                                                                                                                                                                                                   \\ \hline
\multicolumn{1}{|c|}{Abstract}                           & \multicolumn{1}{c|}{2}    & \multicolumn{1}{c|}{See the PRISMA 2020 for Abstracts Checklist.}                                                                                                                                                                                                                                                                                                             & Page \pageref{abstract2}                                  \\ \hline
\multicolumn{4}{|l|}{INTRODUCTION}                                                                                                                                                                                                                                                                                                                                                                                                                                                                                               \\ \hline
\multicolumn{1}{|c|}{Rationale}                          & \multicolumn{1}{c|}{3}    & \multicolumn{1}{c|}{\begin{tabular}[c]{@{}c@{}}Describe the rationale for the review \\ in the context of existing knowledge.\end{tabular}}                                                                                                                                                                                                                                   & Section \ref{introduction3}, Page \pageref{introduction3} \\ \hline
\multicolumn{1}{|c|}{Objectives}                         & \multicolumn{1}{c|}{4}    & \multicolumn{1}{c|}{\begin{tabular}[c]{@{}c@{}}Provide an explicit statement of the objective(s) \\ or question(s) the review addresses.\end{tabular}}                                                                                                                                                                                                                        & Section \ref{introduction4}, Page \pageref{introduction4} \\ \hline
\multicolumn{4}{|l|}{METHODS}                                                                                                                                                                                                                                                                                                                                                                                                                                                                                                    \\ \hline
\multicolumn{1}{|c|}{Eligibility Criteria}               & \multicolumn{1}{c|}{5}    & \multicolumn{1}{c|}{\begin{tabular}[c]{@{}c@{}}Specify the inclusion and exclusion criteria for the review \\ and how studies were grouped for the syntheses\end{tabular}}                                                                                                                                                                                                    & Section \ref{methods5}, Page \pageref{methods5}           \\ \hline
\multicolumn{1}{|c|}{Information sources}                & \multicolumn{1}{c|}{6}    & \multicolumn{1}{c|}{\begin{tabular}[c]{@{}c@{}}Specify all databases, registers, websites, organisations, \\ reference lists and other sources searched \\ or consulted to identify studies. \\ Specify the date when each source was last searched or consulted.\end{tabular}}                                                                                               & Section \ref{methods6}, Page \pageref{methods6}           \\ \hline
\multicolumn{1}{|c|}{Search strategy}                    & \multicolumn{1}{c|}{7}    & \multicolumn{1}{c|}{\begin{tabular}[c]{@{}c@{}}Present the full search strategies for all databases, \\ registers and websites, including any filters and limits used.\end{tabular}}                                                                                                                                                                                          & Section \ref{methods6}, Page \pageref{methods6}           \\ \hline
\multicolumn{1}{|c|}{Selection process}                  & \multicolumn{1}{c|}{8}    & \multicolumn{1}{c|}{\begin{tabular}[c]{@{}c@{}}Specify the methods used to decide whether \\ a study met the inclusion criteria of the review, \\ including how many reviewers screened \\ each record and each report retrieved, \\ whether they worked independently, and if applicable, \\ details of automation tools used in the process.\end{tabular}}                  & Section \ref{methods8}, Page \pageref{methods8}           \\ \hline
\multicolumn{1}{|c|}{Data collection process}            & \multicolumn{1}{c|}{9}    & \multicolumn{1}{c|}{\begin{tabular}[c]{@{}c@{}}Specify the methods used to collect data from reports, \\ including how many reviewers collected data from each report, \\ whether they worked independently, \\ any processes for obtaining or confirming data from study investigators, \\ and if applicable, details of automation tools used in the process.\end{tabular}} & Section \ref{methods9}, Page \pageref{methods9}           \\ \hline
\multicolumn{1}{|c|}{\multirow{2}{*}{Data items}}        & \multicolumn{1}{c|}{10a}  & \multicolumn{1}{c|}{\begin{tabular}[c]{@{}c@{}}List and define all outcomes for which data were sought. \\ Specify whether all results that were compatible \\ with each outcome domain in each study were sought \\ (e.g. for all measures, time points, analyses), \\ and if not, the methods used to decide which results to collect.\end{tabular}}                        & Section \ref{methods9}, Page \pageref{methods9}           \\ \cline{2-4} 
\multicolumn{1}{|c|}{}                                   & \multicolumn{1}{c|}{10b}  & \multicolumn{1}{c|}{\begin{tabular}[c]{@{}c@{}}List and define all other variables for which data were sought \\ (e.g. participant and intervention characteristics, funding sources). \\ Describe any assumptions made about any missing \\ or unclear information.\end{tabular}}                                                                                            & Section \ref{methods9}, Page \pageref{methods9}           \\ \hline
\multicolumn{1}{|c|}{Study risk of bias assessment}      & \multicolumn{1}{c|}{11}   & \multicolumn{1}{c|}{\begin{tabular}[c]{@{}c@{}}Specify the methods used to assess risk of bias \\ in the included studies, including details of the tool(s) used, \\ how many reviewers assessed each study \\ and whether they worked independently, \\ and if applicable, details of automation tools used in the process.\end{tabular}}                                    & Section \ref{methods11}, Page \pageref{methods11}         \\ \hline
\multicolumn{1}{|c|}{Effect measures}                    & \multicolumn{1}{c|}{12}   & \multicolumn{1}{c|}{\begin{tabular}[c]{@{}c@{}}Specify for each outcome the effect measure(s) \\ (e.g. risk ratio, mean difference) \\ used in the synthesis or presentation of results\end{tabular}}                                                                                                                                                                         & Section \ref{methods9}, Page \pageref{methods9}           \\ \hline
\multicolumn{1}{|c|}{\multirow{6}{*}{Synthesis methods}} & \multicolumn{1}{c|}{13a}  & \multicolumn{1}{c|}{\begin{tabular}[c]{@{}c@{}}Describe the processes used to decide \\ which studies were eligible for each synthesis \\ (e.g. tabulating the study intervention characteristics and \\ comparing against the planned groups for each synthesis (item \#5)).\end{tabular}}                                                                                   & Section \ref{methods5}, Page \pageref{methods5}           \\ \cline{2-4} 
\multicolumn{1}{|c|}{}                                   & \multicolumn{1}{c|}{13b}  & \multicolumn{1}{c|}{\begin{tabular}[c]{@{}c@{}}Describe any methods required to prepare the data \\ for presentation or synthesis, such as handling of missing \\ summary statistics, or data conversions.\end{tabular}}                                                                                                                                                      & Section \ref{methods11}, Page \pageref{methods11}         \\ \cline{2-4} 
\multicolumn{1}{|c|}{}                                   & \multicolumn{1}{c|}{13c}  & \multicolumn{1}{c|}{\begin{tabular}[c]{@{}c@{}}Describe any methods used to tabulate or \\ visually display results of individual studies and syntheses.\end{tabular}}                                                                                                                                                                                                        & Section \ref{methods11}, Page \pageref{methods11}         \\ \cline{2-4} 
\multicolumn{1}{|c|}{}                                   & \multicolumn{1}{c|}{13d}  & \multicolumn{1}{c|}{\begin{tabular}[c]{@{}c@{}}Describe any methods used to synthesize results \\ and provide a rationale for the choice(s). \\ If meta-analysis was performed, describe the model(s), method(s) to identify \\ the presence and extent of statistical heterogeneity, and software package(s) used.\end{tabular}}                                             & Section \ref{methods11}, Page \pageref{methods11}         \\ \cline{2-4} 
\multicolumn{1}{|c|}{}                                   & \multicolumn{1}{c|}{13e}  & \multicolumn{1}{c|}{\begin{tabular}[c]{@{}c@{}}Describe any methods used to explore possible causes of \\ heterogeneity among study results (e.g. subgroup analysis, meta-regression).\end{tabular}}                                                                                                                                                                          & Section \ref{methods11}, Page \pageref{methods11}         \\ \cline{2-4} 
\multicolumn{1}{|c|}{}                                   & \multicolumn{1}{c|}{13f}  & \multicolumn{1}{c|}{\begin{tabular}[c]{@{}c@{}}Describe any sensitivity analyses conducted \\ to assess robustness of the synthesized results.\end{tabular}}                                                                                                                                                                                                                  & Section \ref{methods11}, Page \pageref{methods11}         \\ \hline
\multicolumn{1}{|c|}{Reporting bias assessment}          & \multicolumn{1}{c|}{14}   & \multicolumn{1}{c|}{\begin{tabular}[c]{@{}c@{}}Describe any methods used to assess risk of bias \\ due to missing results in a synthesis (arising from reporting biases).\end{tabular}}                                                                                                                                                                                       & Section \ref{methods11}, Page \pageref{methods11}         \\ \hline
\multicolumn{1}{|c|}{Certainty assessment}               & \multicolumn{1}{c|}{15}   & \multicolumn{1}{c|}{\begin{tabular}[c]{@{}c@{}}Describe any methods used to assess certainty \\ (or confidence) in the body of evidence for an outcome.\end{tabular}}                                                                                                                                                                                                         & Section \ref{methods11}, Page \pageref{methods11}         \\ \hline
\end{tabular}%
}
\end{table}

%% file: table-prisma2.tex
% \usepackage{multirow}
% \usepackage{graphicx}
\begin{table}[]
\centering
\caption{PRISMA 2020 Checklist, Cont'd.}
\label{tab:prisma2}
\resizebox{\columnwidth}{!}{%
\begin{tabular}{|cccc|}
\hline
\multicolumn{1}{|c|}{Section and Topic}                              & \multicolumn{1}{c|}{Item} & \multicolumn{1}{c|}{Checklist Item}                                                                                                                                                                                                                                                                                                                           & Location where item is reported                                                                                                                         \\ \hline
\multicolumn{4}{|l|}{RESULTS}                                                                                                                                                                                                                                                                                                                                                                                                                                                                                                                                                                                              \\ \hline
\multicolumn{1}{|c|}{\multirow{2}{*}{Study selection}}               & \multicolumn{1}{c|}{16a}  & \multicolumn{1}{c|}{\begin{tabular}[c]{@{}c@{}}Describe the results of the search and selection process, \\ from the number of records identified \\ in the search to the number of studies \\ included in the review, ideally using a flow diagram.\end{tabular}}                                                                                            & Section \ref{results16a}, Page \pageref{results16a}                                                                                                     \\ \cline{2-4} 
\multicolumn{1}{|c|}{}                                               & \multicolumn{1}{c|}{16b}  & \multicolumn{1}{c|}{\begin{tabular}[c]{@{}c@{}}Cite studies that might appear to meet the inclusion criteria, \\ but which were excluded, and explain why they were excluded.\end{tabular}}                                                                                                                                                                   & Section \ref{results16a}, Page \pageref{results16a}                                                                                                     \\ \hline
\multicolumn{1}{|c|}{Study Characteristics}                          & \multicolumn{1}{c|}{17}   & \multicolumn{1}{c|}{Cite each included study and present its characteristics.}                                                                                                                                                                                                                                                                                & \begin{tabular}[c]{@{}c@{}}\ref{appendix:summary}, Page \pageref{appendix:summary}\\ and Section \ref{results17}, Page \pageref{results17}\end{tabular} \\ \hline
\multicolumn{1}{|c|}{Risk of bias in studies}                        & \multicolumn{1}{c|}{18}   & \multicolumn{1}{c|}{Present assessments of risk of bias for each included study.}                                                                                                                                                                                                                                                                             & \begin{tabular}[c]{@{}c@{}}\ref{appendix:probast}, Page \pageref{appendix:probast}\\ and Section \ref{results17}, Page \pageref{results17}\end{tabular} \\ \hline
\multicolumn{1}{|c|}{Results of individual studies}                  & \multicolumn{1}{c|}{19}   & \multicolumn{1}{c|}{\begin{tabular}[c]{@{}c@{}}For all outcomes, present, for each study: \\ (a) summary statistics for each group (where appropriate) and \\ (b) an effect estimate and its precision \\ (e.g. confidence/credible interval), \\ ideally using structured tables or plots.\end{tabular}}                                                     & Section \ref{results19}, Page \pageref{results19}                                                                                                       \\ \hline
\multicolumn{1}{|c|}{\multirow{4}{*}{Results of syntheses}}          & \multicolumn{1}{c|}{20a}  & \multicolumn{1}{c|}{\begin{tabular}[c]{@{}c@{}}For each synthesis, briefly summarise the characteristics \\ and risk of bias among contributing studies\end{tabular}}                                                                                                                                                                                         & \begin{tabular}[c]{@{}c@{}}Section \ref{results17}, Page \pageref{results17}\\ and Section \ref{results20a}, Page \pageref{results20a}\end{tabular}     \\ \cline{2-4} 
\multicolumn{1}{|c|}{}                                               & \multicolumn{1}{c|}{20b}  & \multicolumn{1}{c|}{\begin{tabular}[c]{@{}c@{}}Present results of all statistical syntheses conducted. \\ If meta-analysis was done, present for each the summary estimate \\ and its precision (e.g. confidence/credible interval) \\ and measures of statistical heterogeneity. \\ If comparing groups, describe the direction of the effect.\end{tabular}} & \begin{tabular}[c]{@{}c@{}}Section \ref{results19}, Page \pageref{results19}\\ and Section \ref{results20b}, Page \pageref{results20b}\end{tabular}     \\ \cline{2-4} 
\multicolumn{1}{|c|}{}                                               & \multicolumn{1}{c|}{20c}  & \multicolumn{1}{c|}{\begin{tabular}[c]{@{}c@{}}Present results of all investigations of possible \\ causes of heterogeneity among study results.\end{tabular}}                                                                                                                                                                                                & Section \ref{results20b}, Page \pageref{results20b}                                                                                                     \\ \cline{2-4} 
\multicolumn{1}{|c|}{}                                               & \multicolumn{1}{c|}{20d}  & \multicolumn{1}{c|}{\begin{tabular}[c]{@{}c@{}}Present results of all sensitivity analyses conducted \\ to assess the robustness of the synthesized results.\end{tabular}}                                                                                                                                                                                    & Section \ref{methods11}, Page \pageref{methods11}                                                                                                       \\ \hline
\multicolumn{1}{|c|}{Reporting biases}                               & \multicolumn{1}{c|}{21}   & \multicolumn{1}{c|}{\begin{tabular}[c]{@{}c@{}}Present assessments of risk of bias due to missing results \\ (arising from reporting biases) for each synthesis assessed.\end{tabular}}                                                                                                                                                                       & \begin{tabular}[c]{@{}c@{}}Section \ref{methods11}, Page \pageref{methods11}\\ and \ref{appendix:funnel}, Page \pageref{appendix:funnel}\end{tabular}   \\ \hline
\multicolumn{1}{|c|}{Certainty of evidence}                          & \multicolumn{1}{c|}{22}   & \multicolumn{1}{c|}{\begin{tabular}[c]{@{}c@{}}Present assessments of certainty (or confidence) \\ in the body of evidence for each outcome assessed.\end{tabular}}                                                                                                                                                                                           & Section \ref{methods11}, Page \pageref{methods11}                                                                                                       \\ \hline
\multicolumn{4}{|l|}{DISCUSSION}                                                                                                                                                                                                                                                                                                                                                                                                                                                                                                                                                                                           \\ \hline
\multicolumn{1}{|c|}{\multirow{4}{*}{Discussion}}                    & \multicolumn{1}{c|}{23a}  & \multicolumn{1}{c|}{Provide a general interpretation of the results in the context of other evidence.}                                                                                                                                                                                                                                                        & Section \ref{discussion23a}, Page \pageref{discussion23a}                                                                                               \\ \cline{2-4} 
\multicolumn{1}{|c|}{}                                               & \multicolumn{1}{c|}{23b}  & \multicolumn{1}{c|}{Discuss any limitations of the evidence included in the review.}                                                                                                                                                                                                                                                                          & Section \ref{discussion23b}, Page \pageref{discussion23b}                                                                                               \\ \cline{2-4} 
\multicolumn{1}{|c|}{}                                               & \multicolumn{1}{c|}{23c}  & \multicolumn{1}{c|}{Discuss any limitations of the review processes used.}                                                                                                                                                                                                                                                                                    & Section \ref{discussion23b}, Page \pageref{discussion23b}                                                                                               \\ \cline{2-4} 
\multicolumn{1}{|c|}{}                                               & \multicolumn{1}{c|}{23d}  & \multicolumn{1}{c|}{Discuss implications of the results for practice, policy, and future research.}                                                                                                                                                                                                                                                           & Section \ref{discussion23d}, Page \pageref{discussion23d}                                                                                               \\ \hline
\multicolumn{4}{|l|}{OTHER INFORMATION}                                                                                                                                                                                                                                                                                                                                                                                                                                                                                                                                                                                    \\ \hline
\multicolumn{1}{|c|}{\multirow{3}{*}{Registration and protocol}}     & \multicolumn{1}{c|}{24a}  & \multicolumn{1}{c|}{\begin{tabular}[c]{@{}c@{}}Provide registration information for the review, \\ including register name and registration number, \\ or state that the review was not registered.\end{tabular}}                                                                                                                                             & Section \ref{other24a}, Page \pageref{other24a}                                                                                                         \\ \cline{2-4} 
\multicolumn{1}{|c|}{}                                               & \multicolumn{1}{c|}{24b}  & \multicolumn{1}{c|}{\begin{tabular}[c]{@{}c@{}}Indicate where the review protocol can be accessed, \\ or state that a protocol was not prepared.\end{tabular}}                                                                                                                                                                                                & Section \ref{other24a}, Page \pageref{other24a}                                                                                                         \\ \cline{2-4} 
\multicolumn{1}{|c|}{}                                               & \multicolumn{1}{c|}{24c}  & \multicolumn{1}{c|}{\begin{tabular}[c]{@{}c@{}}Describe and explain any amendments \\ to information provided at registration or in the protocol.\end{tabular}}                                                                                                                                                                                               & Section \ref{methods9}, Page \pageref{methods9}                                                                                                         \\ \hline
\multicolumn{1}{|c|}{Support}                                        & \multicolumn{1}{c|}{25}   & \multicolumn{1}{c|}{\begin{tabular}[c]{@{}c@{}}Describe sources of financial or non-financial support for the review, \\ and the role of the funders or sponsors in the review.\end{tabular}}                                                                                                                                                                 & Section \ref{other24a}, Page \pageref{other24a}                                                                                                         \\ \hline
\multicolumn{1}{|c|}{Competing interests}                            & \multicolumn{1}{c|}{26}   & \multicolumn{1}{c|}{Declare any competing interests of review authors.}                                                                                                                                                                                                                                                                                       & Section \ref{other24a}, Page \pageref{other24a}                                                                                                         \\ \hline
\multicolumn{1}{|c|}{Availability of data, code and other materials} & \multicolumn{1}{c|}{27}   & \multicolumn{1}{c|}{\begin{tabular}[c]{@{}c@{}}Report which of the following are publicly \\ available and where they can be found: \\ template data collection forms; \\ data extracted from included studies; \\ data used for all analyses; analytic code; \\ any other materials used in the review.\end{tabular}}                                        & \begin{tabular}[c]{@{}c@{}}No data collection forms, extracted data, analytic code, \\ or other review materials are publicly available.\end{tabular}   \\ \hline
\end{tabular}%
}
\end{table}

%% file: table-prismaabstract.tex
% \usepackage{graphicx}
\begin{table}[]
\centering
\caption{PRISMA 2020 Abstract Checklist}
\label{tab:prismaabstract}
\resizebox{\columnwidth}{!}{%
\begin{tabular}{|cccc|}
\hline
\multicolumn{1}{|c|}{Section and Topic}       & \multicolumn{1}{c|}{Item} & \multicolumn{1}{c|}{Checklist Item}                                                                                                                                                                                                                                                                                                                                                  & Reported (Yes/No) \\ \hline
\multicolumn{4}{|l|}{TITLE}                                                                                                                                                                                                                                                                                                                                                                                                                                                          \\ \hline
\multicolumn{1}{|c|}{Title}                   & \multicolumn{1}{c|}{1}    & \multicolumn{1}{c|}{Identify the report as a systematic review.}                                                                                                                                                                                                                                                                                                                     & Yes               \\ \hline
\multicolumn{4}{|l|}{BACKGROUND}                                                                                                                                                                                                                                                                                                                                                                                                                                                     \\ \hline
\multicolumn{1}{|c|}{Objectives}              & \multicolumn{1}{c|}{2}    & \multicolumn{1}{c|}{\begin{tabular}[c]{@{}c@{}}Provide an explicit statement of the main objective(s) \\ or question(s) the review addresses.\end{tabular}}                                                                                                                                                                                                                          & Yes               \\ \hline
\multicolumn{4}{|l|}{METHODS}                                                                                                                                                                                                                                                                                                                                                                                                                                                        \\ \hline
\multicolumn{1}{|c|}{Eligibility criteria}    & \multicolumn{1}{c|}{3}    & \multicolumn{1}{c|}{Specify the inclusion and exclusion criteria for the review.}                                                                                                                                                                                                                                                                                                    & Yes               \\ \hline
\multicolumn{1}{|c|}{Information sources}     & \multicolumn{1}{c|}{4}    & \multicolumn{1}{c|}{\begin{tabular}[c]{@{}c@{}}Specify the information sources (e.g. databases, registers) \\ used to identify studies and the date when each was last searched.\end{tabular}}                                                                                                                                                                                       & Yes               \\ \hline
\multicolumn{1}{|c|}{Risk of bias}            & \multicolumn{1}{c|}{5}    & \multicolumn{1}{c|}{Specify the methods used to assess risk of bias in the included studies.}                                                                                                                                                                                                                                                                                        & Yes               \\ \hline
\multicolumn{1}{|c|}{Synthesis of results}    & \multicolumn{1}{c|}{6}    & \multicolumn{1}{c|}{Specify the methods used to present and synthesise results.}                                                                                                                                                                                                                                                                                                     & Yes               \\ \hline
\multicolumn{4}{|l|}{RESULTS}                                                                                                                                                                                                                                                                                                                                                                                                                                                        \\ \hline
\multicolumn{1}{|c|}{Included studies}        & \multicolumn{1}{c|}{7}    & \multicolumn{1}{c|}{\begin{tabular}[c]{@{}c@{}}Give the total number of included studies and participants \\ and summarise relevant characteristics of studies.\end{tabular}}                                                                                                                                                                                                        & Yes               \\ \hline
\multicolumn{1}{|c|}{Synthesis of results}    & \multicolumn{1}{c|}{8}    & \multicolumn{1}{c|}{\begin{tabular}[c]{@{}c@{}}Present results for main outcomes, \\ preferably indicating the number \\ of included studies and participants for each. \\ If meta-analysis was done, \\ report the summary estimate and confidence/credible interval. \\ If comparing groups, indicate the direction of the effect \\ (i.e. which group is favoured).\end{tabular}} & Yes               \\ \hline
\multicolumn{4}{|l|}{DISCUSSION}                                                                                                                                                                                                                                                                                                                                                                                                                                                     \\ \hline
\multicolumn{1}{|c|}{Limitations of evidence} & \multicolumn{1}{c|}{9}    & \multicolumn{1}{c|}{\begin{tabular}[c]{@{}c@{}}Provide a brief summary of the limitations of the evidence \\ included in the review \\ (e.g. study risk of bias, inconsistency and imprecision).\end{tabular}}                                                                                                                                                                       & Yes               \\ \hline
\multicolumn{1}{|c|}{Interpretation}          & \multicolumn{1}{c|}{10}   & \multicolumn{1}{c|}{Provide a general interpretation of the results and important implications.}                                                                                                                                                                                                                                                                                     & Yes               \\ \hline
\multicolumn{4}{|l|}{OTHER}                                                                                                                                                                                                                                                                                                                                                                                                                                                          \\ \hline
\multicolumn{1}{|c|}{Funding}                 & \multicolumn{1}{c|}{11}   & \multicolumn{1}{c|}{Specify the primary source of funding for the review.}                                                                                                                                                                                                                                                                                                           & Yes               \\ \hline
\multicolumn{1}{|c|}{Registration}            & \multicolumn{1}{c|}{12}   & \multicolumn{1}{c|}{Provide the register name and registration number.}                                                                                                                                                                                                                                                                                                              & Yes               \\ \hline
\end{tabular}%
}
\end{table}